\documentclass[12pt,a4paper]{article}
\pdfoutput=1
\usepackage[utf8]{inputenc}
\usepackage[T1]{fontenc}
\usepackage{style}
\usepackage[english]{babel}
\usepackage{url}
\usepackage{footnote}
\usepackage{multicol}
\usepackage{subcaption}

\usepackage{amsmath}
\usepackage{graphicx} 
\usepackage{verbatim} 

\newcommand{\Mpc}{\text{Mpc}}

\newcommand{\fid}{{\rm fid}}

\renewcommand{\L}{\mathcal{L}}
\renewcommand{\O}{\Omega}
\renewcommand{\l}{\lambda}
\renewcommand{\d}{\partial}
\renewcommand{\a}{\alpha}

\newcommand{\be}{\begin{equation}}
\newcommand{\ee}{\end{equation}}
\newcommand{\beqa}{\begin{eqnarray}}
\newcommand{\eeqa}{\end{eqnarray}}
\newcommand{\bsm}{\begin{smallmatrix}}
\newcommand{\esm}{\end{smallmatrix}}

\newcommand\m{\mu}
\newcommand\p{{\bf p}}
\renewcommand\O{\Omega}

\newcommand\G{\Gamma}

\renewcommand\r{\rho}

\renewcommand\a{\alpha}

\renewcommand\l{\lambda}

\newcommand\x{{\bf x}}
\renewcommand\k{{\bf k}}
\newcommand\q{{\bf q}}

\newcommand{\e}{\eta}

\def\e{{\rm e}}
\def\d{\partial}
\newcommand{\bseq}{\begin{subequations}}
\newcommand{\eseq}{\end{subequations}}

\renewcommand{\ln}{\mathop{\rm ln}\nolimits}

\renewcommand{\L}{\Lambda}

\renewcommand{\k}{{\bf k}}

\newcommand{\z}{{\bf z}}

\newcommand{\eV}{\,\text{eV}}

\renewcommand{\d}{\partial}

\newcommand{\lin}{\mathrm{lin}}

\newcommand{\tot}{{\rm tot}}
\newcommand{\obs}{{\rm obs}}

\newcommand{\true}{{\rm true}}

\def\l{\left(}
\def\r{\right)}

\usepackage[textwidth=3cm,textsize=footnotesize]{todonotes} 

\vspace{-2.0cm}

\title{
Cosmological Parameters from the BOSS Galaxy Power Spectrum
}

\author[a,b]{Mikhail M. Ivanov\footnote{\texttt{mi1271@nyu.edu}}}
\author[c]{Marko Simonovi\'c\footnote{\texttt{marko.simonovic@cern.ch}}} 
\author[d]{Matias Zaldarriaga\footnote{\texttt{matiasz@ias.edu}}}

\affiliation[a]{Center for Cosmology and Particle Physics, Department of Physics,
New York University,\\
New York, NY 10003, USA}
\affiliation[b]{Institute for Nuclear Research of the
Russian Academy of Sciences, \\ 
\normalsize \it  60th October Anniversary Prospect, 7a, 117312
Moscow, Russia}
\affiliation[c]{Theoretical Physics Department, CERN,\\1 Esplanade des Particules, Geneva 23, CH-1211, Switzerland}
\affiliation[d]{School of Natural Sciences, Institute for Advanced Study,\\1 Einstein Drive, Princeton, NJ 08540, USA}

\abstract{
We present cosmological parameter measurements from the publicly
available  
Baryon Oscillation Spectroscopic Survey (BOSS) data 
on anisotropic galaxy clustering in Fourier space. 
Compared to previous studies, our analysis has two main novel features. 
First, we use a complete perturbation theory model 
that properly 
takes into account the non-linear effects of dark matter clustering, 
short-scale physics, galaxy bias, 
redshift-space distortions, and large-scale bulk flows.
Second, we employ a Markov-Chain Monte-Carlo technique and 
consistently reevaluate the full power spectrum likelihood 
as we scan over different cosmologies. 
Our baseline analysis assumes minimal $\Lambda$CDM, 
varies the neutrino masses within a reasonably tight range,
fixes the primordial power spectrum tilt,
and uses the big bang nucleosynthesis prior on the physical baryon density $\omega_b$.
In this setup, we find the following late-Universe parameters: Hubble constant $H_0=(67.9\pm 1.1)$ km$\,$s$^{-1}$Mpc$^{-1}$, matter density fraction $\Omega_m=0.295\pm 0.010$, and the mass fluctuation
amplitude $\sigma_8=0.721\pm 0.043$. 
These parameters were measured directly from the BOSS data
and independently of the Planck cosmic microwave background observations.
Scanning over the power spectrum tilt or 
relaxing the other priors do
not significantly alter our main conclusions.
Finally, we discuss the information content of the 
BOSS power spectrum and show that it is dominated by 
the location of the baryon acoustic oscillations and the 
power spectrum shape. 
We argue that the contribution of the Alcock-Paczynski effect is marginal in 
$\Lambda$CDM, but becomes important for non-minimal cosmological models.
}

\begin{document}

  \begin{minipage}{.45\linewidth}
    \begin{flushleft}
    \end{flushleft}
  \end{minipage}
\begin{minipage}{.45\linewidth}
\begin{flushright}
 {INR-TH-2019-016\\
 CERN-TH-2019-132}
 \end{flushright}
 \end{minipage}

\maketitle
\flushbottom

%%%%%%%%%%%%%%%%%%%%%%%%%%%%%%%%%%%%%%%%
\section{Introduction}
Density fluctuations traced by galaxies are an important source of information about our Universe. They can be used to probe perturbations on scales similar to those measured in the cosmic microwave background (CMB) observations, but at a very different epoch of cosmic evolution and in a very different physical environment. Future galaxy surveys with their increasingly larger volumes have a great potential to provide the most stringent tests of $\Lambda$CDM and possibly lead to new discoveries~\cite{Abell:2009aa,Dodelson:2016wal,Dore:2014cca,Laureijs:2011gra}.

The increasing precision of the large-scale structure (LSS)
surveys calls for a consistent and accurate theoretical 
modeling which is easy to implement in the data analysis pipeline. 
In this paper we focus on some aspects of this problem. 
In particular, we use a rigorous 
perturbation theory model for the redshift-space galaxy power spectrum 
(PS) to measure cosmological parameters from the publicly-available\footnote{We use the data that can be accessed via
\href{https://fbeutler.github.io/hub/hub.html}{
\textcolor{blue}{https://fbeutler.github.io/hub/hub.html}},
see also 
\href{http://www.sdss3.org/science/boss_publications.php}{
\textcolor{blue}{http://www.sdss3.org/science/boss\_publications.php}}\,.
}
Baryon Oscillation Spectroscopic Survey (BOSS) Data Release 12 (DR12) published in 2016.\footnote{We use directly the 
power spectrum multipoles provided by the BOSS collaboration. The details of the data are given in Sec.~\ref{sec:method}.
}
Similar full-shape (FS) analyses of the power spectrum multipole moments \cite{Gil-Marin:2015sqa,Beutler:2016arn} or position-space correlation function and redshift-space wedges 
\cite{Grieb:2016uuo,Satpathy:2016tct,Sanchez:2016sas} have been already done in the past. Our motivation to repeat this exercise is twofold.

First, despite significant progress in understanding the nonlinear evolution of large-scale structure and biased tracers, many recently developed theoretical tools are not routinely used in the data analysis. These new results can be roughly split into two categories. 
The first category comprises consistent perturbative descriptions developed to improve 
matter clustering modeling in the mildly nonlinear regime. 
This includes nonlinearities in the dark matter fluid \cite{Baumann:2010tm,Carrasco:2012cv}, the bias model \cite{McDonald:2009dh,Assassi:2014fva,Senatore:2014eva,Lewandowski:2014rca} (for a review see~\cite{Desjacques:2016bnm}), 
and redshift space distortions \cite{Senatore:2014vja,Perko:2016puo}. 
In all these cases one can systematically, order by order in perturbation theory, write down all independent contributions to the nonlinear density field. 
These contributions are derived using equations of motion and general symmetry arguments, such as mass and momentum conservation, and the equivalence principle. 
The functional form of these contributions is entirely fixed by these arguments,
but the amplitudes are unknown.
These contributions are related to the familiar bias parameters and less popular ``counterterms,'' whose purpose is to capture the impact of unknown small-scale physics on the long-wavelength fluctuations. 
Any consistent theoretical model has to keep all these parameters in the fit 
in order to obtain unbiased estimates of cosmological parameters.

The second category of analytical results is related to the accurate description of the baryon acoustic oscillations (BAO). It has long been known that the shape of the BAO peak is very sensitive to large displacements or bulk flows \cite{Eisenstein:2006nj,Crocce:2007dt,Sugiyama:2013gza}. 
Their effect on the density field can be significant 
since the typical displacements of galaxies are of order $\sim 10$~Mpc. 
However, the basic formulation of Eulerian Perturbation Theory \cite{Bernardeau:2001qr} 
treats bulk flows only perturbatively.\footnote{In Lagrangian Perturbation Theory (LPT) this is not the case since the bulk flows correspond to the linear displacement and they are resummed by construction. This is the reason why even in the Zel'dovich approximation the shape of the BAO peak is described rather well. For some more recent progress in modeling the BAO peak for dark matter and biased tracers in real and redshift space using models based on LPT see \cite{Vlah:2015sea, Vlah:2016bcl}. One practical disadvantage of LPT-based models is that evaluation of power spectra is numerically rather demanding.} This problem has recently been resolved in a number of works within
different but equivalent perturbation theory schemes and in various approximations~\cite{Senatore:2014via,Baldauf:2015xfa,Vlah:2015zda,Blas:2016sfa,Senatore:2017pbn,Ivanov:2018gjr,Lewandowski:2018ywf}. 
In a nutshell, large bulk flows are induced by the long-wavelength or 
``infrared'' modes, whose dominant physical effect is a simple translation of matter. 
This allows for an exact treatment beyond perturbation theory,
which was called infrared (IR) resummation \cite{Senatore:2014via}. 
Using IR-resummation the shape of the 
BAO wiggles can be predicted to very high precision (including higher order loops if necessary).
Crucially, this procedure requires no fitting parameters.
This is very different from usual, more phenomenological methods to predict the spread of the BAO peak and this difference is relevant even for the analysis of data from current surveys. 
We implement all these novel results in our theoretical model for the power spectrum.

Let us stress that the theoretical description of non-linear 
BAO damping may not be the most optimal way to extract cosmological information. 
Rather than modeling the damping of the BAO peak, one can undo this damping by 
means of BAO reconstruction at the catalog level~\cite{Eisenstein:2006nk,Padmanabhan:2008dd}. 
This procedure effectively transfers information from higher order $n$-point functions to the 2-point function. 
The standard BAO reconstruction does sharpen the BAO wiggles, but it also
introduces some distortions in the broadband part, which are hard to model analytically. 
Even though some progress towards a consistent reconstruction of the full initial density field
has recently been made~\cite{Feng:2018for,Schmidt:2018bkr,Elsner:2019rql}, the available methods have not been extensively tested for biased tracers in redshift space and we leave exploration of this direction for future work. 

Our second motivation to reanalyze the BOSS data is to perform a consistent Markov-Chain-Monte-Carlo 
(MCMC) study that samples all relevant cosmological and nuisance parameters without assuming the CMB priors.
This is not a standard practice in the FS studies, in part due to a relatively high computational cost of a 
direct numerical evaluation of perturbation theory loop integrals. 
Some exceptions are BOSS analyses of the position-space correlation function and redshift-space wedges 
\cite{Grieb:2016uuo,Satpathy:2016tct,Sanchez:2016sas} where all relevant parameters in the MCMC chains were varied,
but only in combination with the Planck CMB likelihood.
A more conventional approach to FS analysis is to 
compute the power spectrum shape for one fiducial cosmology 
and parametrize deviations from it by means of the following scaling parameters:
\be 
\label{eq:alphas}
\alpha_\parallel \equiv \frac{H_{\rm fid} }{ H_{\rm true}}\Bigg|_{z_{\text{eff}}}\frac{r_{d,\,\text{fid}}}{r_{d,\,\text{true}}} \;, 
\qquad \alpha_\perp \equiv \frac{D_{A,{\rm true}}}{D_{A,{\rm fid}}}\Bigg|_{z_{\text{eff}}}\frac{r_{d,\,\text{fid}}}{r_{d,\,\text{true}}}\,,\qquad f\sigma_8(z_{\text{eff}})\,,
\ee 
where $z_{\text{eff}}$ is the effective redshift of the data,
$r_d$ is the sound horizon at the drag epoch (which sets the BAO frequency),
$H$ is the Hubble parameter and $D_A$ the angular diameter distance,
$f$ is the logarithmic growth rate ($f\equiv d\ln D/d\ln a$, where $D$ is the linear growth 
factor and $a$ is the scale factor), $\sigma_8$ is the late-time rms mass fluctuation 
in the spheres of comoving radius $8~\Mpc/h$.
The parametrization above is motivated by the BAO studies, in which $r_d/D_A$ 
and $r_d H$ are the most relevant parameters.
However, the use of these scaling parameters 
is not entirely correct in the case of the full-shape analysis. 
To see this, let us consider a variation of the physical dark matter density $\omega_{cdm}$
with all other parameter fixed.
This variation will have an impact not only on $r_d$, 
but also on the amount of the short-scale baryon suppression
and the position of the PS peak.
This argument suggests that if the PS shape is fixed, $r_d$ must be fixed as well for consistency.
In this case the parameterization \eqref{eq:alphas} becomes a 
correct description of the Alcock-Paczynski effect \cite{Alcock:1979mp}, which does not 
assume any priors on the radial and angular distances.

A rational behind the scaling parameter analysis is that ultimately one intends to combine 
LSS and CMB data to constrain a class of non-minimal cosmological models 
that are described by the standard physics at early times but modify the late-time 
expansion, e.g. dynamical dark energy. 
% modifying late-time expansion, 
% e.g. dynamical dark energy. 
The CMB data provide us with (sub-)percent priors on the physical densities of baryons
and dark matter,
which nearly fix the PS shape in the combined analysis. 
In that case the PS complements the CMB with the geometric and distance information 
that is indeed captured by the $\alpha$-parameters in Eq.~\eqref{eq:alphas}. 
The standard analysis thus assumes strong priors on the early physics, i.e. the physical densities 
of baryons and dark matter, which are the most relevant parameters defining the power spectrum shape.
These priors will be referred to as  ``shape priors'' in what follows.

In practice, one may face situations that require a more general treatment. 
These cases include the use of different priors, 
the study of degeneracies between cosmological and nuisance parameters,
the information content of the power spectrum shape,
and exploring various extensions of the minimal $\Lambda$CDM that include physics, which is not 
captured by a simple change
of the late-time expansion and scale-independent growth factor. These include, e.g. massive neutrinos or 
models with non-standard early physics.
In all these cases the common approach becomes inadequate. 
Besides, the future LSS data may supersede Planck, which also calls for a reassessment 
of the standard analysis pipeline.

In the most general setup it would be ideal to measure cosmological parameters directly
from the shape of the observed multipoles, independently of the chosen priors. 
In that case one would have to model the whole evolution of  
perturbations for a given cosmological model in the same way as it is usually done 
in the CMB data analysis.
In this paper we analyze the BOSS data in this general way. 
This task requires a numerical routine able to generate 
theoretical templates for the non-linear spectrum quickly enough for MCMC 
parameter estimation.
A crucial ingredient to achieve this goal is a fast and reliable method for evaluating perturbation theory loop integrals. Fortunately, significant progress has recently been made in this direction \cite{Schmittfull:2016jsw, McEwen:2016fjn, Schmittfull:2016yqx, Fang:2016wcf, Simonovic:2017mhp}. 
For the purposes of our analysis we implement the FFTLog method described in~\cite{Simonovic:2017mhp} 
as a module in the publicly available code \texttt{CLASS} \cite{Blas:2011rf}.
This new module inputs the linear transfer functions computed by \texttt{CLASS} 
and calculates the multipole moments of the one-loop power 
spectrum for biased tracers in redshift space 
for a given set of cosmological parameters. 
The details of the code, performance studies, 
and tests on simulations will be presented in 
a separate publication. 
The code will soon become publicly available.

To summarize, our goal in this paper is to analyze the BOSS power spectrum data using a consistent perturbation theory model, varying all relevant bias parameters and counterterms, 
and including IR-resummation to predict the shape of the BAO wiggles properly. 
In this paper we mostly focus on base $\L$CDM and 
analyze several different priors to explore how they affect our final results.
We point out that our MCMC chains consistently include all the most important 
cosmological and nuisance parameters.

This paper is structured as follows. In Sec.~\ref{sec:results} we brief our main results.
Sec.~\ref{sec:method} summarizes our likelihood. It discusses the theoretical model, data,
covariance matrices, survey geometry, parameters and prior used in this work.
In Sec.~\ref{sec:base} we present a more detailed account of different analyses we ran
to explore the parameter space of the base $\L$CDM with massive neutrinos. 
In Sec.~\ref{sec:info}
we scrutinize the sources of information encoded in the power spectrum data.
Sec.~\ref{sec:dist} focuses on distance measurements and establishes the relation 
between our work and the methods used in the previous full-shape analyses.
There we case study $\L$CDM with shape priors and the model of dynamical dark energy. 
The study of this Section suggests that
a consistent application of the standard analysis requires an accurate
implementation of proper physical priors for a given cosmological model.  
Section~\ref{sec:concl} draws conclusions and points out directions of future research.
Some additional material is collected in Appendices. App.~\ref{app:model} contains the details 
of our theoretical model. App.~\ref{app:mocks} presents the tests of our pipeline on mock
catalogs. 
Some additional supplementary material and various tests are collected in Appendix~\ref{app:sup}.
The extended triangle plots and marginalized limits for cosmological and nuisance parameters
are presented in App.~\ref{app:full}. 
App.~\ref{app:alphas} describes our 
implementation of the standard scaling parameter analysis. 

%%%%%%%%%%%%%%%%%%%%%%%%%%%%%%%%%%%%%%%%
\section{Summary of Main Results}
\label{sec:results}

Let us briefly summarize our main results
before going into the technical details of the analysis. 
First, we test our pipeline on mock catalogs and find that our theoretical model 
can be used reliably up to $k_{\rm max}=0.25\; h/{\rm Mpc}$ with the BOSS survey covariance. 
We found that the residual modeling uncertainty 
coupled with parameter projection effects 
may bias our 1d marginalized constraints for individual parameters at most by $1\sigma$.
It should be stressed, however, that the shifts for different parameters are correlated, 
and the actual bias in the full (unmarginalized) parameter space is much lower than $1\sigma$.
This systematic error should be borne in mind when interpreting our results.

Our main analysis models four independent BOSS power spectrum datasets 
across two redshift bins ($z_{\text{eff}}=0.38,\,0.61$)
in flat $\L$CDM, marginalizing over 7 nuisance parameters for each dataset (28 in total)
and varying 5 cosmological parameters ($\omega_b,\omega_{cdm},H_0,A_s,\sum m_\nu$).\footnote{Here $\omega_b=\Omega_b h^2$ and $\omega_{cdm}=\Omega_{cdm}h^2$ stand for the physical densities of baryons 
and dark matter, respectively, $A_s$ is the amplitude of the primordial spectrum 
of scalar perturbations, $H_0$ is the present-day value of the Hubble parameter in units [km/s/Mpc],
and $\sum m_\nu$ is the sum of neutrino masses (to be quoted in eV units).
}
% We fix the primordial power spectrum tilt $n_s$ to the Planck best-fit value \cite{Aghanim:2018eyx}, 
% but this choice does not impact our main conclusions.
% We also impose two different priors on $\omega_b$ 
% from the Planck CMB measurements and big bang nucleosynthesis (BBN), 
% which yield identical results. 
% We stress that our baseline analysis assumes a fixed $n_s$
% and a narrow prior for neutrino masses (0.06,\,0.18) eV.
We stress that the baseline constraints derived in this work are model-dependent and should be interpreted in conjunction with the priors and assumptions we made:
\begin{itemize}
	\item The Universe is described by the flat $\L$CDM, i.e. 
	it has the standard thermal history and its late-time 
	expansion is controlled by the cosmological constant. 
	\item The spectrum of primordial scalar fluctuations has a simple power-law form dictated by 
	 basic inflationary scenarios. It is fully characterized by two parameters: amplitude $A_s$
	and tilt $n_s$: $P_\zeta = A_s (k/k_{\rm pivot})^{n_s}$.
	The initial conditions are assumed to be adiabatic. We fix the power spectrum tilt to the Planck best-fit value \cite{Aghanim:2018eyx}. This can also be seen as a theoretical prior motivated by inflation, which predicts that the deviations from scale-invariance must be small. 
	\item We assume an informative prior on the current physical baryon density $\omega_b$, which can be obtained either from Planck or from the BBN.\footnote{It is worth mentioning that 
	the measurement of $\omega_b$ from the shape of the CMB acoustic peaks 
	is nearly model-independent (see \cite{Audren:2012wb,Audren:2013nwa}
	and also Table 5 of Ref.~\cite{Aghanim:2018eyx}).
	It is almost not sensitive to the late-time expansion and early-time physics.}
	\item We vary the neutrino mass in the reasonably narrow range (0.06,\,0.18) eV, which is motivated  
 by particle physics\footnote{It is natural to expect that 
the individual masses are of the same order as the mass splittings inferred from 
oscillation experiments $\sim 0.05$ eV \cite{Tanabashi:2018oca}. 
Generating masses of this order of magnitude is a common benchmark
of many particle physics models, see e.g.~\cite{Abazajian:2012ys} for a review.  
% This is a standard situation is many particle physics models
% see e.g.~\cite{Abazajian:2012ys} for a review. 
% Moreover, having large masses for different eigenstates 
% with very small differences between them requires rather exotic 
} and by other cosmological measurements, e.g. of the Ly$\alpha$ forest \cite{Palanque-Delabrouille:2019iyz}.
	\end{itemize}
Results obtained beyond these base assumptions will be discussed at the end 
of this Section and in several Appendices.
In particular, in App.~\ref{app:sup} we show what 
our main conclusions, e.g. the low prediction $H_0$, hold true in the extended analyses too. 
It is important to emphasize that our baseline analysis treats $\omega_{cdm}$ as a completely free parameter,
i.e. our priors do not entirely fix the shape 
of the matter power spectrum.
This can be contrasted with the previous full-shape studies, which kept the shape totally fixed. 

In principle, the priors on $\omega_b$, $n_s$, and $\sum m_\nu$ are not necessary for our analysis.
However, given that the BOSS data are not very sensitive to these parameters, 
we prefer to fix, or nearly fix them by priors, which are ultimately CMB-motivated.
This is reasonable keeping in mind an eventual combination of BOSS with other cosmological probes
in order to pin down one correct model that would explain all the observed phenomena in the Universe.

% We have checked that our
% constraints do not strongly rely on these priors and in principle, we could waive them.
% These extensions are discussed in Sec.~\ref{sec:base}
% and App.~\ref{app:tilt}.
% However, we prefer to stick to these priors realistically keeping in mind an eventual combination 
% of our BOSS likelihood with Planck. 

\begin{figure}[h!]
% \centering
% \begin{minipage}{1.0\textwidth}
% \centering
  % \begin{subfigure}{\textwidth}
\includegraphics[width=0.52\textwidth]{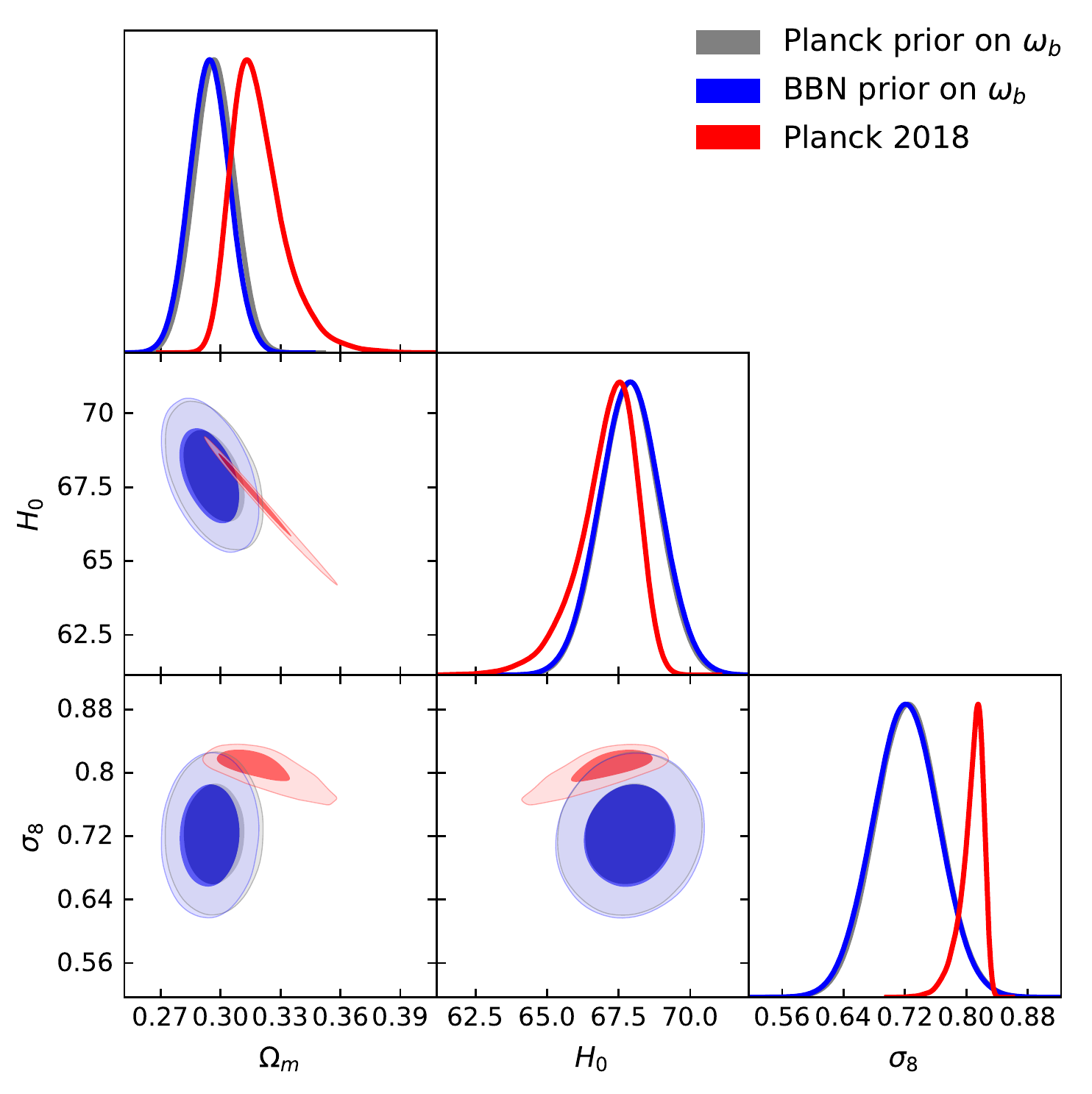}
 % \end{subfigure}
   % \end{minipage}
  \hspace{0.6cm}
 \begin{minipage}{1.0\textwidth}
 % \centering
 \vspace{-7.5cm}
  \begin{subfigure}{\textwidth}
\includegraphics[width=0.4\textwidth]{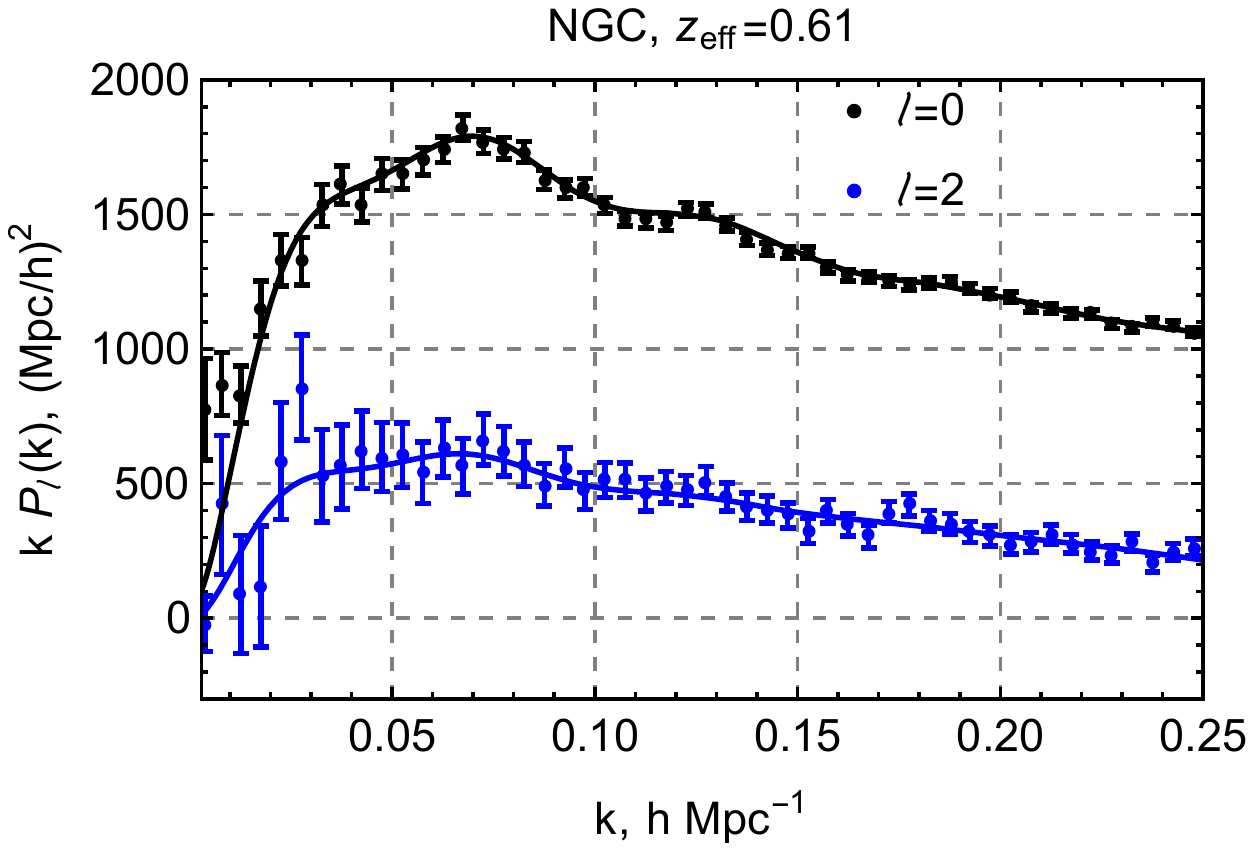}\vfill
\includegraphics[width=0.4\textwidth]{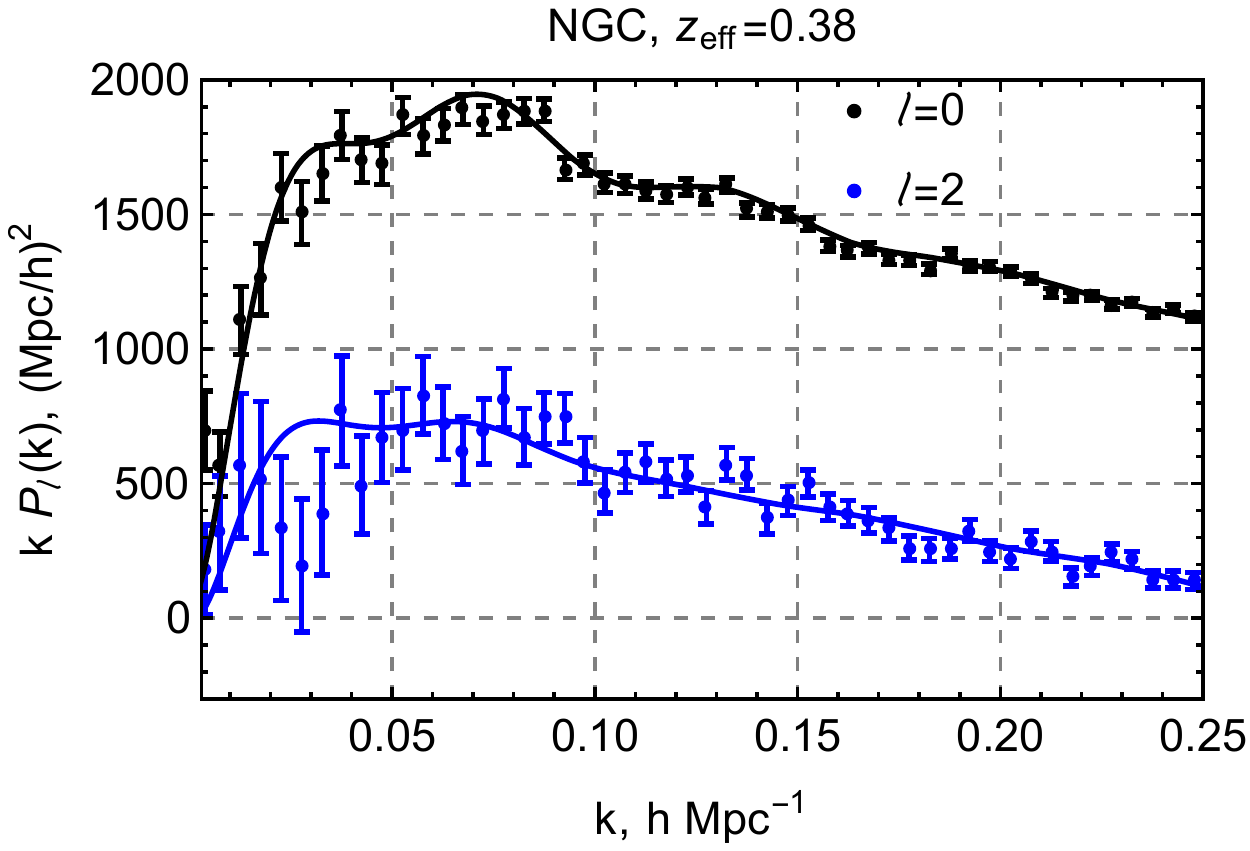}
 \end{subfigure}
 \end{minipage}
\caption{
\textit{Left panel}: The posterior distribution for the late-Universe parameters $H_0,\O_m$
and $\sigma_8$ obtained with priors on $\omega_b$ from Planck (gray contours) and BBN (blue contours). 
For comparison we also show the Planck 2018 posterior (red contours)
for the same model (flat $\L$CDM with 
massive neutrinos).
\textit{Right panel}: The monopole (black dots) and quadrupole (blue dots) 
power spectra moments of the BOSS data 
for high-z (upper panel) and low-z (lower panel) 
north galactic cap (NGC) samples, along with the best-fit theoretical model curves.
The corresponding best-fit theoretical spectra 
are plotted in solid black and blue.
$H_0$ is quoted in units [km/s/Mpc].
}    
\label{fig:final}
\end{figure}
\begin{table}[ht!]
\begin{center}
\begin{tabular}{|c|c|c|} 
 \hline
BBN $\omega_b$ & best-fit & mean  $\pm 1\sigma$  \\ [0.5ex] 
 \hline\hline
 $\omega_{cdm}$  & $0.1117$ & $0.1113\pm 0.0046$  \\ \hline
 $H_0$  & $67.93$ & $67.89\pm 1.06$  \\ \hline \hline
    $\Omega_{m}$ & $ 0.2950$ & $0.2945\pm 0.0100$  \\ \hline
 $\sigma_8$ & $ 0.722$ & $ 0.721 \pm 0.043$  \\ \hline
 $f\sigma_8(z_{\text{eff},\,1})$ & $ 0.431$ & $ 0.434 \pm 0.038$  \\ \hline
  $f\sigma_8(z_{\text{eff},\,2})$ & $ 0.393$ & $ 0.394 \pm 0.034$  \\ \hline
\end{tabular}
\begin{tabular}{|c|c|c|} 
 \hline
Planck 2018& best-fit & mean  $\pm 1\sigma$  \\ [0.5ex] 
 \hline\hline
 $\omega_{cdm}$  & $0.1197$ & $0.1201 \pm 0.0013$  \\ \hline
 $H_0$  & $68.03$ & $67.1^{+1.2}_{-0.67}$  \\ \hline \hline
   $\Omega_{m}$ & $ 0.3071$ & $0.3191^{+0.0085}_{-0.016}$  \\ \hline
  $\sigma_8$ & $ 0.8224$ & $0.807^{+0.018}_{-0.0079}$  \\ \hline
    $f\sigma_8(z_{\text{eff},\,1})$ & $ 0.4769$ & $ 0.4766^{+0.0062}_{-0.0053}$  \\ \hline
       $f\sigma_8(z_{\text{eff},\,2})$ & $ 0.4714$ & $0.4689^{+0.0070}_{-0.0045}$  \\ \hline
\end{tabular}
\caption{
The results of our analysis for the combined likelihood with the BBN prior on $\omega_b$
(left panel). 
For comparison we also show the results from the final Planck data release \cite{Aghanim:2018eyx}
(right table)
for the same cosmological model 
as used in our analysis 
($\L$CDM with varied neutrino masses).
Note that the first two parameters were used as actual sampled parameters in our chains, 
while the 
last four are derived from them and from other parameters, which we do not display here 
(see Sec.~\ref{sec:method} and App.~\ref{app:full} for the full set of sampled parameters and corresponding limits). 
The effective redshifts of the samples are
$z_{\text{eff},\,1}=0.38$ and $z_{\text{eff},\,2}=0.61$. 
% The top group of two rows are the cosmological parameters which were sampled in the MCMC analysis
% with flat uninformative priors. The other parameters are the derived ones. We show only the cosmological parameters whose limits are not driven by the priors. Full tables with all parameters can be found in App.~\ref{app:full}.
}
 \label{tab:final}
\end{center}
\end{table}

The outcome of our analyses is shown in Fig.~\ref{fig:final}, 
where we display the final triangle plot (left panel) and best-fit spectra
for two BOSS data samples with the biggest volume\footnote{These are high-z and low-z north galactic cap (NGC) samples.} (right panel).
The inferred cosmological parameters are given in Table~\ref{tab:final}.
We have chosen to present the parameters $H_0$, $\Omega_m$ and $\sigma_8$ 
as our main results
because they are more common in the LSS literature and because they are close 
to the actual principal components of the BOSS data.

Our constraints on $\Omega_m$ and $H_0$ are competitive with the Planck measurements
for the same cosmological model with varied neutrino masses.\footnote{There are several caveats that should be mentioned at this point. First, we approximate the neutrino sector with one massive eigenstate, which should be contrasted with the approximation of three degenerate 
eigenstates used in Planck 2018. The difference between these two approaches is a few percent at the matter power spectrum level, and hence can be neglected for our purposes. Second, the Planck Legacy contours
that we show roughly correspond to the variation of the total neutrino mass in the range ($0-0.24$) eV, which is somewhat different from our prior ($0.06-0.18$) eV. 
However, the effect of weighting the Planck posterior with our prior on $\sum m_\nu$ is marginal. 
We show the original Planck contours for clarity.} Moreover, the use of the full parameter likelihood adopted in this work allows for a clear comparison 
between the two experiments at the level of the fundamental $\L$CDM parameters. 
Our measurement of $H_0$ is driven by the geometric location of the BAO peaks, whereas 
the limits on $\Omega_m$ result from the combination of both the geometric (distance) and shape information. 
$\sigma_8$ is measured through redshift-space distortions. 
We performed several tests to ensure that our constraints are saturated with these three effects, 
and confirmed that distance ratio measurements implemented through the Alcock-Paczynski effect
can only marginally affect the cosmological parameters of $\Lambda$CDM.
However, the situation changes in its extensions that modify the late-time evolution, 
in which the Alcock-Paczynski effect
becomes a significant source of information to constrain the parameters of these models.

In order to explore the relation with the previous works on the galaxy power spectrum we ran 
an analysis with very tight shape priors and obtained essentially the same results as in
Tab.~\ref{tab:final}. However, in that case $\Omega_m$ cannot be viewed as an independently
measured parameter, since the shape priors completely 
fix the relation between $\Omega_m$ and $H_0$ in $\Lambda$CDM.
This suggests that the shape priors are 
not necessary for the parameter estimation from the BOSS data. 
Moreover, the power spectrum shape itself can be a source of  
independent measurements of $\omega_{cdm}$ and $\Omega_m$, whose precision rivals that 
of the Planck CMB data. This happens because of two main reason. 
First, the parameter $\omega_{cdm}$ can be measured directly from the shape of the galaxy 
power spectrum with $5\%$ precision. 
Second, the degeneracy direction corresponding to the angular acoustic scale of 
the galaxy power spectrum happened to be more orthogonal to $H_0$ 
than the angular acoustic scale of the CMB. 
Thus, even though the later is measured with Planck much 
more precisely than the former one, their projections onto the $H_0$ plane happened to be 
comparable. In Section \ref{sec:info} we give some further details on this effect.

Our results agree with the Dark Energy Survey (DES) data on weak lensing and 
photometric galaxy clustering \cite{Abbott:2017wau}. 
The combination best constrained by DES $S_8=\sigma_8(\O_m/0.3)^{0.5}=0.773^{+0.026}_{-0.020}$
is within 2$\sigma$ of our limit \mbox{$S_8=0.703\pm 0.045$}.

Let us comment on the neutrino masses. 
Our analysis shows that the BOSS data itself can only rule out very large neutrino masses \mbox{$\sim 1$ eV}, which produce
significant scale-dependent modifications to the matter power spectrum.
These modifications are not degenerate with effects of other cosmological and nuisance parameters.
Smaller neutrino masses cannot be constrained with the BOSS data 
mainly because of the degeneracy between galaxy bias and $\sum m_\nu$, which persists 
even if we use the Planck 2018 prior on $A_s$. 
Naively, this degeneracy may be broken 
by the quadrupole moment, but its large statistical error
along with the strong sensitivity to the finger-of-God uncertainties  
do not allow us to derive constraints on the neutrino mass that could be competitive with 
the CMB data.
Note that the degeneracy between $A_s$, $\sum m_\nu$
and the galaxy bias 
can, in principle, be alleviated by the bispectrum data \cite{Gil-Marin:2014sta,Gil-Marin:2016wya}.

% $\sum m_\nu$, and $A_s$ that roughly 
% correspond to a fixed late-time fluctuation amplitude $\sigma_8$.
% This suggests that the BOSS
% data cannot give satisfactory measurements without an input from
% external datasets. 
% However, even if we use the Planck 2018 prior on $A_s$,
% the neutrino masses cannot be 

% mainly due to the degeneracy between $\sigma_8$
% and the galaxy bias, which can be broken if we use, e.g. the Planck measurement of $\sigma_8$.

To estimate results from a combined Planck + BOSS likelihood 
we analyzed the BOSS data with a multi-variate Gaussian prior 
on all cosmological parameters of the minimal $\L$CDM (not including the neutrino mass)
from the final Planck data release \cite{Aghanim:2018eyx}.
We obtained the following limit from the combination of the two biggest BOSS data samples:
\be
\label{eq:mnu}
\begin{split}
% \text{high-z sample:} \qquad & m_{\nu} < 1.00 \,\eV \qquad (95\%\,\text{C.L.})\,,\\
% \text{low-z sample:} \qquad & m_{\nu} < 1.52 \,\eV \qquad (95\%\,\text{C.L.})\,.
\sum m_{\nu} < 0.84 \,\eV \qquad (95\%\,\text{CL})\,.
\end{split}
\ee
This suggests that the BOSS data can only improve the current neutrino mass bounds 
by breaking degeneracies internal to the CMB data (e.g. the degeneracy between
$m_\nu$ and $H_0$), and not by actually 
observing the free-streaming short-scale suppression 
of the galaxy power spectrum \cite{Lesgourgues:2006nd}.
It would be curious to see if the full-shape BOSS + Planck data 
will give better constraints than the Planck + BAO likelihood.

Finally, we tested some simple extensions of our baseline analysis, which assumes 
a BBN prior on $\omega_b$ and fixes the power spectrum tilt $n_s$. 
To that end we explored the whole likelihood with all 
relevant parameters \mbox{($\omega_b,\omega_{cdm},n_s,H_0,A_s,\sum m_\nu$)}. 
We have found that 
varying the tilt in chains with the BBN priors on  $\omega_b$
degrades the $\Omega_m$ constraint but does not significantly alter the $H_0$ and $\sigma_8$
limits.
Moreover, one can
obtain a constraint on $n_s$, which is independent of the Planck CMB data,
\be 
n_s = 0.88\pm 0.08\,.
\ee
Instead, if we keep $n_s$ fixed but use a very wide prior on $\omega_b$, 
the constraints on $H_0$ worsen by a factor of two, but the limits on $\Omega_m$ and $\sigma_8$ remain essentially intact.
This suggests that our main conclusions are stable w.r.t. different prior 
choices.

%%%%%%%%%%%%%%%%%%%%%%%%%%%%%%%%%%%%%%%%%%%%
\section{Methodology and Likelihood}
\label{sec:method}

In this Section we discuss technical aspects of our analysis: the theoretical model,
window function treatment, covariance matrices and model parameterization.

\subsection{Theoretical Model} 
Our model for multipole moments of the redshift-space
galaxy power spectrum is based on 
one-loop perturbation theory. 
Schematically, it can be written as a sum of four pieces,\footnote{We use the following convention: 
$ \langle \delta_\k \delta_{\k'}\rangle = (2\pi)^3 P(k)\delta_{\text{D}}^{(3)}(\k+\k')$, 
where we introduced the density (contrast) field $\delta\equiv \rho(\x,t)/\bar{\rho}(t)-1$
($\rho$ and $\bar{\rho}$ are the local and background densities, respectively), and $\langle...\rangle$
denotes the averaging over the cosmological ensemble. 
}   
\be
\label{eq:model}
P_{g,\ell}(k) = P_{g,\ell}^{\rm tree}(k) + P_{g,\ell} ^{\rm 1-loop} (k) + P_{g,\ell}^{\rm noise} (k) + P_{g,\ell}^{\rm ctr}(k)\;.
\ee
In this work we limit ourselves to the monopole and quadrupole moments ($\ell=0,2$).
All multipoles are computed from the 2D anisotropic galaxy power spectrum $P_g(k,\mu)$,
\be
\label{eq:mult}
P_{g,\ell}(k) \equiv \frac{2\ell+1}{2} \int_{-1}^1 d\mu \; P_g(k,\mu) \mathcal P_\ell(\mu) \;,
\ee
where $\mu \equiv \hat\k \cdot \hat \z$ is cosine of the angle between a Fourier mode $\k$ and the line-of-sight direction $\hat \z$, whereas $\mathcal P_\ell(\mu)$ are Legendre polynomials of order $\ell$. For example, the tree-level contribution to the multipoles $P_{g,\ell}^{\rm tree}(k)$ are given by the Kaiser formula \cite{Kaiser:1987qv},
\be
P_g^{\rm tree} (k,\mu) = (b_1+f\mu^2)^2 P_{\rm lin}(k) \;,
\ee
where $b_1$ is the scale-independent linear bias coefficient. 
For compactness, we suppress
explicit time dependence in all formulas of this section 
assuming that all relevant quantities are evaluated at the effective redshift $z_{\text{eff}}$
of a given data sample. 
For clarity, all the expressions of this section are presented 
without IR-resummation and the Alcock-Paczynski effect,
which are properly taken into account, 
see Appendix~\ref{app:model} for more detail.

The next important ingredient of our analytic model 
is one-loop corrections $P_{g,\ell} ^{\rm 1-loop} (k)$ that encapsulate 
the non-linear redshift-space mapping along with
non-linearities due to dark matter clustering and bias.
This model has been described in detail in Refs.~~\cite{Perko:2016puo,delaBella:2018fdb,Chudaykin:2019ock} 
and is summarized in Appendix~\ref{app:model}. 
We use the following basis of bias operators\footnote{As pointed out in~\cite{Senatore:2014eva,Mirbabayi:2014zca} the evolution of biased tracers is non-local in time, which leads to appearance of bias operators that cannot be written in terms of tidal tensor $\partial_i\partial_j\Phi$ at a
finite time slice. However, these operators appear only at fourth order in perturbation theory and this important subtlety is not relevant for the one-loop power spectrum that we consider.} 
\be
\delta_g = b_1 \delta + \frac{b_2}{2} \delta^2 + b_{\mathcal G_2} \mathcal G_2 \;,
\ee
where $\delta$ is the nonlinear matter density field and the Fourier representation of 
the tidal field operator $\mathcal G_2$ is given by
\be
\mathcal G_2 (\k) = \int \frac{d^3\p}{(2\pi)^3} \left[ \frac{(\p\cdot(\k-\p))^2}{p^2|\k-\p|^2} - 1 \right] \delta_{\rm lin}(\p) \delta_{\rm lin}(\k-\p) \;,
\ee
where $\delta_{\rm lin}$ is the linear theory density field.
Note that there is one extra bias parameter that contributes 
to the one-loop power spectrum, $b_{\Gamma_3}$. We have found that this parameter is 
very degenerate with other nuisance parameters and the BOSS data are not accurate enough 
to break this degeneracy. For the purposes of this paper we have fixed it to zero. 
This choice still allows for a sufficient freedom in the parameter space exploration. 
We have checked that fixing $b_{\Gamma_3}$ or varying it within some priors has no effect on the cosmological parameter estimates.

The stochastic contribution 
is modeled as a simple Poisson shot noise 
with the constant power spectrum in Fourier space and a free amplitude. Note that in the absence of the window function only the monopole moment
has a constant shot noise power, i.e.
\be
P_{g,0}^{\rm noise} (k) = P_{\text{shot}}\,,\qquad P_{g,2}^{\rm noise} (k) = 0\,.
\ee

Finally, the last part of our model are 
the so-called ultraviolet (UV) counterterms $P_{g,\ell}^{\rm ctr}(k)$. 
The counterterms were not included in theoretical models used in the previous data analyses. 
For this reason, we discuss them in more detail here. 
The purpose of the counterterms is to fix the dependence of the one-loop power spectrum on the complicated unknown short-scale physics, 
which cannot be modeled by means of perturbation 
theory.
To understand qualitatively why these corrections are needed let us note that a part of the loop integral comes from integrating over high-$k$ Fourier modes for which perturbation theory does not apply. 
This means that results of loop calculations are necessarily wrong, 
even though they converge to some finite values. 
For the theory to be consistent, 
there must be counterterms to cancel the spurious UV-dependence. 
Besides, standard perturbation theory does not 
correctly capture the backreaction of 
short scales on long-wavelength fluctuations.
These effects are taken into account by the so-called ``finite'' part of the UV counterterms,
which describes physical effects missing in standard perturbation theory.
Since the loop integrals converge for the $\L$CDM linear power spectrum,
there is no practical need to distinguish between these two physically different parts 
of the counterterms.
Hence, every counterterm can be parametrized by a single free coefficient to be fitted 
from the data. 
Note that the scale-dependence of the counterterms is not free. 
It is fully fixed by symmetry arguments at any order in perturbation theory. 
This statement holds true for pure dark matter~\cite{Carrasco:2012cv}, dark matter halos~\cite{Senatore:2014eva,Desjacques:2016bnm}, and galaxies in redshift space~\cite{Senatore:2014vja,Perko:2016puo}. 
  
At first non-trivial order in the gradient and field power expansion
there are two counterterms needed for the one-loop monopole and quadrupole moments~\cite{Senatore:2014vja,Perko:2016puo}, which can be cast in the following form:
\be
\label{eq:LO_counter}
\begin{split}
P_\ell^{\rm ctr, LO}(k) & \equiv -2\,c^2_\ell \, k^2 \, P_{\rm lin}(k) \;,\qquad \ell=0,2\,.
% P_2^{\rm ctr,LO}(k)  & \equiv -2\, c^2_2\, k^2 \, P_{\rm lin}(k) \,.
\end{split}
\ee
The reason to keep two different free coefficients is that they fix different loops and 
capture different physical effects. 
For instance, the monopole counterterm includes the contribution of 
the higher-derivative bias term $b_{\nabla^2} \nabla^2\delta$, which is absent for higher moments.
This should be contrasted with the quadrupole counterterm, which is dominated by 
the fingers-of-God effect \cite{Jackson:2008yv}.
Indeed, neglecting other nonlinearities, the $c^2_2$-contribution 
can be related to the short-scale galaxy velocity dispersion $\sigma_v^2$,
\be
c_2^2 = \frac{f(5f^2+12f b_1 +7 b_1^2)}{14} \sigma_v^2 \approx 2.5 \, \sigma_v^2 \,,
\ee
where we assumed $b_1=2$ and $f=0.75$ typical for the high-z BOSS sample. 
This formula is derived by expanding the velocity field into the short and long-wavelength
contributions and averaging the redshift-space power spectrum over the short-scale modes, 
\be
\label{eq:P^FoG}
P^{\rm FoG}(k,\mu) \approx -(\mu f k \sigma_v)^2 P_g^{\rm tree} (k,\mu) + \text{higher orders}\,,
\ee
which is then matched to our expression for $P_2^{\rm ctr,LO}(k)$. Note that a similar expression
can be obtained upon Taylor-expanding some simple phenomenological models 
for the fingers-of-God with a Gaussian or Lorentzian damping, e.g. \cite{Hand:2017ilm,Beutler:2016arn,Gil-Marin:2015sqa}.
The typical value for the velocity dispersion
for the BOSS-like sample $\sigma_v \sim 5$ Mpc$/h$ yields $c_2^2 \sim 60 \, {\rm Mpc}^2/h^2$. 
We emphasize that this is just a simple order-of-magnitude estimate 
and that the true amplitude (and even the sign) of the counterterms cannot be predicted. 
  
 The one-loop perturbation theory model \eqref{eq:model} is sufficient to describe the statistics of biased tracers in real space up to $k_{\rm max} = 0.3\; h/{\rm Mpc}$
 for the volume and redshifts typical to the BOSS survey~\cite{Schmittfull:2018yuk}. 
 While two-loop contributions due to dark matter clustering may be sufficiently small, 
 the mapping from real to redshift space can produce significant 
 correction to the one-loop result because 
 of higher order short-scale velocity cumulants, whose characteristic momentum 
 scale $\sigma_v^{-1}$
 can be significantly lower than the non-linear scale $k_{\text{NL}}$ controlling 
 gravitational non-linearities.
This implies that the usual one-loop power spectrum model \cite{Senatore:2014vja,Perko:2016puo} 
is not sufficient for an accurate description of the data even on large scales.
One option to get around is to use some phenomenological model for the fingers-of-God. 
However, these models are not derived from first principles 
and can introduce uncontrollable biases in cosmological parameter estimations.
To proceed, we choose a different strategy which fits the spirit of perturbation theory.
We introduce an additional counterterm to capture the redshift space non-linearities at next-to-leading order (NLO),
\be
\label{eq:NLO}
P^{\rm ctr, NLO}(k,\mu)  \equiv \tilde{c}\, k^4\, \mu^4 \, f^4 \, (b_1+f\mu^2)^2 P_{\rm lin}(k) \,.
\ee 
Let us discuss the form of this expression. 
As argued above, the non-linear scale for the 
velocity dispersion $\sim \sigma^{-1}_v$ is smaller than 
the dark matter nonlinear scale $k_{\rm NL}$,
but the stochastic velocity 
field couples with the large-scale density dominantly along the line-of-sight.
Thus, the redshift-space mapping effectively generates an expansion in powers of $(\mu k \sigma_v)^2$.
The standard one-loop counterterms in Eq.~\eqref{eq:LO_counter} correspond to 
the term $\nabla_z^2 \delta(k,\mu)$ in this expansion.
From this point of view, the NLO counterterm in Eq.~\eqref{eq:NLO} 
can be naturally viewed as a next-to-leading term in this expansion, 
i.e. $\nabla_z^4 \delta(k,\mu)$ contribution. 

It should be stressed that the main objective of introducing the new counterterm \eqref{eq:NLO}
is to capture the NLO sensitivity to fingers-of-God.
The contributions from other physical effects (higher-derivative bias etc.) 
are expected to be sub-dominant since they
have the same order of magnitude as the two-loop corrections to the real-space matter density.
Thus, they can be neglected at the one-loop order that we use in this paper.
Given this reason, we choose the NLO contribution \eqref{eq:NLO}
to be universal for all multipole moments, as expected from the redshift-space mapping. 

Another way to understand role of the NLO counterterm is to view it as a simple model for the theoretical error. Marginalizing over the amplitude $\tilde{c}$, we are marginalizing over the estimated uncertainty due to the fingers-of-God modeling. 
While in principle a more elaborate procedure is needed to ensure that the 
results of the analysis are unbiased~\cite{Baldauf:2016sjb}, this simple prescription is sufficient
given the BOSS survey volume.
  
In summary, our model
for the power spectrum is based on one-loop perturbation theory for galaxies in redshift-space
supplemented with LO and NLO counterterms. 
It includes seven free nuisance parameters: three bias coefficients ($b_1,b_2,b_{\mathcal{G}_2}$), 
three redshift-space counterterms ($c_0^2,c_2^2,\tilde{c}$) and 
the shot noise amplitude $P_{\text{shot}}$.

\subsection{Power Spectra and Covariance Matrices}
\label{sec:data}

The BOSS survey has measured the spectroscopic redshifts of 1 198 006 galaxies using the SDSS multi-fibre spectrographs and multi-color SDSS imaging (see~\cite{Alam:2016hwk} and references therein). The BOSS-DR 12 galaxy sample spans over the redshift range $0.2<z<0.75$. 
The data include four different selections: LOWZ, LOWZE2, LOWZE3, CMASS.
They are combined into two non-overlapping redshift bins with $z_{\text{eff}}=0.38$ and $z_{\text{eff}}=0.61$. 
Each redshift bin sample is additionally divided into two sub-samples
depending on the Galactic hemisphere where the galaxies are observed.
These are called ``South and North Galactic Cap'' (SGC and NGC). 
To avoid confusion with the previous selections analyzed, e.g.~in~\cite{Gil-Marin:2015sqa}, we will call the two redshift bins simply ``low-z'' and ``high-z''. 
Note that each of the four data chunks has a different selection function and therefore represents 
a different galaxy population \cite{Alam:2016hwk}. 
The comoving and effective volumes of the BOSS data samples are listed in Table~\ref{tab:volume}.
To obtain these numbers, the observed angles and redshifts were converted into comoving distances
assuming the following fiducial parameters: \mbox{$h=0.676$, $\O_m=0.31$},
which were also used to create galaxy catalogs.\footnote{Throughout this paper we will use $h$ and the present day Hubble parameter 
\mbox{$H_0=h\cdot 100$ km s$^{-1} $Mpc$^{-1}$} interchangeably.}
Any departure of the real cosmology from the fiducial one is accounted for by explicitly including the Alcock-Pazcynski effect
in our theoretical model.
The mean number density of each sample is approximately~$\bar n\sim 3\times 10^{-4} \,(h/{\rm Mpc})^3$, implying that the shot noise is not a dominant contribution to the galaxy power spectrum on 
the mildly non-linear scales.

We use the redshift space power spectrum monopole ($\ell=0$) and quadrupole ($\ell=2$) of the publicly available data from BOSS DR12. The spectra are binned with the bin size~$\Delta k = 0.005~h/{\rm Mpc}$ in the wavenumber range~$[0.0025,0.25]~h/{\rm Mpc}$. Our baseline analysis is performed for~$k_{\rm max}=0.25\,h/$Mpc, which contains~50~$k$-bins in each multipole. 
We have checked that our method can recover the correct cosmology from 
mock catalogs for this choice of $k_{\text{max}}$
(see Appendix~\ref{app:mocks}).

\begin{table}
\begin{center}
 \begin{tabular}{|c c c|} 
 \hline
Data \quad & $\quad$ $V_{\text{eff}}$ $[(\text{Gpc}/h)^3]$  & $\quad$ $V$ $[(\text{Gpc}/h)^3]$ $\quad$ \\ [0.5ex] 
 \hline\hline low-z NGC\quad & $0.84$ &  $1.46$ \\ 
 \hline low-z SGC \quad & $0.31 $ & $0.53$  \\
 \hline high-z NGC\quad & $0.93$  & $2.8$ \\
 \hline high-z SGC \quad & $0.34$ & $1.03$  \\
 \hline
\end{tabular}
\caption{Effective and comoving volumes for four independent samples of BOSS DR12.}
\label{tab:volume}
\end{center}
\end{table}
  
\vskip 5pt
\noindent 
{\bf Window function.} We incorporate the effects of the survey geometry following the procedure described in~\cite{Beutler:2016arn}. 
The theory multipoles are first transformed to position space via
\be 
\label{eq:xi}
\xi_\ell (r)=i^\ell \int \frac{dk\,k^2}{2\pi^2}j_\ell (kr)P_\ell (k)\,,
\ee
and then the corresponding correlation function multipoles 
are convolved with the appropriate window functions,
\be 
\begin{split}
& \hat{\xi}_0(r)=\xi_0 W_0^2(r)+ \frac{1}{5}\xi_2(r)W_2^2(r)\,,\\
& \hat{\xi}_2(r)=\xi_0 W_2^2(r) +\xi_2(r)\left[W_0^2(r)+\frac{2}{7}W_2^2(r)\right]\,.
\end{split}
\ee
The windowed power spectrum multipoles 
are then simply obtained by means of an inverse Fourier transform,
\be
\label{eq:Pi}
\hat P_\ell(k)=(-i)^\ell (4\pi)\int dr\,r^2 j_\ell (kr)\hat{\xi}_\ell (r)\,.
\ee
The integrals in Eqs.~(\ref{eq:xi}) and~(\ref{eq:Pi}) are computed with the FFTLog method~\cite{Hamilton:1999uv}.

\vskip 5pt
\noindent 
{\bf Covariance matrix.}  We extract the covariance matrix from \textsc{patchy} mock catalogs, which are described in detail in Ref.~\cite{Kitaura:2015uqa}. The \textsc{patchy} algorithm is based on extended Lagrangian perturbation theory and a stochastic halo biasing scheme calibrated on high-resolution N-body MultiDark simulations run for a $\Lambda$CDM cosmology with the following fiducial parameters:
\be
\begin{split}
&\Omega_m=0.307115\,,\quad \Omega_b=0.048206\,,\quad h=0.6777\,,\\
&\sigma_8 = 0.8288\,,\quad n_s = 0.9611\,.
\end{split} 
\ee
The \textsc{patchy} algorithm uses halo occupation distribution (HOD) to construct catalogs which match the BOSS galaxy clustering and its redshift evolution. The \textsc{patchy} mocks were generated for every data chunk separately. In each case, they were assigned the same selection function and survey geometry as the real data.

We are using the covariance matrix extracted from the corresponding mocks,
\be 
C^{(\ell \ell')}_{ij}=\frac{1}{N_m-1}\sum_{n=1}^{N_m}
\left[P_{\ell,n}(k_i) - \bar{P}_{\ell}(k_i)\right]
\left[P_{\ell',n}(k_j) - \bar{P}_{\ell'}(k_j)\right]\,,
\ee
where $N_m=2048$ is the number of mock catalogs and $\bar{P}_{\ell}(k)$ is the mean power spectrum,
\be 
\bar{P}_{\ell}(k)\equiv\frac{1}{N_m}\sum_{n=1}^{N_m}P_{\ell,n}(k) \;.
\ee
In our analysis we neglect the Hartlap factor correction \cite{Hartlap:2006kj} which affects the covariance matrix at the level of~$\sim 1\%$. For simplicity we will also defer from the standard practice of rescaling the parameter variances to account for the difference between the extracted values and the ones used in the mock catalogs \cite{Percival:2013sga}. A more accurate treatement of the covariance matrix would require its recalculation for the best-fit cosmology, 
which can be done analytically along the lines of \cite{Li:2018scc}.

\subsection{Parameters and Priors} 

In all our analyses for the base flat $\Lambda$CDM we vary 5 cosmological and 7 nuisance parameters:
\be
\label{eq:set_of_priors}
(\omega_b,\omega_{cdm},h,A^{1/2},\sum m_\nu)\times (b_1 A^{1/2},b_2 A^{1/2},b_{\mathcal{G}_2} A^{1/2},P_{\text{shot}},c_0^2,c_2^2,\tilde{c})\,, 
\ee
where $m_\nu$ is the sum of neutrino masses, $A$ is defined as 
\be
A\equiv \frac{A_s}{A_{s,\,\text{Planck}}} \,,
\ee
and $A_{s,\,\text{Planck}}=2.099\cdot 10^{-9}$. 
Since each BOSS data sample has its own selection function, we allow biases, $P_{\rm shot}$ and counterterms for each data chunk to be different. 

Let us discuss the choice of parameters and the corresponding priors. First, the initial conditions 
for fluctuations
are described by two parameters, the amplitude of the power spectrum $A$ and the spectral index $n_s$. The BOSS data can constrain the amplitude at~$\mathcal O(10\%)$ level and the tilt cannot be measured with a reasonable accuracy. For this reason we fix the spectral index to be
\be
\label{eq:ns}
n_s=0.9649 \;,
\ee 
as measured by Planck~\cite{Aghanim:2018eyx}, and we do not vary it in the MCMC chains. This is why this parameter does not appear in~\eqref{eq:set_of_priors}. 
Since we cannot probe the amplitude of the primordial power spectrum accurately,
our eventual results are not very sensitive to variations of the fiducial value of $n_s$
in a reasonable range around $n_s=1$. 
In particular, all main results of our study would remain the same had we chosen the flat 
Harrison-Zel'dovich spectrum 
instead of \eqref{eq:ns}.
% \footnote{Alternatively, one could impose some tight priors on the tilt
% and marginalize the posterior over its value. 
% This approach was adopted in the analysis of the Dark Energy Survey (DES) data, see Ref.~\cite{Abbott:2017wau}.}
In App.~\ref{app:tilt} we analyze the full power spectrum likelihood and show 
the effect of varying the tilt. 
As for the relative amplitude $A$,
we choose its prior to be uniform in the range~$(0.04,4)$. 

Our final results will be presented in terms of the late-time mass fluctuation amplitude $\sigma_8$
because (a) this parameter is better constrained than $A_s$, 
(b) it is close to the actual 
principal component of the BOSS data and hence is less sensitive to prior choices, 
(c) it is more common in the large-scale structure literature. In Appendix~\ref{app:full}
we show results for both the rescaled primordial amplitude $A$ and $\sigma_8$.

As far as the neutrino sector is concerned, 
we approximate it with one state of mass~$m_\nu$
and two massless states.\footnote{This approximation is accurate for the matter power spectrum within $\sim 10\%$ precision for highest neutrino masses considered in this paper, which is sufficient for our purposes.} 
Therefore, we will use $m_\nu$ and $\sum m_\nu$ interchangeably in what follows.
We assume a flat prior on~$m_\nu$ in the range 
\be 
\label{eq:mnuprior}
m_\nu \in (0.06,0.18)\,\text{eV}.
\ee
The lower limit is inferred from the neutrino oscillation experiments and the upper limit is the 3$\sigma$ constraint obtained 
from the combination of the Planck 2018 \texttt{TTTEEE+lowE+lensing} data and the BAO scale measurements
\cite{Aghanim:2018eyx}. 
The BOSS data are not accurate enough to improve the measurement of the neutrino mass, hence we marginalize the final results over it. 
Nevertheless, it is important to keep this parameter in the chains since the neutrino mass is very degenerate with the amplitude of the power spectrum. Varying~$m_\nu$ in the allowed range can bias the amplitude $A$ by the amount comparable to the $1\sigma$ error on this parameter. We have found that $m_\nu$ does not affect significantly the limits on $H_0, \O_m$ and $\sigma_8$, which will be quoted as our final results.\footnote{Note that our analysis constrains the late time fluctuation amplitude $\sigma_8$ more directly than $A_s$ and this is why it is less sensitive to the neutrino mass.}
Specifically, we have repeated our analysis with no priors on the neutrino mass ($m_\nu \in (0,\infty)$), and found very similar results for the cosmological parameters, see App.~\ref{app:mnu} for more detail. Even if we impose the Planck priors 
on all cosmological parameters, the neutrino mass can only be constrained at the level of 
\mbox{$\sim$ 1 eV (95$\%$ CL)},
which is not competitive with other cosmological probes. Given this reason, we prefer to stick to the 
realistic prior allowed by other experiments and/or motivated by particle physics.
The use of a somewhat wider prior $m_\nu\in(0,0.24)$ eV
matching the Planck 2018 2$\sigma$-allowed region has a negligible impact on our results.

Finally, assuming the flat $\Lambda$CDM, the only additional cosmological parameters that are needed to describe the matter content of the Universe are physical densities of baryons and cold dark matter, $\omega_b$ and $\omega_{cdm}$. 
The baryons have very distinctive effect on the CMB power spectrum, which allows one to measure their physical 
density with sub-percent accuracy~\cite{Aghanim:2018eyx}
(assuming standard physics before and during recombination),
\be
\label{eq:omega_b_CMB}
\omega_b = 0.02237 \pm 0.00015  \qquad (\omega_b{\rm -CMB \; prior}) \;.
\ee
More conservatively, with minimal assumptions about the thermal and expansion history, the physical baryon density can be inferred using 
the BBN predictions and the measurement of helium and deuterium primordial abundances 
\cite{Aver:2015iza,Cooke:2017cwo,Aghanim:2018eyx,Schoneberg:2019wmt},\footnote{One may find different limits depending on nuclear rate predictions.
Below we present constraints obtained using the helium data from \cite{Aver:2015iza},
deuterium data from \cite{Cooke:2017cwo}
and assuming $N_{\text{eff}}=3.046$,
\be
\label{eq:bbn}
\begin{split}
(\text{standard})\qquad & \omega_b = 0.02268 \pm 0.00038\qquad (68\%)\,,\\
(\text{Marcucci et al.})\qquad & \omega_b = 0.02197 \pm 0.00022\qquad (68\%)\,,\\
(\text{\texttt{PRIMAT}})\qquad & \omega_b =  0.02188 \pm 0.00023\qquad (68\%)\,.
\end{split}
\ee
The fist limit is obtained using the $d(p,\gamma)^3$ He nuclear rate from \cite{Adelberger:2010qa}
and the \texttt{PArthENoPE} code \cite{Pisanti:2007hk}. 
The same code but a different nuclear rate taken from \cite{Marcucci:2015yla} 
yield the second constraint. Finally, using nuclear rates from \cite{Pitrou:2018cgg}
and the \texttt{PRIMAT} code (introduced in the same paper) gives the third constraint. 
In all the limits quoted above the systematic error is added in quadratures. 
We prefer to use the ``standard'' case in our analysis,
although any other choice from \eqref{eq:bbn} would produce very similar results.
We are grateful to Julien Lesgourgues for sharing with us the limits \eqref{eq:bbn}.
} 
\be
\label{eq:omega_b_BBN}
\omega_b = 0.02268 \pm 0.00038 \qquad (\omega_b{\rm -BBN \; prior}) \;. 
\ee
We will see momentarily that both priors yield identical constraints for the BOSS data.

The physical density of cold dark matter $\omega_{cdm}$ can be also inferred from the 
shape of the CMB spectra with percent accuracy~\cite{Aghanim:2018eyx},
\be
\label{eq:omcdm}
\omega_{cdm} = 0.1200  \pm 0.0012 \;.
\ee
We will not use this prior in our main analysis, and vary $\omega_{cdm}$ in the flat 
range $(0.05,0.2)$. 
The prior \eqref{eq:omcdm} will only be imposed 
in a side analysis that compares our method with the previous
BOSS FS pipeline which also fixes $\omega_{cdm}$.

As already pointed out, using the tight CMB priors on~$\omega_b$ and~$\omega_{cdm}$ effectively fixes the shape of the matter power spectrum and in this case our analysis reduces to the standard BOSS analysis. The only remaining difference is in the theoretical model used. 
This allows us to investigate the relation between our constraints on cosmological parameters  
and the previous BOSS results. 
It is worth noting that this choice of priors is equivalent to fixing a prior on the sound horizon at decoupling, since it can be approximated as~\cite{Aubourg:2014yra},
\be 
\label{eq:rd}
r_d\approx \frac{55.154\, \e^{-72.3(\omega_\nu+0.0006)^2}}{(\omega_{cdm}+\omega_{b})^{0.25351}\omega_b^{0.12807} }\,\text{Mpc}\,,
\ee
where $\omega_\nu\equiv m_\nu/(93.14\,\eV)$.
Note that the sound horizon at the drag epoch is insensitive to the late-Universe physics~\cite{Audren:2012wb, Lemos:2018smw}. Planck gives a sub-percent measurement of this scale~\cite{Aghanim:2018eyx},
\be
r_d = (147.09 \pm 0.24) \, \Mpc\,. 
\ee

\begin{table}[h!]
\begin{center}
 \begin{tabular}{|c|c|} 
 \hline
Parameter & Prior \\ [0.5ex] \hline\hline
 \multicolumn{2}{|c|}{Cosmology}\\  \hline
 $\quad$ $n_s$ (not varied) $\quad$ & $n_s=0.9649$  \\ \hline
  ${\omega_{b}}$ & different for each analysis  \\ \hline
 $A^{1/2}$ & flat$(0.02, 2)$  \\ \hline
 ${ h}$ & flat$(0.4, 1)$  \\ \hline
 ${\omega_{cdm}}$ & flat$(0.05, 0.2)$  \\ \hline
 ${ m_\nu}$ & flat$(0.06, 0.18)$ eV   \\ \hline
 \multicolumn{2}{|c|}{Biases and shot noise}\\  \hline
 ${ b_1\times A^{1/2}}$ & flat$(1, 4)$  \\ \hline
 ${ b_2\times A^{1/2}} $& flat$(-4, 2)$  \\ \hline
 ${ b_{\mathcal{G}_2}\times A^{1/2}}$ & flat$(-3, 3)$  \\ \hline
 $b_{{\Gamma}_3}$ (not varied) & $b_{{\Gamma}_3}=0$  \\ \hline
 ${ P_{\text{shot}} }$& flat$(0, 10^4)$  $\Mpc^3/h^3$  \\ \hline
  \multicolumn{2}{|c|}{Counterterms}\\  \hline
 ${ c_0^2,\, c_2^2}$ & flat$(-\infty, \infty)$ $\Mpc^2/h^2$  \\ \hline
 ${ \tilde{c}}$& $\quad$ flat$(-\infty, \infty)$  $\Mpc^4/h^4$ $\quad$ \\ \hline
\end{tabular}
\caption{
Priors that are common to all baseline $\Lambda$CDM analyses. 
The analyses of these paper use different priors on $\omega_b$, 
which will be specified separately in each case. 
In this table ``flat$(min,max)$'' stands for a flat prior in the range $(min,max)$. By (not varied) we denote the parameters that were not varied in our MCMC chains.}
\label{tab:priors}
\end{center}
\end{table}

Regarding the bias parameters, we adopt flat priors centered around 
the values expected from N-body simulations.
The previous BOSS analyses have already measured $b_1\simeq 2$, for which we use
a flat prior in the range~$(1,4)$.
The second order biases are varied in the range
\be
b_2 \in  (-4,2)\,,\quad b_{\mathcal{G}_2}\in(-3,3)\,.
\ee
These intervals are motivated by the measurements of biases for dark matter halos with masses similar to typical hosts for BOSS galaxies~\cite{Lazeyras:2015lgp}. These measurements roughly predict\footnote{Note that~\cite{Lazeyras:2015lgp} use a different basis of biased operators. Their values have been appropriately converted to match our bias prescription.}
\be 
b_2\approx -0.6\, , \quad {\rm and} \quad b_{\mathcal G_2} \approx -0.3\, , \quad {\rm for} \quad b_1\approx 2\, .
\ee
The halo bias is in general different from galaxy bias, but given that the satellite fraction is relatively small in the BOSS samples \cite{Rodriguez-Torres:2015vqa}, we expect these estimates not to be too far from the truth. In all analyses we set $b_{\Gamma_3}= 0$ and we do not vary it. The reason for this choice is that $b_{\Gamma_3}$ is very degenerate with the counterterms~$c_0^2$,~$c_2^2$ and $b_{\mathcal G_2}$. The data are not accurate enough to break this degeneracy. We have verified this using the mock catalogs.

Finally, let us discuss the amplitude of the shot noise. The number density of the galaxies in the BOSS samples is known, and it is roughly~$\bar n\sim 3\times 10^{-4} \,(h/{\rm Mpc})^3$. However, one might expect some deviations from the Poisson value for the shot noise amplitude due to effects like exclusions~\cite{Baldauf:2013hka}. Detailed comparisons to simulations~\cite{Schmittfull:2018yuk} show that this deviation for BOSS galaxy number density is not expected to be very large (at most $50\%$). For this reason we will vary $P_{\text{shot}}$ in the chains within the flat prior in the following range:
\be
P_{\text{shot}}\in (0,10^4)\, \Mpc^3/h^3\,.
\ee

Another reason to vary the constant $P_{\text{shot}}$ in our analysis 
is to capture the fiber collision
effect. Indeed, this is a common 
practice to correct for the fiber collision residual contributions 
left after applying the 
nearest neighbor method \cite{Gil-Marin:2015sqa,Beutler:2016arn}.
Ref.~\cite{Hahn:2016kiy} pointed out that this practice 
is not sufficient for the quadrupole, which does not have a constant
shot noise contribution. This reference showed that the problem
can be alleviated by applying the effective window function
supplemented with additional nuisance parameters, which correspond to a stochastic 
constant contribution for the monopole and a $k^2$-contribution to the quadrupole.
While the first term is accounted for precisely by $P_{\text{shot}}$, the second contribution 
happened to be fully degenerate with our NLO $k^4 P_{\text{lin}}$ counterterm.
We have checked that, to a precision of $0.5\%$, 
the difference between the spectra with and without 
the effective window function can be absorbed into the nuisance parameters of our theory model.
This difference is much below the statistical error and can be safely neglected, which is why we 
proceed without the effective mask.

All nuisance parameters, $A$, $h$ and $m_\nu$ have the same priors in all our analyses. We summarize them in Table.~\ref{tab:priors}. 
We use different combinations of priors on $\omega_b$ and $\omega_{cdm}$ 
in our analyses and we will specify them in each example separately.

\vskip 5pt
\noindent 
{\bf Software.} 
Our analysis is based on a modification of the publicly 
available \texttt{CLASS} code \cite{Blas:2011rf}
that incorporates the \texttt{FFTLog} method \cite{Simonovic:2017mhp} for fast evaluation of one-loop perturbation theory integrals. 
The parameter constraints are obtained with the April 2018 version of the \texttt{Montepython} code \cite{Audren:2012wb,Brinckmann:2018cvx}. 
Plots with posterior densities and marginalized limits are produced with the latest version of the \texttt{getdist} package\footnote{\href{https://getdist.readthedocs.io/en/latest/}{
\textcolor{blue}{https://getdist.readthedocs.io/en/latest/}}
},
which is part of the \texttt{CosmoMC} code \cite{Lewis:2002ah,Lewis:2013hha}.
We monitor the convergence of our MCMC chains with the 
Brooks-Gelman and Gelman-Rubin criteria \cite{Gelman:1992zz,Brooks:1997me}

%%%%%%%%%%%%%%%%%%%%%%%%%%%%%%%%%%%%%%%%%%%%%%%%%%%%%%%
\section{Constraints on Base ${\bf \Lambda}$CDM }
\label{sec:base}

In this section we present measurements of parameters of the minimal flat~$\Lambda$CDM with massive neutrinos. 
Our final results are quoted in terms of $\sigma_8$, $H_0$ and $\Omega_{m}$
since these parameters are most common in the large-scale structure literature. 
Another reason for the use of these particular parameters is that they are close 
to the actual principal components of the BOSS data\footnote{E.g. the amplitude $A_s$ is 
very correlated with the neutrino mass, which degrades the relative error on $A_s$ compared to $\sigma_8$.
Moreover, the asymmetric priors on $m_\nu$ make the posterior for $A_s$ very
asymmetric as well.
}.
Our main analysis does not assume CMB priors on $\omega_{cdm}$ (equivalently, $r_d$).
We use several different priors on $\omega_b$. 
These are the CMB prior~\eqref{eq:omega_b_CMB}, a slightly weaker BBN prior~\eqref{eq:omega_b_BBN}, 
and the CMB prior with a $30$-times bigger variance. We impose the latter prior in order to check
to what extent the $\omega_b$ prior is crucial for our results.

We start with the first case (the CMB prior on $\omega_b$).
The reduced triangle plot with the relevant cosmological parameters for four different 
BOSS datasamples are shown in the left panel of Fig.~\ref{fig:noprComb}. 
The full triangle plot and the 1d marginalized limits are given 
in Appendix~\ref{app:full}. There we also present results for parameters 
$f\sigma_8(z_{\text{eff}})$, $H(z_{\text{eff}})$, $D_A(z_{\text{eff}})$ and $D_V(z_{\text{eff}})$,
derived from our MCMC chains.

\begin{figure}[ht!]
\includegraphics[width=0.49\textwidth]{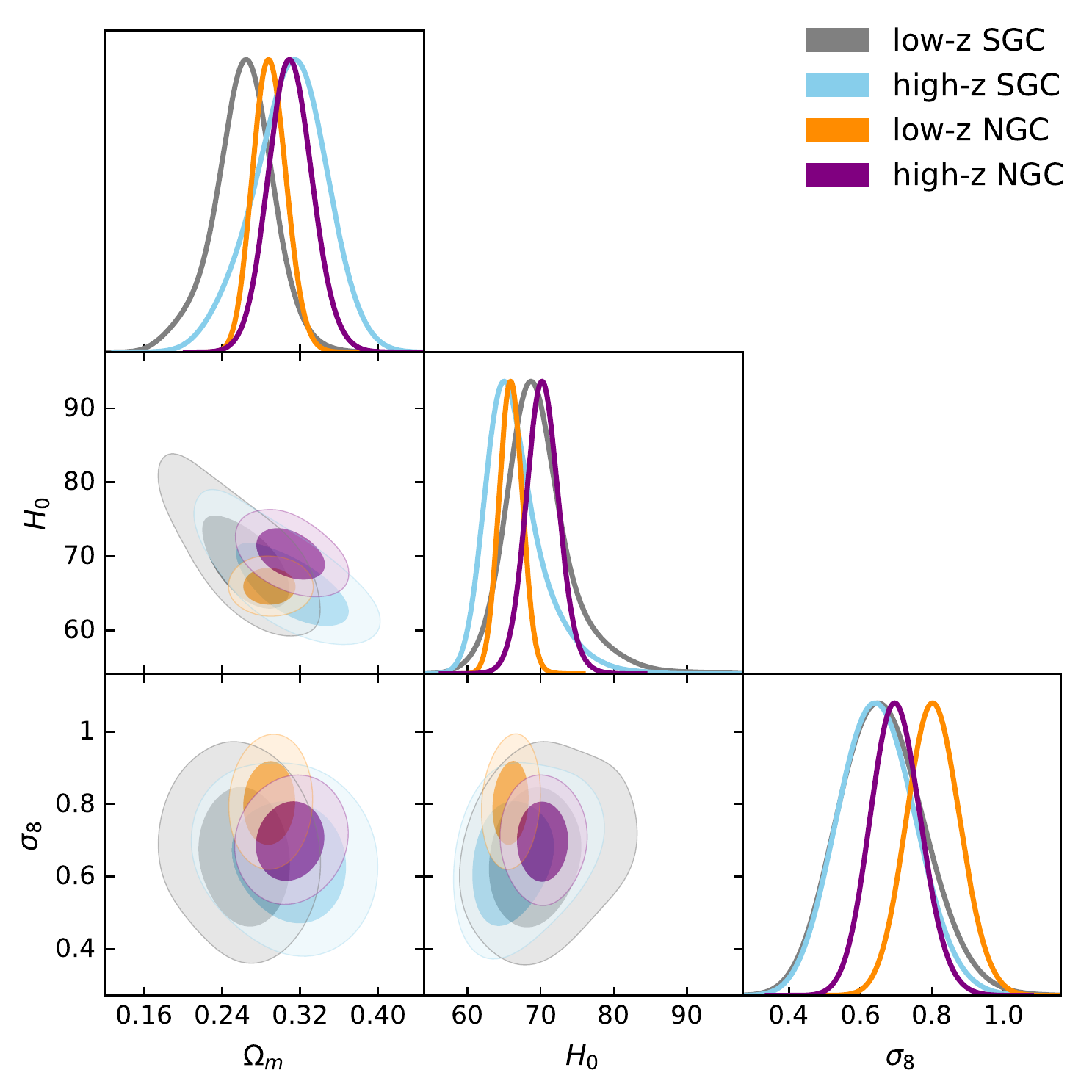}
\includegraphics[width=0.49\textwidth]{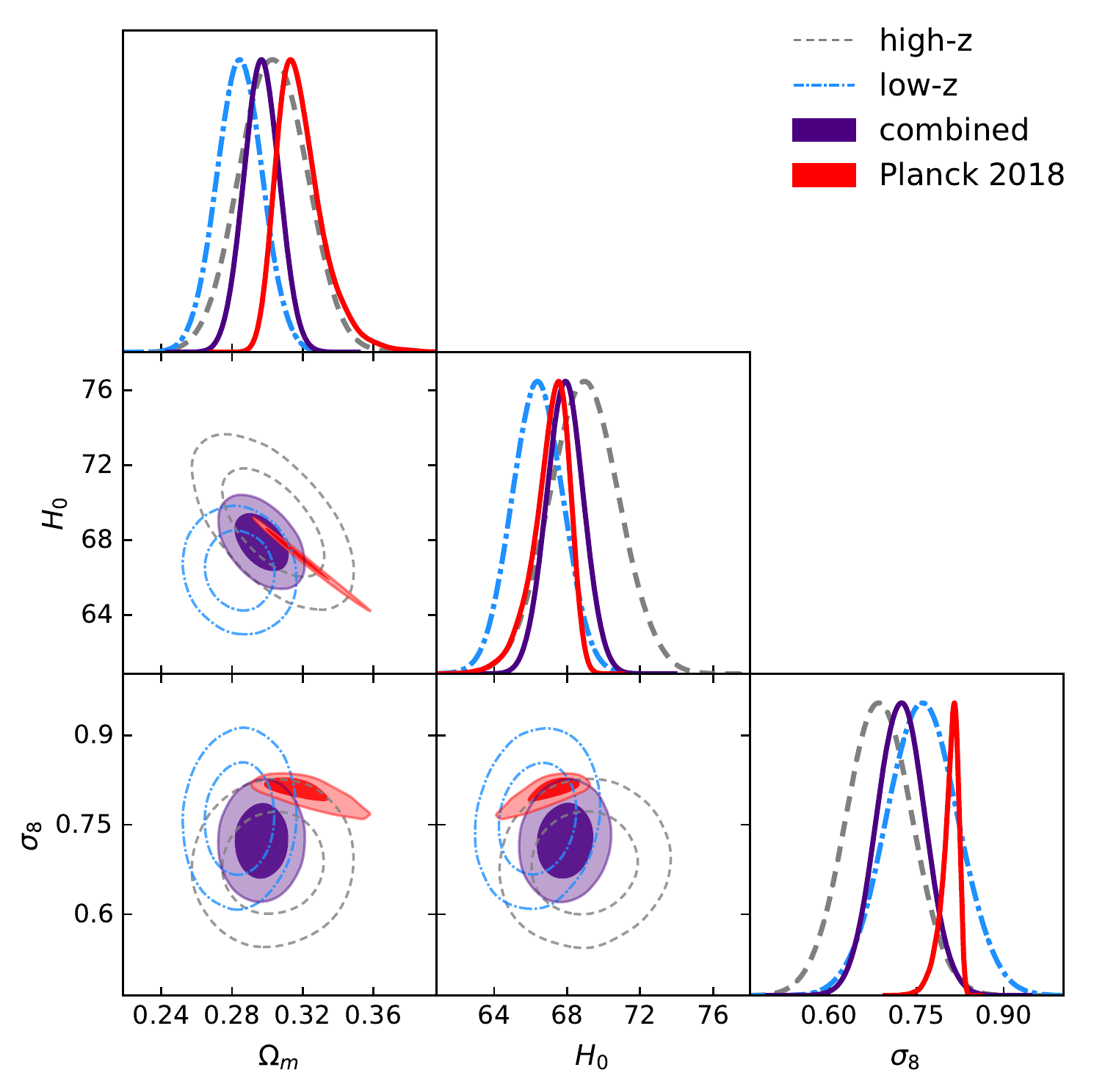}
\caption{
The 2d posterior distribution for 
cosmological parameters extracted 
from the BOSS DR12 power spectrum likelihood. 
We show results for four independent samples of the BOSS data separately 
(left panel)
and the combined likelihoods (right panel).
In the latter case we also plot the posterior distribution 
for the parameters of a similar model ($\L$CDM with massive neutrinos) 
measured from the
final Planck 2018 CMB data. 
$H_0$ is quoted in units [km/s/Mpc].
}   
\label{fig:noprComb} 
\end{figure}

Let us first discuss the consistency of our results. 
The posterior distributions seen in the left panel of Fig.~\ref{fig:noprComb} overlap within 1$\sigma$ 
regions. The observed scatter is compatible with the hypothesis that all the 
independent samples are drawn from a single set of cosmological parameters. 
This suggests that these samples can be combined. 
The combinations of low-z, high-z, and all four samples are shown in the right panel in Fig.~\ref{fig:noprComb}, while the corresponding 1d marginalized intervals are presented in Table \ref{tab:comb}.
For comparison, we also show the Planck 2018 results from the 
TT,TE,EE+lowE+lowl+lensing data\footnote{The MCMC chains
for the 
\texttt{base\_mnu\_plikHM\_TTTEEE\_lowl\_lowE\_lensing} likelihood
were downloaded from the Planck Legacy Archive \href{http://pla.esac.esa.int/pla/\#cosmology}{\textcolor{blue}{http://pla.esac.esa.int/pla/\#cosmology}}.}, 
which were derived for $\L$CDM with the varied neutrino mass. Overall, we observe good consistency between BOSS and Planck. The mean values of $H_0$ and $\Omega_{m}$ inferred from different BOSS redshift bins are within $1\sigma$ distance from each other and from the Planck posterior mean values. 
One can notice that the high-z data prefer smaller $\sigma_8$ than Planck. This tendency has already been observed in the previous BOSS full-shape analyses \cite{Beutler:2013yhm,Beutler:2016arn}. 
However, the obtained 
difference between the Planck and our BOSS measurements 
is still consistent with a statistical
fluctuation.

\begin{table}[ht!]
\begin{center}
\vspace{-0.1cm}
 \begin{tabular}{|c|c|c|} 
 \hline
\text{\small{Pl}}$_{\omega_b}$ high-z & best-fit & mean  $\pm 1\sigma$  \\ [0.5ex]  \hline\hline
 $\omega_{cdm}$  & $0.1199$ & $0.1201 \pm 0.0082$  \\\hline 
 $H_0$  & $68.92$ & $68.96 \pm 1.94$  \\\hline \hline
   $\Omega_{m}$  & $0.3030$ & $0.3033 \pm 0.0194$  \\\hline 
 $\sigma_8$ & $ 0.6844$ & $0.6862 \pm  0.0589$  \\ \hline
\end{tabular}
\vspace{0.2cm}
\begin{tabular}{|c|c|c|} 
 \hline
\text{\small{Pl}}$_{\omega_b}$ low-z & best-fit & mean  $\pm 1\sigma$  \\ [0.5ex] 
 \hline
  $\omega_{cdm}$  & $0.1013$ & $0.1014\pm 0.0075$  \\ \hline 
 $H_0$  & $66.34$ & $66.38 \pm 1.44$  \\ \hline \hline
      $\Omega_{m}$  & $0.2842$ & $0.2846 \pm 0.0144$  \\\hline 
  $\sigma_8$ & $ 0.7552$ & $0.7604  \pm  0.0634$  \\ \hline
\end{tabular}
\begin{tabular}{|c|c|c|} 
 \hline
\text{\small{Pl}}$_{\omega_b}$ comb.& best-fit & mean  $\pm 1\sigma$  \\ [0.5ex] 
 \hline\hline
 $\omega_{cdm}$  & $0.1125$ & $0.1127 \pm 0.0046$  \\ \hline
 $H_0$  & $67.86$ & $67.88 \pm 1.06$  \\ \hline\hline
   $\Omega_{m}$ & $ 0.2965$ & $0.2967 \pm  0.0103$  \\ \hline
  $\sigma_8$ & $ 0.723$ & $0.723 \pm  0.043$  \\ \hline
\end{tabular}
 \begin{tabular}{|c|c|c|} 
 \hline
\text{\small{Pl}}$_{\omega_b+\omega_{cdm}}$& best-fit & mean  $\pm 1\sigma$  \\ [0.5ex] \hline\hline
 $\omega_{cdm}$ & $ 0.1200$ & $ 0.1195 \pm  0.0012$  \\ \hline
 $H_0$ & $ 69.64$ & $ 68.57 \pm  0.93$  \\ \hline \hline
  $\Omega_m$ & $ 0.2979$ & $ 0.3057 \pm  0.0082$  \\ \hline
 $\sigma_8$ & $ 0.721$ & $ 0.731 \pm  0.042$  \\ \hline
\end{tabular}
\caption{
The results for cosmological parameters from the combined likelihoods. 
We assume Planck priors on $\omega_{b}$ everywhere, whereas the 
results from the lower right table were derived upon additionally imposing the
Planck prior on $\omega_{cdm}$. $H_0$ is quoted in units [km/s/Mpc].
The group of first two parameters ($\omega_{cdm}$ and $H_0$) 
display the parameters which were sampled with flat uninformative
priors. The second two parameters ($\Omega_m$ and $\sigma_8$) are derived ones.
}
 \label{tab:comb}
\end{center}
\end{table}

The statistical errors of our $H_0$ and $\Omega_m$ measurement 
are comparable with Planck errorbars for the parameters of 
the same cosmological model with massive neutrinos. 
Note that these parameters 
do not form principal components for the Planck data, and hence are 
relatively poorly
measured, e.g. compared to the combination $\Omega_m h^3$, which controls
the angular position of acoustic oscillations in the CMB temperature power 
spectrum \cite{Percival:2002gq}. 
This fact is reflected in a well-known degeneracy between $H_0$ and $\Omega_m$, which can be
clearly observed in the Planck contours shown in the right panel of Fig.~\ref{fig:noprComb}.
On the contrary, the degeneracy between these two parameters is not very strong 
in the BOSS data, which provide us with 
more direct measurements of $H_0$ and $\Omega_m$
than Planck.

Our main conclusions remain exactly the same if we use the BBN prior on $\omega_b$. 
Even in this case one can measure $H_0$ and $\Omega_m$ quite well using no information from 
CMB whatsoever. Remarkably, our $\sim3\%$ limit on the late-time matter density fraction $\Omega_m$
is one of the best measurements of this parameter from the LSS data.
We emphasize that this constraint is driven by the shape of the power spectrum. 
Since there is no difference between our measurements
in the case of Planck and BBN priors on $\omega_b$, we prefer to quote the latter ones 
as our final results
because they use no input from the CMB data. 
The corresponding posteriors are shown in Fig.~\ref{fig:final},
and limits are displayed in Tab.~\ref{tab:final}.

To test the stability of our results we have run the same analysis assuming
the Planck Gaussian prior on $\omega_b$ with a $33$ times bigger error ($\omega_b=0.02237\pm 0.005$). 
In that case the BOSS data 
are able to deliver an independent constraint on $\omega_b$.
Still, this limit is by far superseded by the BBN, and will not be
quoted here.
Upon marginalizing over $\omega_b$, we obtain the following constraints: $\O_m=0.293\pm 0.012$, $H_0=66.6\pm2.1$ km/s/Mpc, $\sigma_8= 0.713\pm 0.045$. 
Remarkably, our measurement of $\O_m$ did not degrade once we relaxed the prior on $\omega_b$,
whereas the measurement of $H_0$ worsened by a factor of 2. 
The stability of $\Omega_m$ is consistent with the observation that upon marginalizing over $\omega_b$
the matter density forms a principle component of the geometric 
information is $\sim \O_m^{-0.5}$ \cite{Chudaykin:2019ock}. 
The degradation of $H_0$ occurs because it is mainly extracted from $r_d/D_V$ by using the power spectrum shape
(which probes $\omega_b$ and $\omega_{cdm}$), 
which has less constraining power without the $\omega_b$ prior.

It is important to stress that so far we have not imposed a prior on $r_d$. Moreover, since $r_d$
depends on $\omega_{cdm}$ and $\omega_b$, our analysis provides an independent measurement of the acoustic horizon at decoupling, which is consistent with Planck,
\be
r_d=  (149.1\pm 1.3)~\text{Mpc}\quad \text{(BOSS FS+BBN $\omega_b$)}\,.
\ee
To see how much this result depends on the $\omega_b$ prior, let us also quote the value obtained in the analysis with a 
loose non-informative Gaussian prior $\omega_b=0.02237\pm 0.005$ described in the previous paragraph,
\be
r_d=  (150.0\pm 4.5)~\text{Mpc}\quad \text{(BOSS FS+loose $\omega_b$)}\,.
\ee

\begin{figure}[h!]
\begin{center}
\includegraphics[width=0.75\textwidth]{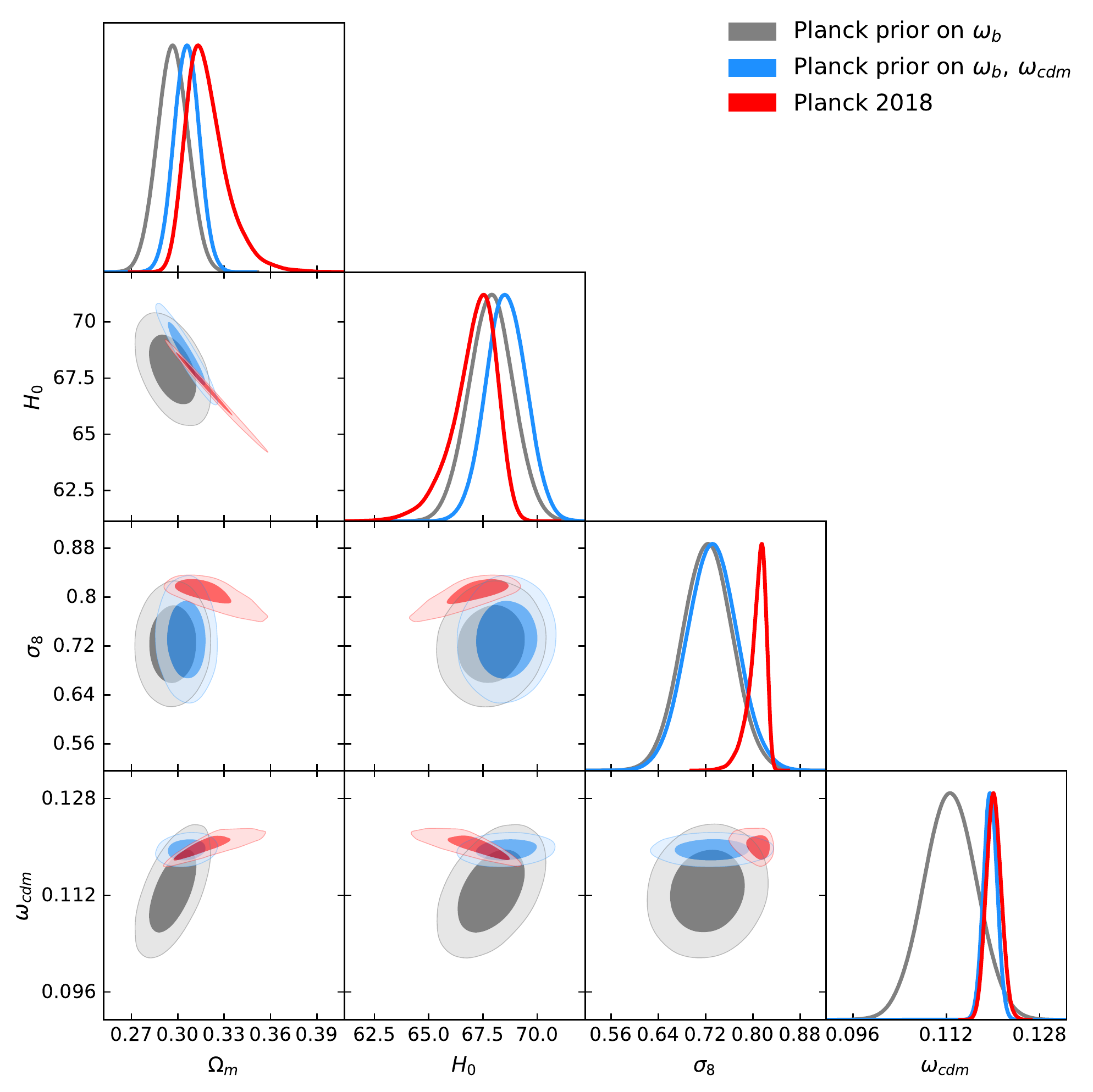}
\caption{The posterior contours for the combined analysis assuming the Planck prior on $\omega_{b}$
(in gray), Planck priors on $\omega_{b}$ and $\omega_{cdm}$ (in light blue). For comparison also shown
are the contours from the Planck CMB data for $\L$CDM with massive neutrinos (in red). 
$H_0$ is quoted in units [km/s/Mpc].
}    
\label{fig:rd}
\end{center}
 \end{figure}

Now let us discuss the constraints obtained with the Planck priors
on both the physical baryon and dark matter densities. 
As argued previously, in this case the shape of the matter power 
spectrum is only allowed to vary within very tight priors, thus for 
practical purposes the shape is effectively fixed. This case corresponds
to the standard FS BOSS analysis.

Our results for the 2d posterior contours are shown in Fig.~\ref{fig:rd}, while the 1d marginalized limits are quoted in the lower right corner of Tab.~\ref{tab:comb}. 
One may notice that $H_0$ and $\Omega_{m}$ have shifted upwards by
$\sim 0.5\sigma$ w.r.t.~our baseline analysis with the $\omega_{b}$ prior only,
while their errorbars reduced only marginally. 
Obviously, in this case $\Omega_m$ is a derived parameter which is  
almost fully correlated with $H_0$. 
This should be contrasted with our baseline analysis without the $\omega_{cdm}$ prior, 
where $\Omega_m$ is a valid degree of freedom. 
Remarkably, the 
$\omega_{cdm}$ prior has a marginal effect on $H_0$ and $\sigma_8$,
which implies that this prior is not necessary for an accurate parameter 
estimation from the LSS data. 
This result is clear from Fig.~\ref{fig:rd}, which shows that $\omega_{cdm}$ is not very degenerate with 
$H_0$ and $\sigma_8$.   
This effect will be discussed in more detail in Section~\ref{sec:info}.

Finally, let us discuss some implications of our results.
Our measurement of $H_0$ is consistent with Planck \cite{Aghanim:2018eyx} 
and the recent BAO + BBN analyses of Refs.~\cite{Cuceu:2019for,Schoneberg:2019wmt}.
However, it is in tension with 
the results of the local astrophysical measurements of SNIa \cite{Riess:2019cxk}
and strong gravitational lensing of distant quasars~\cite{Wong:2019kwg}. 
Our study shows that the full-shape power spectrum information
constrains $H_0$ at $1.6\%$ level, which is comparable to the SNIa limits.
Since there is no $r_d$ prior in our analysis, it disfavors explanations for the 
``tension'' based on modified expansion history at high redshifts 
which preserve the shape of the power spectrum, e.g decaying DM.

As for our constraint on $\omega_{cdm}$, it is $\sim 4$ times worse than the Plank limit, but can be used to discriminate various proposals for the resolution of the $H_0$ tension that involve modifications of the linear power spectrum, such as early dark energy. 
We will explore this in more detail in a separate publication.

\section{Geometric, Shape and Alcock-Paczynski Information}
\label{sec:info}

In this Section we quantify the information content of various effects relevant 
for the galaxy clustering data. 
To that end, we will first roughly classify all the relevant effect and then
give some theoretical background on the difference between the geometric and the shape information.
In the second part 
of this Section we will analyze several mock likelihoods mimicking the BOSS data in order to explicitly
see how much different effects contribute to the final constrains.
Throughout this Section, we will be working within base $\L$CDM and for simplicity assume that all neutrinos are massless.
% After a short review of these effects, we will discuss in detail the shape and 
% the geometric information, which are most important for our analysis.
% Finally, we will present analyses of mock datasamples that are designed to 
% shed some light on the origin of our parameter constraints. 

The sources of cosmological information can be roughly classified into four categories:

\begin{itemize}
\item Distance-free shape information. 
For a fixed $n_s$, the power spectrum shape mostly 
depends on $\omega_{b}$ and $\omega_{cdm}$ (and $\omega_\nu$, to a lesser extent), 
which control the relative amplitude of the BAO wiggles (through $\omega_b/\omega_{cdm}$), 
their frequency (through $r_d$),
the amount of the short-scale suppression due to baryons (through $\omega_b/\omega_{cdm}$), 
and the relative position of the BAO wiggles 
and the baryon Jeans scale w.r.t~the power spectrum peak (via\footnote{We introduced an obvious 
notation $\omega_{cb}=\omega_{cdm}+\omega_{b}$.} $r_d\omega_{cb}$). 
The relative shape does not depend on the choice of rulers (i.e.~$H_0$). 
\item Geometric (or distance) information. The features discussed above (e.g.~the BAO frequency) 
can be assigned a (comoving) length scale 
for a given cosmological model, which constrains parameters of this model.
Indeed, the position of the 
BAO wiggles in momentum 
space as extracted from the monopole is set by $r_d(\omega_{cdm},\omega_b)/D_V$, 
where the effective ``volume-averaged'' distance $D_V$ is defined as\footnote{We work in the unit system with $c=1$.}
\begin{align}
\label{eq:Dvdef}
& D_V(z)\equiv ((1+z)^2D^2_A(z)z/H(z))^{1/3}\,,\\
\label{eq:Dadef}
& D_A(z)\equiv \frac{1}{1+z}\int_0^z\frac{dz'}{H(z')}\,.
\end{align}
Analogously, the location of the monopole power spectrum peak
is set by $k_{\rm eq}D_V\sim \omega_{cb}D_V$.
In $\Lambda$CDM the physical densities of baryons and dark matter 
are fixed by the transfer functions, 
thus there is only one parameter $H_0$, which controls the location of the power spectrum 
features.
\item Alcock-Pazcynski information.\footnote{It should be pointed out that the division into 
``geometric''
vs.~``AP'' information is somewhat artificial as these two effects cannot be isolated in a real survey.
Alternatively, one may discuss the monopole vs. quadrupole distance information, see e.~g.~\cite{Kobayashi:2019jrn}.
} 
The radial and angular distances can be measured separately through the AP effect \cite{Alcock:1979mp}, parameterized by 
\[
F_{\text{AP}}=(1+z)D_A(z)H(z)\,.
\]
This parameter in encoded in the power spectrum quadrupole.
We will see that in $\L$CDM these distances are fixed by the shape and geometric information,
but they can measured independently of this information in the extensions of $\Lambda$CDM.
\item Redshift-space distortions. RSD help to measure the velocity power spectrum 
from the quadrupole power spectrum moment,
which constrains $f\sigma_8$. 
The shape and geometric information breaks the 
degeneracy between $\sigma_8$ and $f$ (which mostly depends only on the background expansion, i.e.~in $\L$CDM $f\simeq \Omega_m^{0.5}(z)$).
\end{itemize}
Our main goal is to show how the first two effects contribute to our new constraints.
Let us focus on them.

\subsection{Shape vs.~Geometry}

In this Section we will discuss in more detail the shape information and its 
distinction from the distance information.
This material will be somewhat pedagogical and has an overlap with old works on the galaxy clustering
that were using the power spectrum shape for cosmological parameter measurements 
independent of CMB~\cite{Tegmark:2006az,Percival:2006gt,Reid:2009xm}. 
Unless otherwise stated, all numerical estimates of this Section will be made for the Planck best-fit $\L$CDM cosmology~\cite{Aghanim:2018eyx}.

It is instructive to review the role of the shape and distance information from the CMB power spectrum of temperature (TT) fluctuations. 
The primary CMB spectrum has three main sources of information, which can be cast into 
the amplitude, shape and geometric distance. The latter two are the relevant ones for our discussion.
They can also be loosely called the ``horizontal'' and ``vertical'' information \cite{Tegmark:2006az}.
Vertical information refers to the relative height of the acoustic peaks, i.e.~their shape, which 
depends only on the physical matter densities
$\omega_m$ and $\omega_b$ and the tilt $n_s$. 
The distinctive physical effects produced by variations of these parameters allow 
to measure them regardless of any late-time physics \cite{Audren:2012wb,Audren:2013nwa}.
By horizontal information we mean the acoustic angular scale, which controls our freedom 
to shift the spectra in the horizontal direction (rescaling of angular multipoles $\ell$'s).
The angular size of the 
sound horizon at the drag epoch is given by
\be 
\label{eq:tcmb}
\theta_{s,\,{\rm CMB}}=\frac{r_s(z_{\rm d})}{(1+z_{\rm d})D_A(z_{\rm d})}\,,
\ee
(where $r_s(z_{\rm d})=r_d$ and $D_A(z_{\rm d})$ are the sound horizon at decoupling and the angular diameter distance corresponding 
to the decoupling redshift $z_{\rm d}$). 
Although this single parameter has been measured by Planck with tremendous precision $0.05\%$ \cite{Aghanim:2018eyx},
it depends on multiple cosmological parameters. 
The numerator of \eqref{eq:tcmb} is a slow function of $\omega_m$ and $\omega_b$ (see Eq.~\eqref{eq:rd}). 
However, the denominator $D_A$ depends sensitively on the 
late-time expansion.
If one expresses the measurement of $\theta_{s,\,{\rm CMB}}$ in terms of the late-time parameters 
$\Omega_m$ and $h$, one finds a strong degeneracy corresponding to fixed $\Omega_m h^3=\omega_m h$, 
with projections onto each separate parameter being much wider than this combination itself.
The geometric degeneracy of the CMB gets eventually broken by the shape information of the power spectrum, i.e.~by $\omega_m$ 
and $\omega_b$ being measured from the relative hight of the CMB peaks.

Analogously to the CMB, the angular position of the BAO 
in the monopole power spectrum of galaxies at some $z_{\rm eff}$ is given by 
\be 
\label{eq:tlss}
\theta_{s,\,{\rm LSS}}=\frac{r_d}{D_V(z_{\rm eff})}\,.
\ee
% Similarly to the CMB, this angle 
% controls our freedom to shift the power spectrum horizontally across
% different wavenumbers $k$.
If one were to measure only the combination \eqref{eq:tlss} just like 
in the BAO analysis, 
the degeneracy between $\omega_m$ and $h$ could not be broken
and one would be left with the horizontal information only.
However, it is precisely the shape (vertical) information
that allows one to decouple $h$ and $\omega_m$.

\begin{figure}[h!]
\begin{center}
\includegraphics[width=0.49\textwidth]{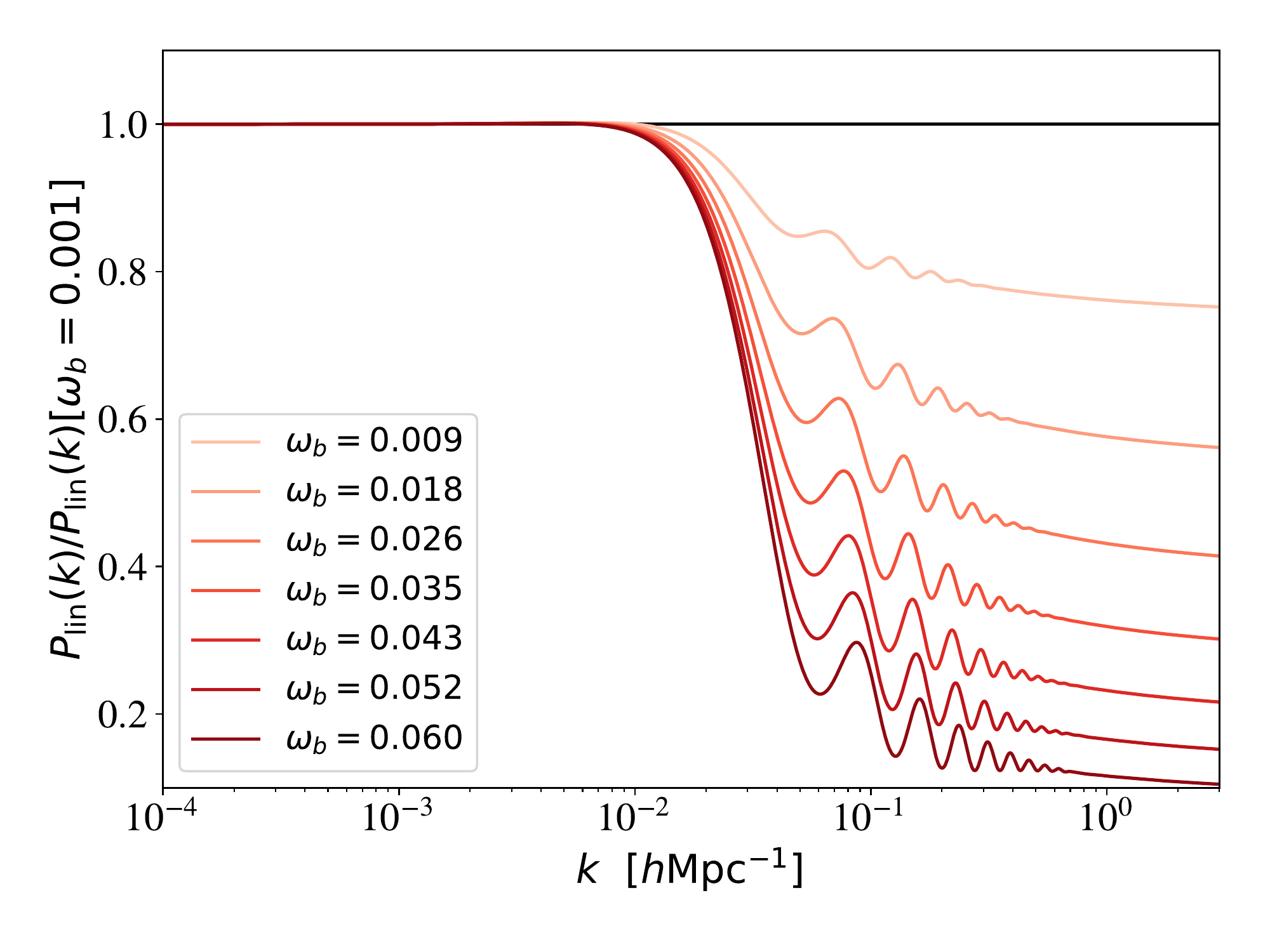}
\includegraphics[width=0.49\textwidth]{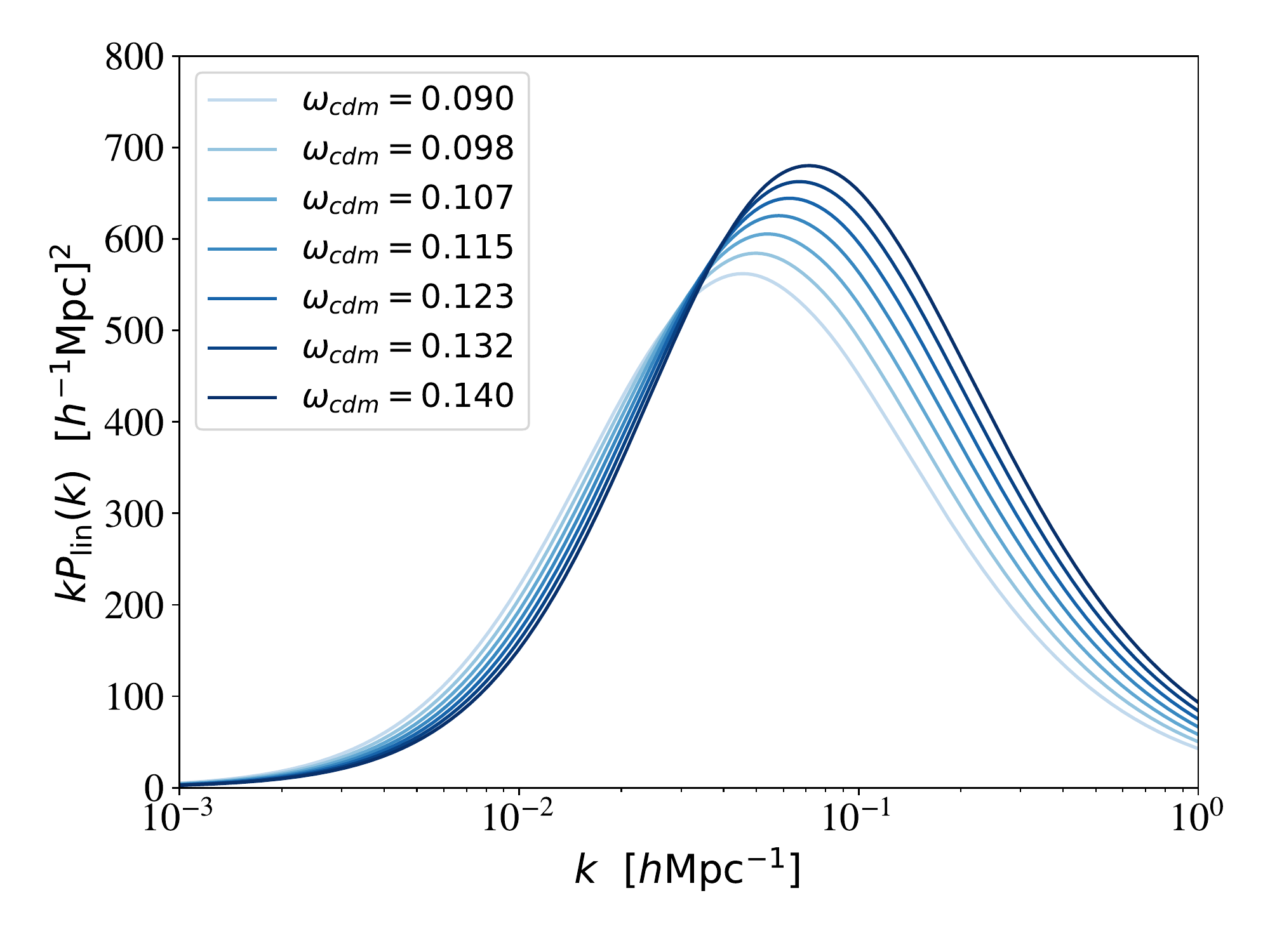}
\caption{The effect of varying the physical baryon (left panel) and cold dark matter (right panel)
densities on the shape of the linear matter
power spectrum (at $z=0$). In the first case we adjust $\omega_{cdm}$ to keep $\omega_{m}$ fixed, 
while in the second case we put $\omega_b\to 0$ to illustrate shape modifications 
exclusively due to $\omega_{cdm}$. All other cosmological parameters 
are fixed to the Planck best-fit values \cite{Aghanim:2018eyx}. 
The scale rage that dominates the constraints presented in this paper is [0.01,~0.25] $h$Mpc$^{-1}$. 
}    
\label{fig:shape}
\end{center}
 \end{figure}

As discussed above, the sound horizon $r_d$
depends on $\omega_b$, $\omega_m$ only (though very weakly, see Eq.~\eqref{eq:rd}). 
These two parameters can be measured directly from the shape of the matter power 
spectrum regardless of the late-time expansion just like in the CMB case. 
To see this, we display in Fig.~\ref{fig:shape}.
the effect of varying these parameters. 
One clearly sees that $\omega_b$ and $\omega_m$ control the frequency of the BAO, the shape of the BAO wiggles,
the amount of the short-scale suppression due to the baryon free-streaming before recombination,
the overall slope of the power spectrum and its turnover.
In the case of our baseline analysis with fixed $\omega_b$ and $n_s$, 
all these effects depend only on one parameter $\omega_{cdm}$, which results in quite tight 
constraints.

It is precisely the shape information on $\omega_m$ (and hence, $r_d$) that allows one to break 
the degeneracy 
between $D_V$ and $r_d$
and measure $D_V$ directly from $\theta_{s,\,{\rm LSS}}$. 
The crucial point is that in $\L$CDM $D_V$ is an extremely slow function of $\omega_m$
at small redshifts relevant for galaxy surveys. Indeed, using Eq.~\eqref{eq:Dvdef} one finds\footnote{The Alcock-Paczynski effect
also allows one to independently measure $D_A(z_{\rm eff})$ from the quadrupole.
However, it turned out to be quite insensitive to $\omega_m$ either, $D_A(z=0.38)\propto h^{-0.83}\omega_m^{-0.08}$. 
This shows that the low-redshift AP effect is a very weak probe of $\omega_m$.
}
\be 
D_V(z=0.38)\propto h^{-0.78}\omega_m^{-0.11}\,.
\ee
Since $\omega_m$ is absolutely fixed by the shape information, 
$D_V$ reduces directly to $H_0$.
Overall, the situation is very similar to 
the CMB temperature fluctuation spectrum,
whose $\omega_m-H_0$ degeneracy gets broken
by the vertical shape information.
Crucially, the degeneracy direction between $\omega_m$ and $H_0$ 
in the galaxy BAO is more perpendicular to $H_0$ than that of the CMB acoustic scale, which results in better 
constraints even though at face value the precision of LSS measurement is worse than that of 
the CMB.
This fact was pointed out long ago in Refs.~\cite{Tegmark:2006az,Percival:2006gt,Reid:2009xm}.
Let us explicitly illustrate this. 
Using the expressions
\eqref{eq:rd} and \eqref{eq:Dvdef},
we get
\be
\label{eq:degts}
\frac{\d \ln \theta_{s,\,{\rm LSS}}}{\d \ln h}\Bigg|_{z=0.38} = 0.78\,,\quad \frac{\d \ln \theta_{s,\,{\rm LSS}}}{\d \ln \omega_{m}}\Bigg|_{z=0.38} = -0.14\,. 
\ee
This implies that the acoustic peaks in the galaxy spectrum constrain the combination $h\omega^{-0.18}_{m}$.
A similar calculation carried out for the CMB acoustic peak \eqref{eq:tcmb} 
gives $h\omega_{m}^{0.8}$ (see Ref.~\cite{Percival:2002gq}).
Clearly, unlike the CMB, the LSS acoustic angle is a very weak function of $\omega_{m}$
and hence it allows one to accurately measure $h$.

Importantly, the galaxy power spectrum contains 
additional geometric information on top of the BAO wiggles. 
The first piece of this information is given by the angular position 
of the power spectrum peak, 
\be 
\label{eq:keq}
\theta_{{\rm eq}}=1/(k_{\rm eq}D_V)\,.
\ee
The second piece of additional information beyond the BAO is given by the same sound horizon scale $\theta_{s,\,{\rm LSS}}$, 
which also marks the 
location of the baryon free-streaming scale (see the left panel of Fig.~\ref{fig:shape}).
Thus, in principle, one could derive constraints
on $H_0$ and $\Omega_m$ even if the BAO were not present in the matter power spectrum.
This point will be illustrated in a mock data analysis of the next subsection.

The power spectrum peak (turnover) itself gives a complementary way to break the 
degeneracy between $\omega_{m}$ and $D_V$. 
Indeed, one can notice that 
the two angular scales \eqref{eq:tlss} and \eqref{eq:keq} have very different sensitivity to $\omega_m$ and $h$. 
Indeed, the BAO angle constrains $h\omega_m^{-0.18}$.
However, the power spectrum turnover fixes a combination $h\omega^{-1.14}_{m}$,
\be 
\frac{\d \ln \theta_{\rm eq}}{\d \ln h}\Bigg|_{z=0.38} = 0.78\,,\quad \frac{\d \ln \theta_{{\rm eq}}}{\d \ln \omega_{m}}\Bigg|_{z=0.38} = -0.89\,. 
\ee
Therefore, the following two combinations of these angles would directly measure $\omega_{m}$ and $h$,
\be
\frac{\theta_{s,\,{\rm LSS}}}{\theta_{{\rm eq}}}\propto \omega_m^{0.75}\,,\quad 
\frac{\theta^{6.4}_{s,\,{\rm LSS}}}{\theta_{{\rm eq}}}\propto h^{4.2}\,.
\ee

 This shows that even in the case where the measurement of $\omega_{m}$ 
 from the slope is complicated by marginalizing over the power spectrum tilt $n_s$ (see App.~\ref{app:tilt}),
 $\omega_m$ can still be inferred from the power spectrum peak.

 \begin{figure}[h!]
\begin{center}
\includegraphics[width=0.49\textwidth]{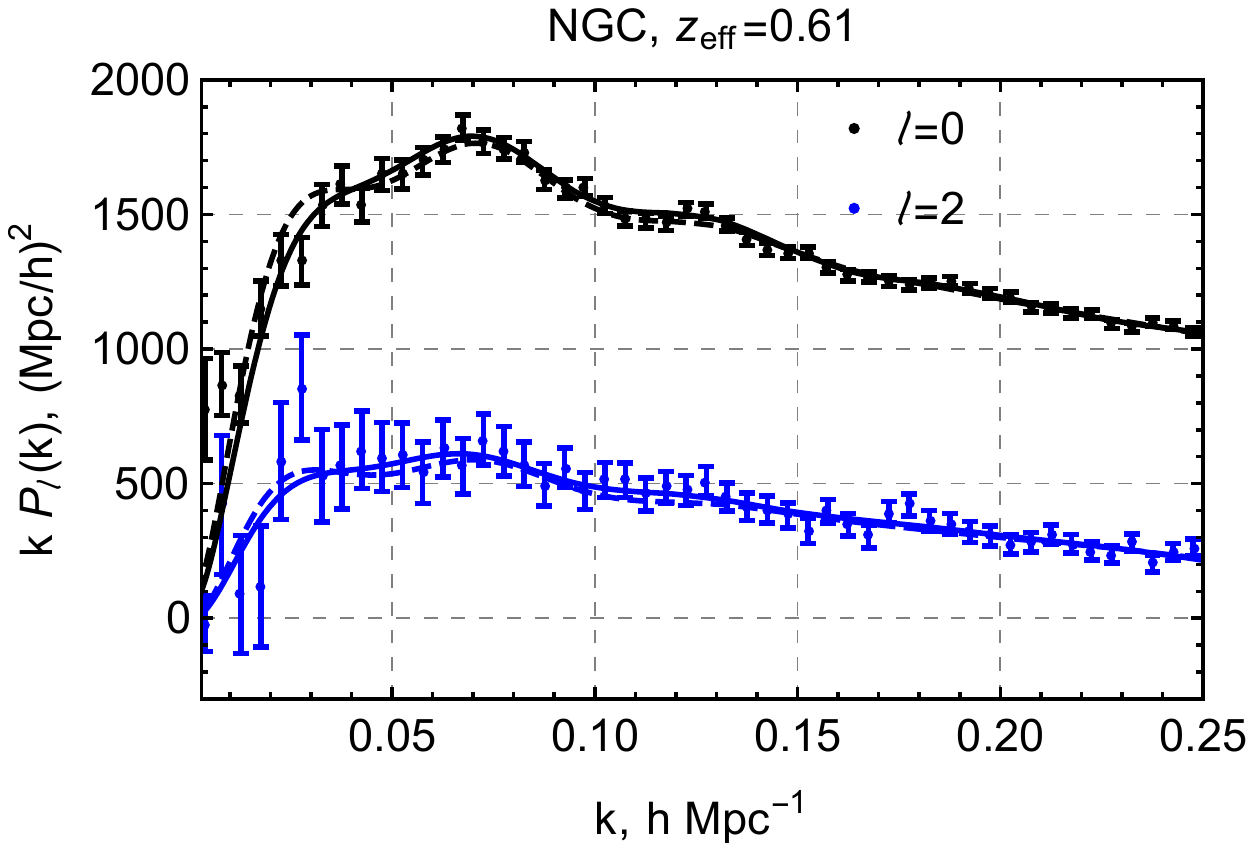}
\includegraphics[width=0.49\textwidth]{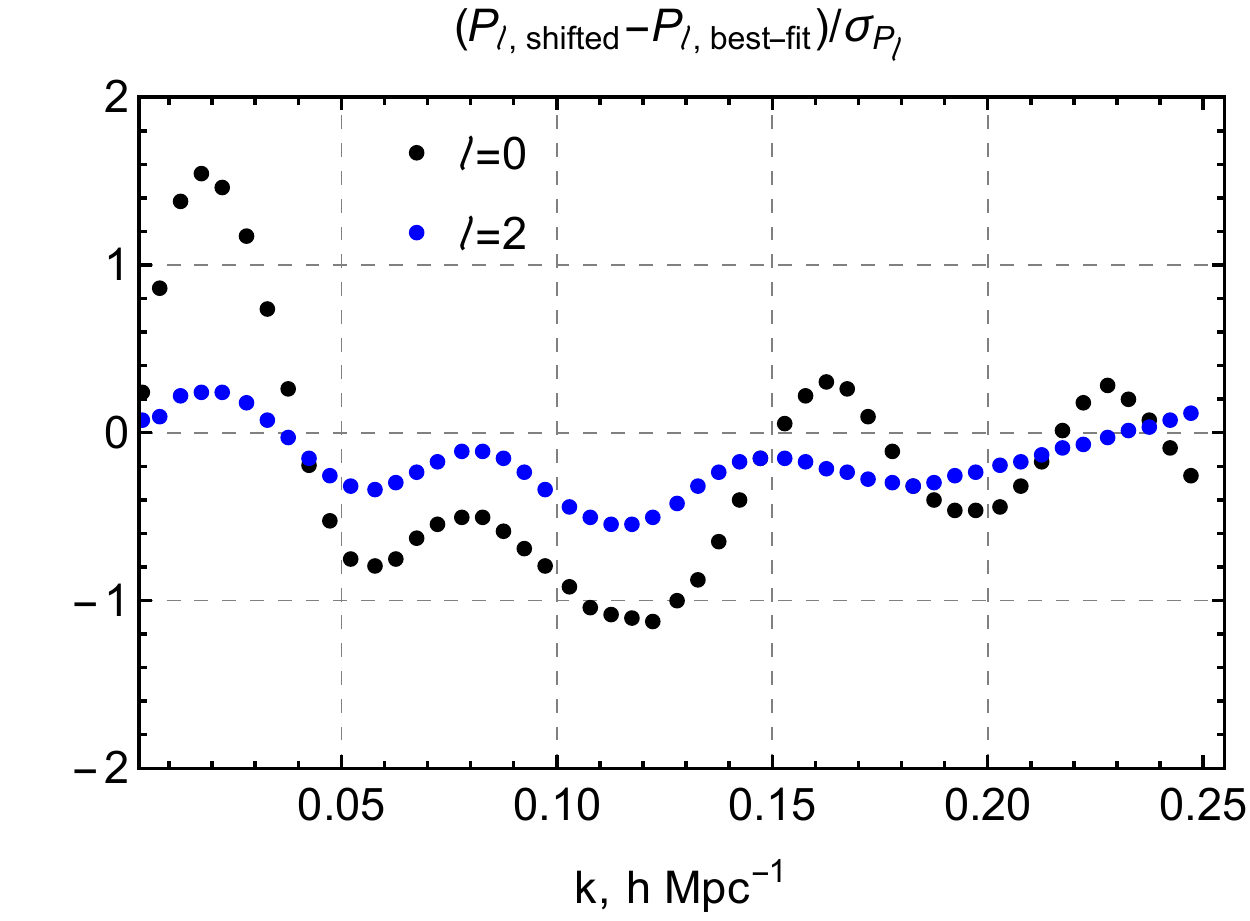}
\includegraphics[width=0.49\textwidth]{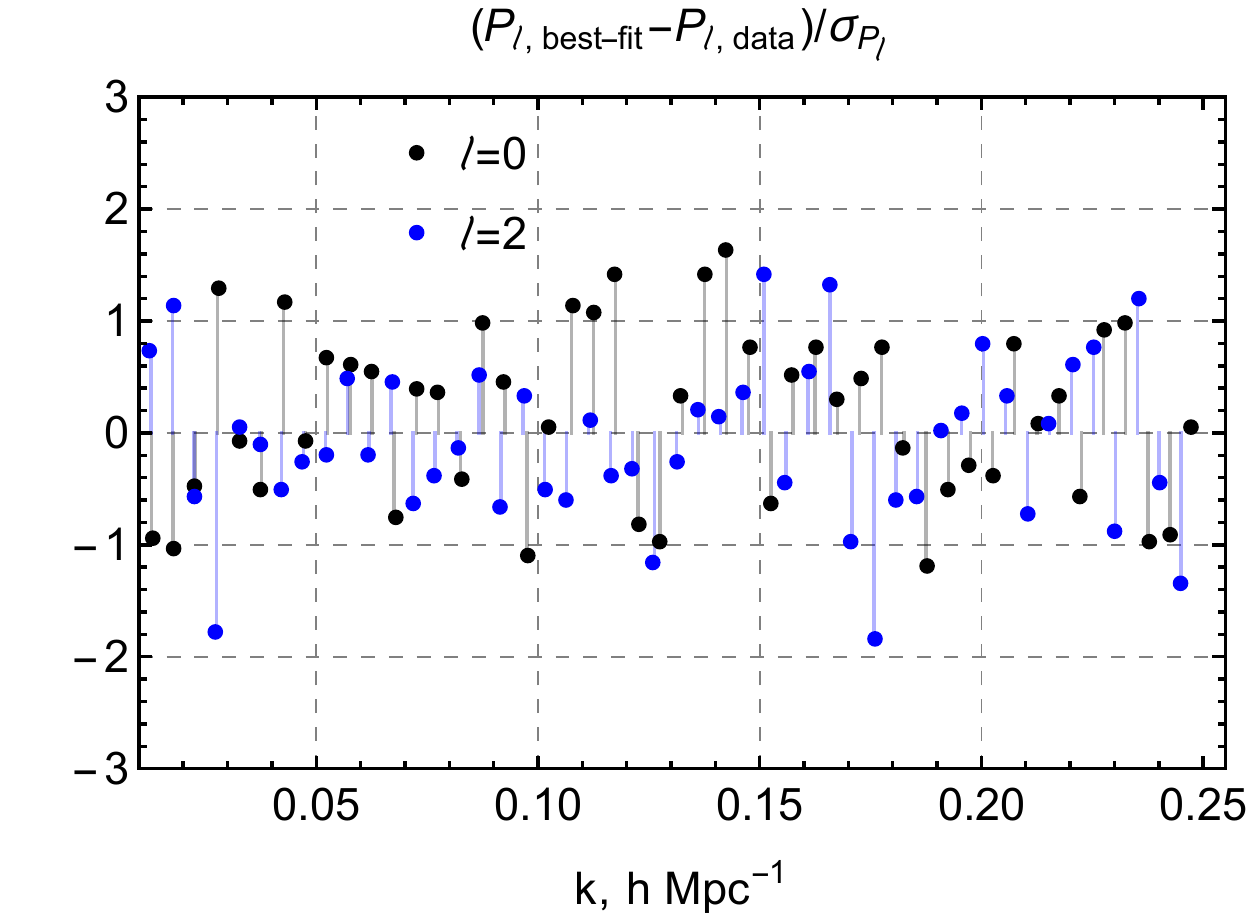}
\includegraphics[width=0.49\textwidth]{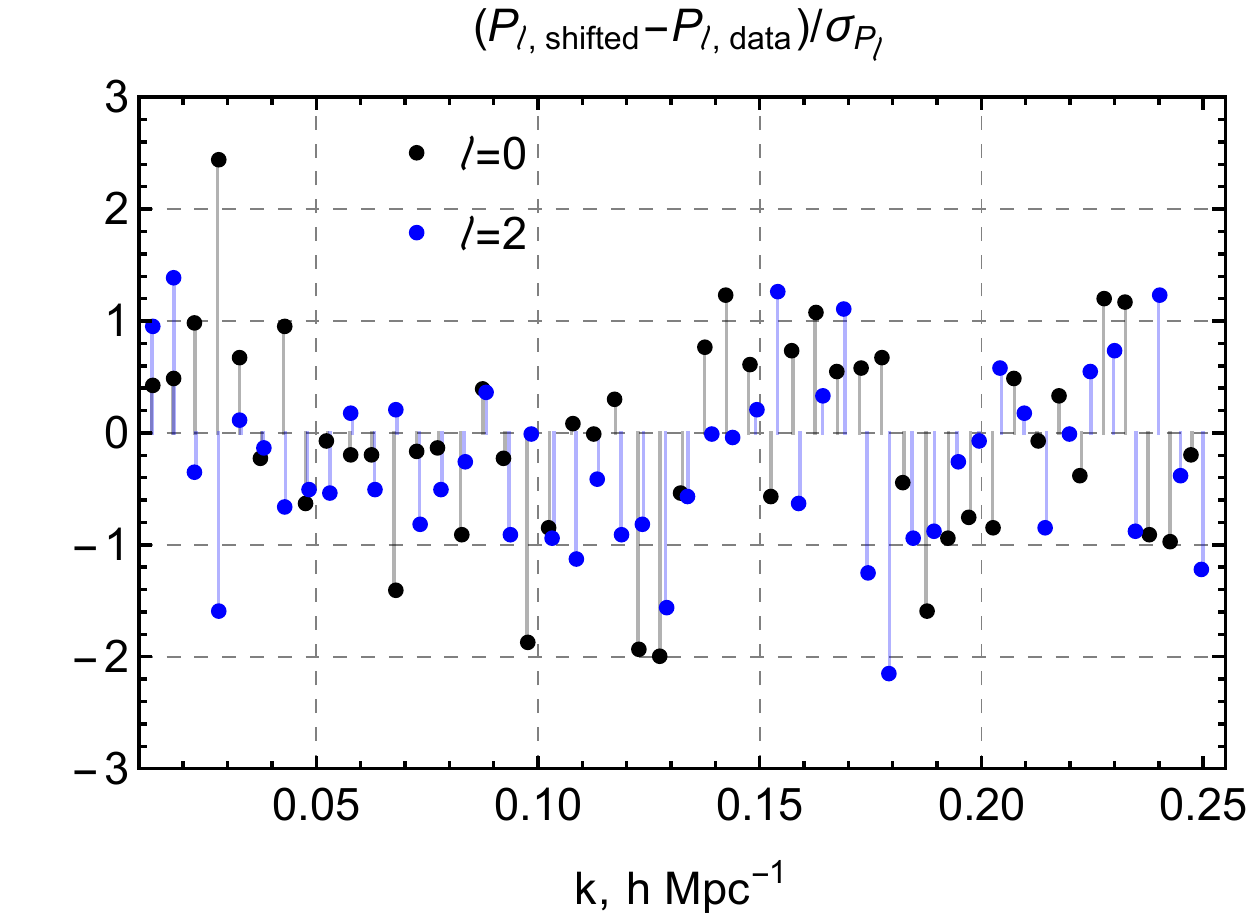}
\caption{\textit{Upper left panel:} the high-z NGC data along with the best fitting theory curves (solid lines) 
and a prediction of the test model with $\omega_{cdm}$ shifted by $3\sigma$ (dotted lines), for which we have refitted the other parameters.
\textit{Upper right panel:} the residuals between the two models $\Delta P_\ell = P_{\ell,\,{\rm shifted}}-P_{\ell,\,{\rm best-fit}}$ divided by the data errors.
\textit{Lower panels:} the residuals between the models 
$P_{\ell,\,{\rm best-fit}}$ (\textit{left panel}),
$P_{\ell,\,{\rm shifted}}$ (\textit{right panel})
and the data.
}    
\label{fig:shift}
\end{center}
 \end{figure}

Finally, in order to get convinced that our constraints are indeed driven by the shape we have performed
the following exercise. We have taken the best-fit power spectrum from the NGC high-z datasample
(which has the biggest volume) and 
compared it to the spectrum computed for a 
model with $\omega_{cdm}$ shifted by $3\sigma$ away from the best-fit value. 
At face value, this leads to an extremely large difference in $\chi^2$ because $\omega_{cdm}$
enters various normalizations. However, much of this difference is 
absorbed into the nuisance parameters and cosmological parameters. 
Thus, we have refitted all the parameters of the ``shifted'' trial model.
The results are shown in Fig.~\ref{fig:shift}, where one can see the two trial spectra 
and the difference between them in terms of the statistical error on the power spectrum $\sigma_{P_\ell}(k)$. 
The difference between $\chi^2$ values of the two models is $\Delta \chi^2=13.6$. 
Clearly, the variation in $\omega_{cdm}$ is detectable. It cannot be undone by a simple shift in $h$: 
either the BAO wiggles or the slope will be wrong.

% depends only on $H_0$ and $\omega_m$ and hence, 
% the geometric location of the BAO yields directly $H_0$. Moreover, $D_V$ is a very slow function of $\Omega_m$
% at small redshifts relevant for galaxy surveys.

% However, the crucial difference w.r.t. the CMB 
% is that $z_{\rm eff}$ and one has much less freedom in the expansion history.

\subsection{Mock data analysis}

\begin{figure}[h!]
\begin{centering}
\includegraphics[width=0.8\textwidth]{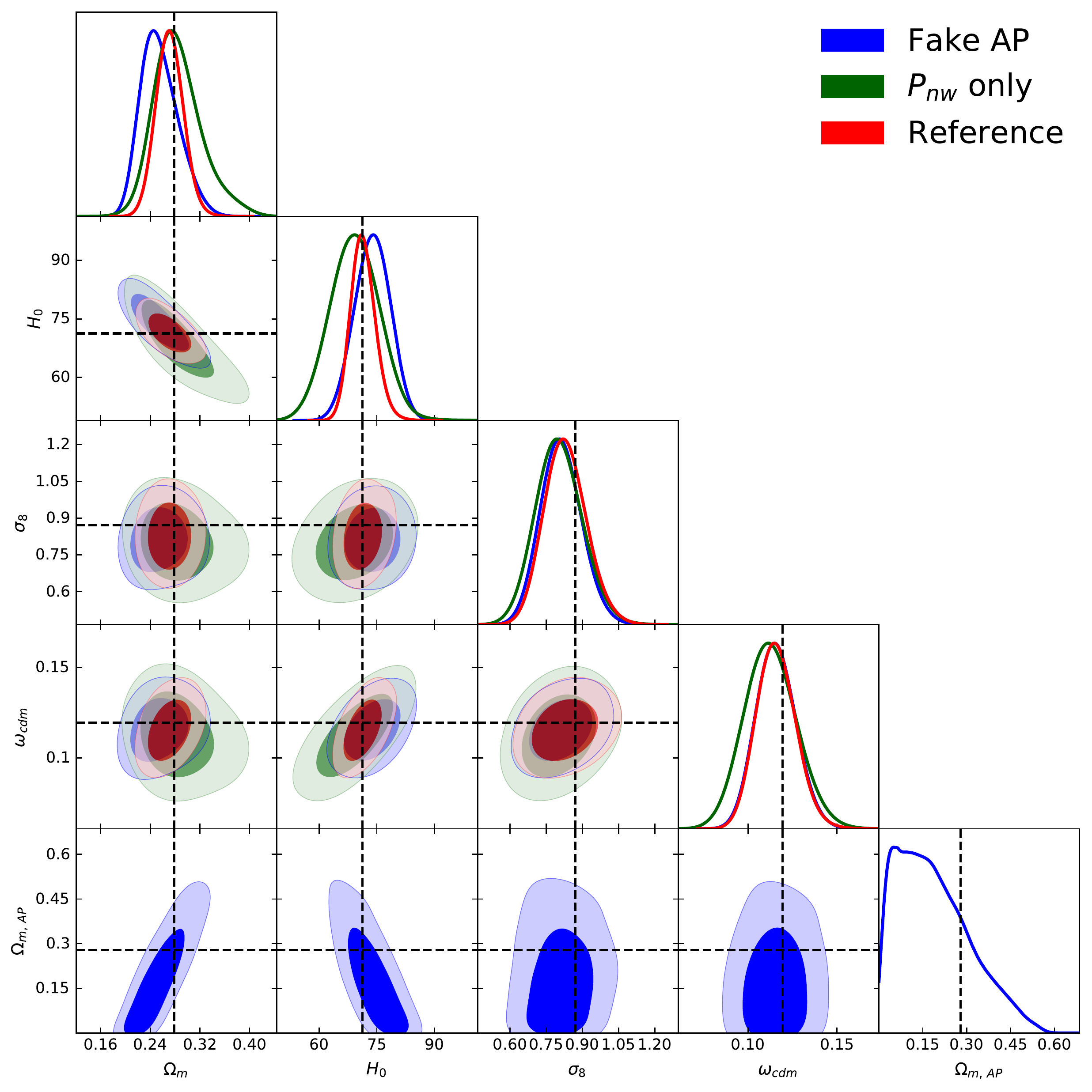}
\caption{
The 2d and 1d posterior distributions for the parameters of the 
mock likelihood analysis. 
Black dashed lines reflect the fiducial values used to generate the mock datavectors.
See the text for further details. $H_0$ is quoted in units [km/s/Mpc].
}  
\label{fig:tests}  
\end{centering}
\end{figure}

In this Section we will be mainly focused on disentangling the shape, geometry and AP information, 
which are most relevant for the constraints on $\Omega_m$
and $H_0$.
To quantify the amount of information coming from them we 
analyze several mock BOSS-like likelihoods. 
We use our theoretical pipeline to 
generate datavectors for a random set of 
cosmological and nuisance parameters extracted from 
the MCMC chains for the low-z NGC mocks.\footnote{These are: $\omega_b=0.02215,\,\omega_{cdm}=0.1194,\,\sigma_8 = 0.867,\,b_1=1.73,\,b_2=-0.34,\,b_{\mathcal{G}_2}=0.06,\,c_0^2=36.7\,[\Mpc/h]^2,\,c_2^2=53.3\,[\Mpc/h]^2,\,P_{\text{shot}}=3.2\cdot 10^3 \,[\Mpc/h]^3,\,\tilde{c}=382\,[\Mpc/h]^4$. Note that these parameters are within 1$\sigma$-distance from the fiducial values used in mock catalogs.}
We analyze these mock spectra using the same pipeline in order 
to obtain the reference posterior distribution.
We assume the same priors as in our main analysis (see Tab.~\ref{tab:priors}), 
and additionally put the following Gaussian prior on $\omega_b$:
\be
\omega_b = (2.214 \pm 0.015)\times 10^{-2} \,,
\ee
which is equivalent to the BBN (or Planck) prior on $\omega_b$ used in our baseline analysis, 
but centered at the fiducial value
used in the mocks. 
The reference posterior contours 
are shown in Fig.~\ref{fig:tests},
the 1d marginalized limits are given in Table.~\ref{tab:fake}.
Note that they match the results of our analysis 
of the mock catalogs and the real data for the same data chunk.

\begin{table}[ht!]
\begin{center}
 \begin{tabular}{|c|c|c|} 
 \hline
Reference & best-fit & mean  $\pm 1\sigma$  \\ [0.5ex] \hline\hline
$\omega_{cdm}$ & $ 0.1154$ & $ 0.1157 \pm  0.0105$  \\ \hline
 $H_0$ & $ 71.26$ & $ 71.39 \pm  3.15$  \\ \hline \hline
 $\Omega_m$ & $ 0.271$ & $ 0.271 \pm  0.021$  \\ \hline
\end{tabular}
 \begin{tabular}{|c|c|c|} 
 \hline
$P_{\text{nw}}$ only & best-fit & mean  $\pm 1\sigma$  \\ [0.5ex] \hline\hline
$\omega_{cdm}$ & $ 0.1207$ & $ 0.1125 \pm  0.0140$  \\ \hline
 $H_0$ & $ 69.06$ & $ 69.28 \pm  6.23$  \\ \hline \hline
 $\Omega_m$ & $ 0.291$ & $ 0.284 \pm  0.038$  \\ \hline
\end{tabular}
 \begin{tabular}{|c|c|c|} 
 \hline
Fake AP & best-fit & mean  $\pm 1\sigma$  \\ [0.5ex] \hline\hline
$\omega_{cdm}$ & $ 0.1191$ & $ 0.1157 \pm  0.0108$  \\ \hline
 $H_0$ & $ 71.26$ & $ 73.84 \pm  4.55$  \\ \hline
 $\Omega_{m,\,\text{AP}}$ & $ 0.277$ & $ 0.189_{-0.165}^{+0.066}$  \\ \hline \hline
  $\Omega_m$ & $ 0.278$ & $ 0.255^{+0.025}_{-0.035}$  \\ \hline 
\end{tabular}
\caption{
The outcomes of our mock data analysis for a fiducial datavector with the
NGC low-z covariance. The shown are: the reference sample (upper left table)
that corresponds to the actual BOSS data, the sample without the BAO wiggles 
(`$P_{\text{nw}}$ only', upper right table), 
and the results obtained in the analysis of the reference sample
assuming that the AP effect is controlled by a separate parameter $\Omega_{m,\,\text{AP}}$,
which has nothing to do with the real $\Omega_{m}$ (`fake AP').
$H_0$ is quoted in units [km/s/Mpc].
}
\label{tab:fake}
\end{center}
\end{table}

To estimate the information content of the BAO wiggles, 
we generate and analyze a datavector without them. 
A similar approach was previously employed in Ref.~\cite{Hamann:2010pw,Chudaykin:2019ock}.
To that end we use 
the same wiggly-smooth decomposition procedure that performs IR resummation. These mock non-wiggly 
data are then analyzed with a modified pipeline that does not have the BAO wiggles in theoretical
template too.\footnote{We emphasize that we only removed the BAO wiggles from the power spectrum templates.
All other baryonic effects, e.g. the Jeans suppression, 
are present in our theory model.} 
The results of this analysis and the reference posteriors 
are shown in Fig.~\ref{fig:tests}. 
The 1d marginalized limits are given in Table.~\ref{tab:fake}.

First, we see that the constraints on $\omega_{cdm}$ are similar in the BAO and no-BAO cases. 
This means that the BAO wiggles represent only a part of the shape information. 
However, their presence is crucial for 
constraining $H_0$ through the geometric information. 
To see this, let us focus on the degeneracy
directions seen in the $\omega_{cdm}-H_0$ panel.
These are $\omega_{cdm}/H_0$ for the no-BAO case and 
$\omega_{cdm}/H_0^{2.5}$ with the BAO. The first one exactly corresponds to the power spectrum shape 
(or the location of the power spectrum peak in units Mpc/$h$). 
The second one is likely a combination of the location of the power spectrum peak and
BAO wiggles (set by $\omega_{cdm}/H_0^5$, see \eqref{eq:degts}). 
As a consequence, in the realistic BAO case the projection of the
degeneracy contour onto the $H_0$ plane is twice more narrow compared to the no-BAO contour.
We point out once again that in the BAO case the principle component of the geometric information happens 
to be quite perpendicular to $\Omega_m$, which explains 
why this combination of $\omega_{cdm}$ and $H_0$ is well constrained. 
Once we remove the BAO, the principal component changes and the projection onto the $\Omega_m$ plane
becomes twice larger than before.

Now let us focus on the Alcock-Paczynski information. 
To quantify its amount we take the reference datavector
with the BAO wiggles and analyze it assuming the matter
density fraction that enters the geometric distances and the AP effect $\Omega_{m,\,\text{AP}}$ to be different 
from the true $\Omega_m$. 
Technically, it is equivalent to considering a model where the late-time geometric expansion
is controlled by an additional parameter, which is not related to the ones fixing the shape of the matter power spectrum. 
We use the following flat prior on $\Omega_{m,\,\text{AP}}$:
\be
\Omega_{m,\,\text{AP}} \in (0,1)\,.
\ee
The outcome of this analysis is also displayed in Fig.~\ref{fig:tests} and in Table.~\ref{tab:fake} (``fake AP'').
One first notices that the constraints on $\omega_{cdm}$ are identical in the
reference and the ``fake AP'' cases, which implies that the shape information is not diluted by the AP distortions.
This result explicitly proves our intuition that $\omega_{cdm}$ is measured directly from the power spectrum shape regardless of the
late-time expansion. 

However, since the location of the BAO wiggles mainly constrains $D_V$,
the presence of an additional parameter entering $D_V$ makes it harder to translate this constraint directly to $H_0$.
This explains why the constraints on the physical $\Omega_m$ 
and $H_0$ degrade by $\sim 50\%$.
These limits, however, are not significantly worse than the reference ones because 
the degeneracy between $H_0$ and $\Omega_{m,\,\text{AP}}$
gets eventually broken by the quadrupole, 
which essentially constrains $\Omega_{m,\,\text{AP}}$ in our example.\footnote{To be more precise, the quadrupole constrains the combination $H(z_{\rm eff})D_A(z_{\rm eff})$ evaluated with $\Omega_{m,\,\text{AP}}$ instead of actual $\Omega_m$.}
The reason why the coupling between $D_V$ and $H_0$ does not dramatically worsen the $H_0$ measurement 
is that $D_V$ has a very weak sensitivity to $\Omega_{m,\,\text{AP}}$
the redshifts of interest,
and at leading order\footnote{At first non-vanishing order in $\Omega_{m,\,\text{AP}}$ one finds $D_V\propto h^{-1}\Omega^{-0.06}_{m,\,\text{AP}}$.} $D_V\sim H_0^{-1}$ even in our unphysical example with 
$\Omega_{m}\neq \Omega_{m,\,\text{AP}}$.
Note that the posterior distribution of $\Omega_{m,\,\text{AP}}$ is highly asymmetric;
its upper limit is set by the quadrupole information (which decouples $H$ and $D_A$ from $D_V$), 
while the lower limit is prior-driven.

The upshot of this discussion is the following. Our constraints on $\omega_{cdm}$ are driven by the 
power spectrum shape, $H_0$ is set by the geometric information (extracted from $D_V$) and $\Omega_m$ is a combination of the two.
As for the AP effect, it is absolutely superseded by the shape
and geometric information in $\Lambda$CDM. 
The situation is different for extensions 
of the minimal $\Lambda$CDM, which we discuss now.

%%%%%%%%%%%%%%%%%%%%%%%%%%%%%%%%%%%%%%%%%%%%%%%%%%%%%%
\section{Distance Measurements}
\label{sec:dist}

This section has three main objectives: 

(a) establish the connection between our method 
and the one commonly used in the previous BOSS full-shape analyses with scaling parameters
($\alpha$-analysis in what follows),

(b)  show that our analysis with the Planck priors on $\omega_b,\omega_{cdm}$
is equivalent to the $\alpha$-analysis if one takes into account that $D_A(z_{\text{eff}})$ and $H(z_{\text{eff}})$
are coupled in $\Lambda$CDM,

(c) show that the $\alpha$-analysis is valid if 
one wants to constrain some generic late-time expansion models, 
for which the distance measurements become a leading source of information.

The analyses performed in this section have a demonstrative character. 
They aim to illustrate the relation between our method and the $\alpha$-analysis
in different settings.
We present results obtained for the BOSS NGC data samples only.

\begin{figure}[h!]
\includegraphics[width=0.49\textwidth]{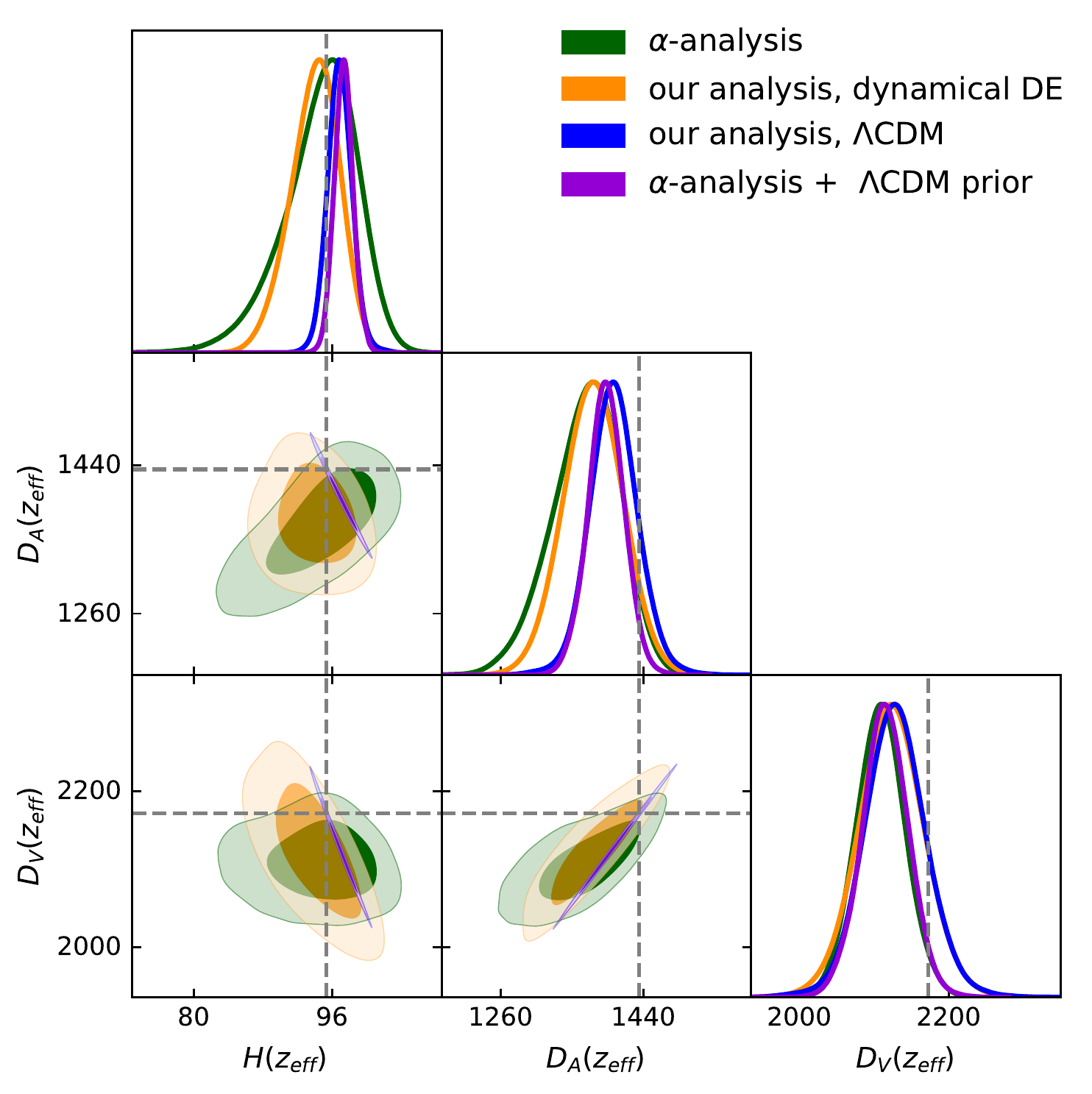}
\includegraphics[width=0.49\textwidth]{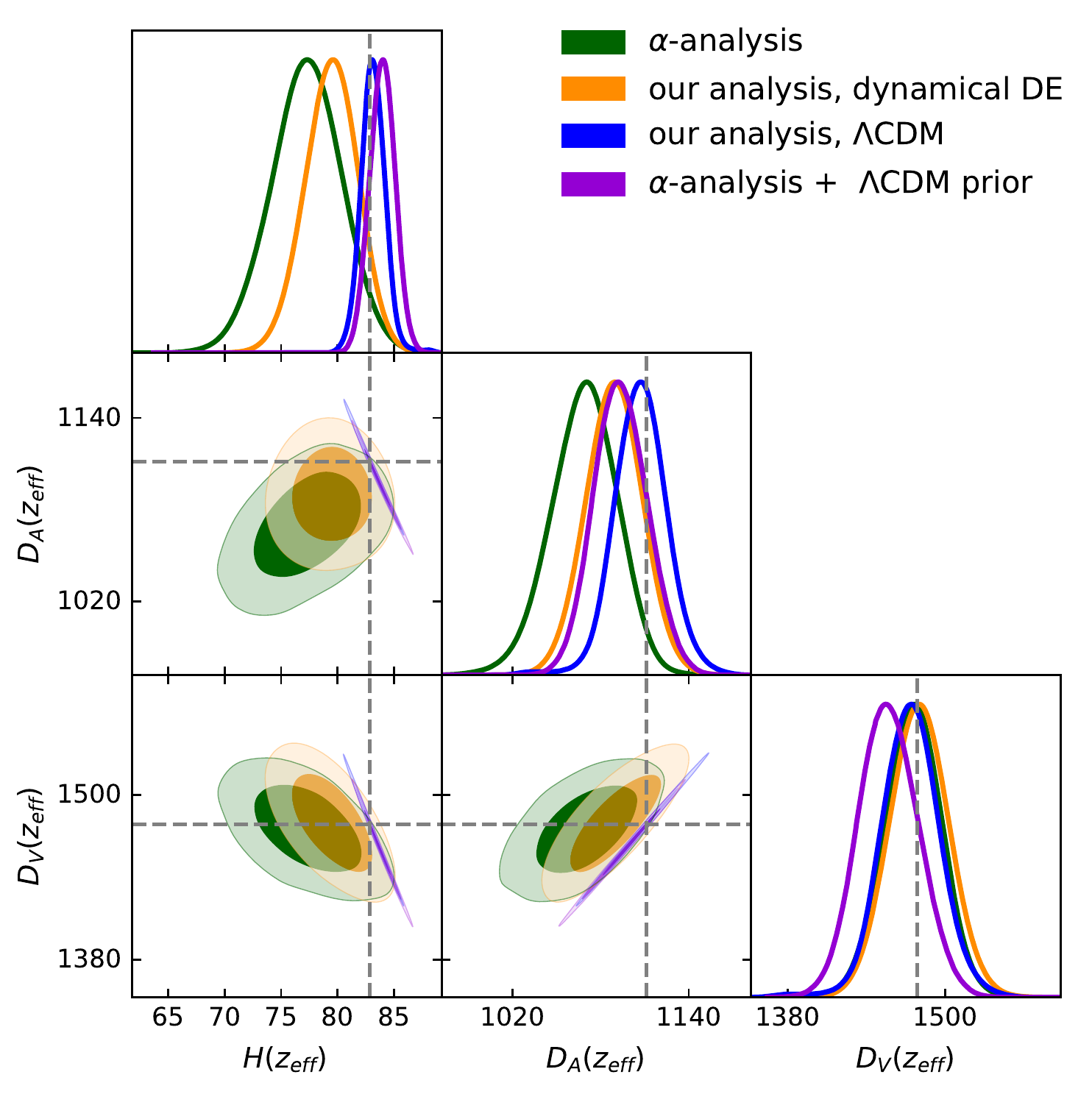}
\caption{
The posterior contours for $H(z_{\text{eff}}),D_A(z_{\text{eff}}),D_V(z_{\text{eff}})$
for the NGC high-z (left panel) and 
low-z (right panel) samples. We show the results of our analysis 
for $\L$CDM and the dynamical dark energy model, both with the Planck priors on
$\omega_b$ and $\omega_{cdm}$. We also show the results of 
the model-independent $\alpha$-analysis without any priors and
with the $\L$CDM prior that reflects the coupling between $D_A$ and $H$. 
Dashed lines represent 
the Planck best-fit values.
The values of $H$ are quoted in units of [km/s/Mpc], $D_A$ and $D_V$
in [Mpc].
\label{fig:distageneric}
}    
\end{figure}
 
For the purposes of this Section we have run an $\alpha$-analysis using the same methodology as 
the previous BOSS FS studies. 
The details of this analysis are given in Appendix~\ref{app:alphas}.
The $\alpha$-analysis computes the PS shape for a Planck-like cosmology and does not vary 
it in the MCMC chains.
The main idea behind the $\alpha$-analysis is that once the physical densities 
of dark matter and baryons are fixed,
the leading response to a change in cosmological
parameters should be through $r_d H(z_{\text{eff}})$ and $r_d/D_A(z_{\text{eff}})$. 
However, fixing the shape is equivalent to fixing $\omega_{cdm}$ and $\omega_b$, 
which also fix $r_d$. 
Hence, the $\alpha$-analysis and our method should technically coincide if we fix $r_d$ 
in the $\alpha$-analysis and $\omega_b,\omega_{cdm}$ in our analysis.
We stress that unlike the pure BAO-studies, fixing the shape and treating $r_d$ as a free parameter in the full-shape 
studies is unphysical. In any realistic model $r_d$ and the transfer functions' shape are 
controlled by the same parameters. Thus, the $\alpha$-analysis of the full-shape power spectrum
actually measures the absolute distances
$D_V$ and $D_A$ and not $r_d/D_V$ or $r_d/D_A$, which would be the case for the BAO-only study.

Another important observation is that 
the $\alpha$-analysis assumes $H(z_{\text{eff}})$ and $D_A(z_{\text{eff}})$
to be completely independent from each other, while in reality they are related by construction,
see Eq.~\eqref{eq:Dadef}.
% \be
% \label{eq:Dadef}
% D_A(z)=\frac{1}{1+z}\int_0^z\frac{dz'}{H(z')}\,. 
% \ee
In $\Lambda$CDM a prior on $\omega_{cdm}$ and $\omega_b$ completely fixes the relation between $D_A$ and $H$ 
at any redshift.
Once we impose this relation,\footnote{To that end 
we have run mock MCMC chains that fitted $D_A$ and $H$ from the Gaussian likelihood for $r_d$
assuming $\L$CDM. 
Then we found the principal component of these variables 
and imposed this as a prior in the MCMC chains 
which sampled $\alpha$ parameters.
} the limits on $H$ and $D_A$ from the $\alpha$-analysis
coincide with the limits obtained with our method 
(modulo some small difference which can be explained by the 
use of slightly different priors and theoretical models, see App.~\ref{app:alphas}
for more detail).
This can be seen in Fig.~\ref{fig:distageneric} and Tabs.~\ref{tab:distz3}, \ref{tab:distz1}.

\begin{table}[ht!]
% \vspace{-1cm}
\begin{center}
 \begin{tabular}{|c|c|c|} 
 \hline
$\alpha$-parm. & best-fit & mean  $\pm 1\sigma$  \\ [0.5ex] \hline\hline
 $H(z_{\text{eff}})$ & $ 96.69$ & $ 94.11 \pm  4.98$  \\ \hline
 $D_A(z_{\text{eff}})$ & $ 1395$ & $ 1364 \pm 47 $  \\ \hline
 $F_{\text{AP}}(z_{\text{eff}})$ & $ 0.723$ & $ 0.690 \pm 0.054 $  \\ \hline
 $D_V(z_{\text{eff}})$ & $ 2118$ & $ 2109 \pm 40 $  \\ \hline
\end{tabular}
\vspace{0.2cm}
\begin{tabular}{|c|c|c|} 
 \hline
$\alpha$-parm.$+\L$CDM & best-fit & mean  $\pm 1\sigma$  \\ [0.5ex] \hline\hline
 $H(z_{\text{eff}})$ & $ 97.28$ & $ 97.34 \pm  1.27$  \\ \hline
 $D_A(z_{\text{eff}})$ & $ 1393$ & $ 1392 \pm 26 $  \\ \hline
 $F_{\text{AP}}(z_{\text{eff}})$ & $ 0.7278$ & $ 0.7275 \pm 0.0045 $  \\ \hline
 $D_V(z_{\text{eff}})$ & $ 2115$ & $ 2113 \pm 36 $  \\ \hline
\end{tabular}
\begin{tabular}{|c|c|c|} 
 \hline
DEE & best-fit & mean  $\pm 1\sigma$  \\ [0.5ex] \hline\hline
 $H(z_{\text{eff}})$ & $ 96.03$ & $ 94.05 \pm  2.81$  \\ \hline
 $D_A(z_{\text{eff}})$ & $ 1379$ & $ 1378 \pm 37 $  \\ \hline
 $F_{\text{AP}}(z_{\text{eff}})$ & $ 0.710$ & $ 0.696 \pm 0.028 $  \\ \hline
 $D_V(z_{\text{eff}})$ & $ 2109$ & $ 2123 \pm 43$  \\ \hline
 $H_0$ & $ 72.9$ & $ 75.9 \pm  6.2$  \\ \hline
\end{tabular}
\vspace{0.2cm}
\begin{tabular}{|c|c|c|} 
 \hline
$\L$CDM & best-fit & mean  $\pm 1\sigma$  \\ [0.5ex] \hline\hline
 $H(z_{\text{eff}})$ & $ 96.32$ & $ 96.85 \pm  1.47$  \\ \hline
 $D_A(z_{\text{eff}})$ & $ 1412$ & $ 1403 \pm 31 $  \\ \hline
 $F_{\text{AP}}(z_{\text{eff}})$ & $ 0.7304$ & $ 0.7293 \pm 0.0053 $  \\ \hline
 $D_V(z_{\text{eff}})$ & $ 2141$ & $ 2128 \pm 42 $  \\ \hline
 $H_0$ & $ 68.8$ & $ 69.4 \pm  2.0$  \\ \hline
\end{tabular}
\caption{
Distance measurements for the high-z NGC sample ($z_{\text{eff}}=0.61$). Upper panel: $\alpha$-analysis 
without and with the $\L$CDM priors, in left and right tables, correspondingly.
Lower panel: our analysis for the dynamical dark energy model (left table)
and $\L$CDM (right table) with the Planck priors on $\omega_b$
and $\omega_{cdm}$. The values of $H$ are quoted in units of [km/s/Mpc], $D_A$ and $D_V$
in [Mpc].
}
\label{tab:distz3}
\vspace{0.5cm}
% \end{center}
% \end{table}
% \begin{table}[ht!]
% \begin{center}
 \begin{tabular}{|c|c|c|} 
 \hline
$\alpha$-parm. & best-fit & mean  $\pm 1\sigma$  \\ [0.5ex] \hline\hline
 $H(z_{\text{eff}})$ & $ 78.04$ & $ 77.24 \pm  3.12$  \\ \hline
 $D_A(z_{\text{eff}})$ & $ 1072$ & $ 1069 \pm 23 $  \\ \hline
 $F_{\text{AP}}(z_{\text{eff}})$ & $ 0.385$ & $ 0.380 \pm 0.020 $  \\ \hline
 $D_V(z_{\text{eff}})$ & $ 1473$ & $ 1475 \pm 22 $  \\ \hline
\end{tabular}
\vspace{0.2cm}
\begin{tabular}{|c|c|c|} 
 \hline
$\alpha$-parm.$+\L$CDM  & best-fit & mean  $\pm 1\sigma$  \\ [0.5ex] \hline\hline
 $H(z_{\text{eff}})$ & $82.92 $ & $ 83.95 \pm  1.07$  \\ \hline
 $D_A(z_{\text{eff}})$ & $ 1111$ & $ 1094 \pm 17 $  \\ \hline
 $F_{\text{AP}}(z_{\text{eff}})$ & $ 0.4240$ & $ 0.4225 \pm 0.0014 $  \\ \hline
 $D_V(z_{\text{eff}})$ & $ 1478$ & $ 1457 \pm 22 $  \\ \hline
\end{tabular}
\vspace{0.2cm}
\begin{tabular}{|c|c|c|} 
 \hline
DDE& best-fit & mean  $\pm 1\sigma$  \\ [0.5ex] \hline\hline
 $H(z_{\text{eff}})$ & $ 79.68$ & $ 79.46 \pm  2.19$  \\ \hline
 $D_A(z_{\text{eff}})$ & $ 1086$ & $ 1089 \pm 18 $  \\ \hline
 $F_{\text{AP}}(z_{\text{eff}})$ & $ 0.398$ & $ 0.398 \pm 0.013 $  \\ \hline
 $D_V(z_{\text{eff}})$ & $ 1475$ & $ 1480 \pm 21 $  \\ \hline
 $H_0$ & $ 77.7$ & $ 75.6 \pm  4.7$  \\ \hline
\end{tabular}
\begin{tabular}{|c|c|c|} 
 \hline
 $\L$CDM & best-fit & mean  $\pm 1\sigma$  \\ [0.5ex] \hline\hline
 $H(z_{\text{eff}})$ & $ 83.89$ & $ 83.16 \pm  1.11$  \\ \hline
 $D_A(z_{\text{eff}})$ & $ 1094$ & $ 1107 \pm 18 $  \\ \hline
 $F_{\text{AP}}(z_{\text{eff}})$ & $ 0.4225$ & $ 0.4236 \pm 0.0015 $  \\ \hline
 $D_V(z_{\text{eff}})$ & $ 1458$ & $ 1473 \pm 23 $  \\ \hline
 $H_0$ & $ 68.6$ & $ 67.7 \pm  1.4$  \\ \hline
\end{tabular}
\caption{
Distance measurements for the low-z NGC sample ($z_{\text{eff}}=0.38$). Upper panel: $\alpha$-analysis 
without and with the $\L$CDM priors, in left and right tables, correspondingly.
Lower panel: our analysis for the dynamical dark energy model (left table)
and $\L$CDM (right table) with the Planck priors on $\omega_b$
and $\omega_{cdm}$.
The values of $H$ are quoted in units of [km/s/Mpc], $D_A$ and $D_V$
in [Mpc].
}
\label{tab:distz1}
\end{center}
\end{table}

One can notice that 
the $\L$CDM priors have a very dramatic effect on the measurements
of $H$ and $D_A$, 
whose errorbars reduce by a factor of few compared to the basic $\alpha$-analysis
without any priors. 
However, the effect on $D_V$ is not very strong.\footnote{It is 
useful to compare our limits with the ones obtained in the main BOSS 
Fourier-space BAO and FS power spectrum analyses,
see Refs.~\cite{Beutler:2016arn,Beutler:2016ixs}:
\begin{align}
% \begin{split}
\nonumber
& D_V(z_{\text{eff}}=0.38) = 1493 \pm 28\;[\Mpc]\,, \qquad  D_V(z_{\text{eff}}=0.61) = 2133 \pm 36\;[\Mpc]\,,\qquad (\text{FS})\,,\\
\nonumber
& D_V(z_{\text{eff}}=0.38) = 1479 \pm 23\;[\Mpc]\,, \qquad  D_V(z_{\text{eff}}=0.61) = 2141 \pm 36\;[\Mpc]\,,\qquad (\text{pre-recon BAO})\,,\\
\nonumber
& D_V(z_{\text{eff}}=0.38) = 1474 \pm 17\;[\Mpc]\,, \qquad  D_V(z_{\text{eff}}=0.61) = 2144 \pm 20\;[\Mpc]\,,\qquad (\text{post-recon BAO})\,.
% \end{split} 
\end{align}
Note that these limits were obtained by using 
slightly different datasamples (NGC+SGC), 
$k_{\text{max}}$ cuts and the theoretical model, and hence should be 
compared to our results shown in this section with some caution.
}  
This reflects the observation that $D_V$ is the best measured combination 
of $D_A$ and $H$, which is extracted directly from the monopole, while $H$ and $D_A$
are measured from the quadrupole, which has significantly larger statistical errors and features 
much less pronounced BAO wiggles.
In other words, our analysis shows that the good constraints on $H$ and $D_A$ obtained in $\L$CDM are prior-driven,
these two parameters are not measured directly. 
$D_V$ is the only one actually measured prior-independent distance in $\L$CDM.

In order to explicitly illustrate that the principal distance best measured from our analysis 
is always given by $D_V$ even in extended cosmological models,
we analyze the BOSS data assuming a generic dynamical dark energy (DDE) model,
described by the following Friedman equation:
\be
\label{eq:dde}
H^2(z) = H_0^2\left( \O_m(1+z)^3+ \Omega_\L + \Omega_{de}(1+z)^{3\left(1+w_0 + w_a \frac{z}{1+z}\right)} \right) \,.
\ee 
We assume the following flat priors on $w_a$ and $w_0$:
\be 
\label{dde:priors}
\Omega_{de}\in (0,1)\,,\quad w_0\in (-2,-0.33)\,,\quad w_a\in (-5,5)\,,
\ee
and keep the Planck priors on $r_d$ and $\omega_b$. 
As far as the other cosmological and nuisance parameters are concerned, 
we use the same priors as in our baseline analysis, see Tab.~\ref{tab:priors}.
Note that a model similar to \eqref{eq:dde} has been constrained 
in the previous BOSS analyses, e.g.~\cite{Alam:2016hwk}. 

The results for the NGC high-z and low-z data are presented in Fig.~\ref{fig:distageneric} and Tables~\ref{tab:distz3},~\ref{tab:distz1}.
The first relevant observation is that the background parametrization \eqref{eq:dde} 
is sufficient to decouple the radial and angular distances, so that the 
errorbars on $D_A$ and $H$ become comparable to the ones
obtained with a generic $\alpha$-analysis, 
and these two distances are not noticeably degenerate.
The second important observation is 
that the limit on $D_V$ is the same as in the $\Lambda$CDM case,
which confirms that $D_V$ is an actually measured distance 
that forms the principal component. 
Importantly, the relative precisions of its measurement 
from separate chunks are $1.4\%$ (low-z NGC) and $2\%$ (high-z NGC), which is comparable to our precision on $H_0$ 
in the $\L$CDM analysis.
The comparison between the DDE and $\L$CDM cases presented
in Tabs.~\ref{tab:distz3}, \ref{tab:distz1}
allows us to conclude that our precision on $H_0$ in $\L$CDM 
indeed originates 
from the precise $D_V$ measurements.

As far as the angular diameter distance $D_A$ is concerned,
its errorbars are the same in two models, 
but the mean values are noticeably shifted compared to the $\Lambda$CDM case. 
This shows that $D_A$ is 
fixed by the shape and geometric information in $\Lambda$CDM, 
but can take different values in the non-minimal extensions of this model.

Our measurement of $H$ and $D_A$ in the DDE model are
prior-driven, as can be deduced from comparing  
the corresponding $D_A-H$ contour with the one obtained from the $\alpha$-analysis, which 
did not assume any priors.\footnote{The limits on the parameters of the DDE model are also prior-dominated, which is why we do not quote them here.} 
Indeed, the $\alpha$-analysis reveals a clear degeneracy between $H$
and $D_A$ that corresponds to fixed $D_V$, while our DDE posterior does not show any significant 
degeneracy between $D_A$ and $H$ whatsoever. 
This merely reflects the fact that the quality of 
the quadrupole measurement is not good enough for 
a decent determination of these distances separately. 
This is why our MCMC sampler hits the prior boundaries before it starts seeing the $D_V$ degeneracy.
Finally, it is worth pointing out that the constraints on $H_0$
degrade significantly in the DDE model compared to the $\L$CDM case
as a consequence of increased parameter space, which cannot be constrained 
using the available distance information.

\begin{figure}[ht!]
% \centering \includegraphics[width=0.49\textwidth]{FAP_prior_all}
\centering \includegraphics[width=0.49\textwidth]{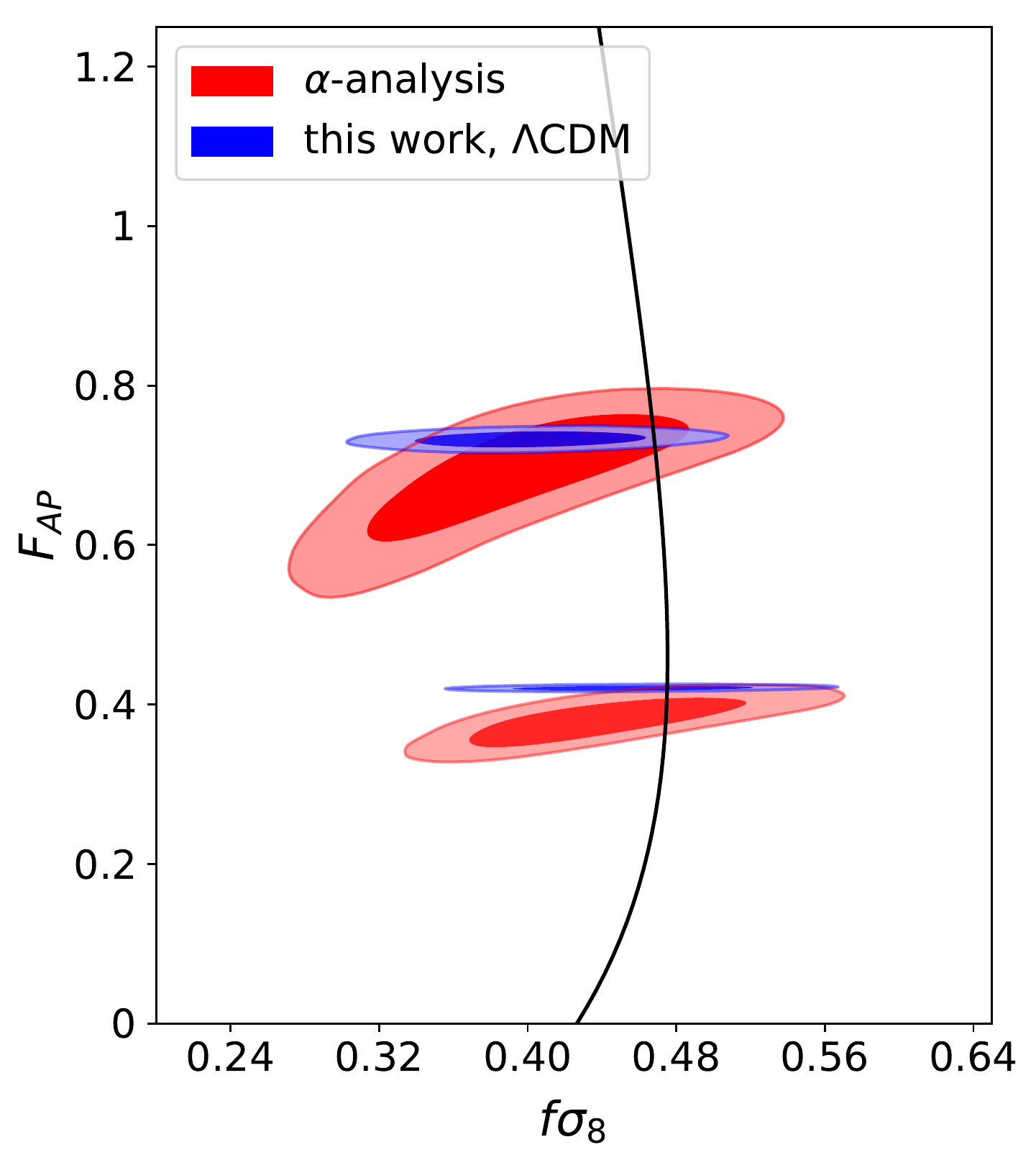}
\caption{ 
The constraints on the AP parameter (y-axis) and the filtered rms velocity fluctuation 
$f\sigma_8$ (x-axis) for two different redshift bins of the NGC BOSS data. 
The solid black line shows the prediction of the best-fit Planck 2018 cosmology \cite{Aghanim:2018eyx}.
We show the results of the generic $\alpha$-analysis (in red) 
and our analysis of the base $\Lambda$CDM, which varies the PS shape (in blue).
}    
\label{fig:fap}
\end{figure}

Our study suggests that the model-independent $\alpha$-parameterization 
might be too generic for some purposes.
Indeed, even in the case of a very general DDE model with quite loose 
priors on its parameters \eqref{dde:priors} we were not able to cover 
all the parameter space sampled by the $\alpha$-analysis.
Hence, one always has to impose proper priors on the $\alpha$-parameters
in order to use the distance information 
for precision constraints on non-minimal cosmological models.
% Otherwise the $\alpha$-analysis would sample large volumes of parameter space that 
% does not correspond to any physical model.
This is important for consistency when combining the BAO/FS data with 
external likelihoods (e.g. CMB, SNe or weak lensing) that assumed 
certain priors on the $\L$CDM extensions, e.g.~\cite{Aghanim:2018eyx}.

Finally, let us briefly comment on the so-called Alcock-Paczynski
parameter,
\be 
F_{\text{AP}}(z)=(1+z)D_A(z)H(z)\,,
\ee
which is often used to present the results of galaxy clustering measurements.
By construction this quantity depends only on $\Omega_m$ in $\Lambda$CDM.
One can easily check that by definition $F_{\text{AP}}$ must be roughly equal to $z$ at low redshifts,
where all cosmology dependence essentially cancels. However, even for the BOSS high-z effective redshift
$F_{\text{AP}}$ has only a logarithmic dependence on $\Omega_m$. 
Thus, the use of this quantity for comparison might be misleading, since
in $\L$CDM it always has a very small error because of a negligible 
small sensitivity to cosmology. 
This is illustrated in Fig.~\ref{fig:fap}, where we show 
the $F_{\text{AP}}-f\sigma_8$ diagram extracted from our MCMC chains for the 
base $\Lambda$CDM.
The situation changes in extensions of $\L$CDM, where $F_{\text{AP}}$
can reflect some non-trivial information.

\section{Conclusions and Outlook}
\label{sec:concl}

We have presented new limits on the cosmological parameters of the minimal $\L$CDM
from the BOSS DR12 on the anisotropic redshift-space galaxy clustering. 
Our study features several important improvements. 
They include the use of a complete theoretical model
for the non-linear power spectrum and the MCMC technique for parameter inference.
In contrast to previous Fourier-space galaxy clustering analyses of the power spectrum
multipoles \cite{Beutler:2013yhm,Gil-Marin:2015sqa,Beutler:2016arn},
we consistently recompute the full likelihood 
as we sample different cosmological and nuisance parameters.

% Note that the previous BOSS analyses of the position-space correlation function and redshift-space wedges 
% \cite{Grieb:2016uuo,Satpathy:2016tct,Sanchez:2016sas} did vary all relevant parameters in the MCMC chains,
% but only in combination with the Planck likelihood. 
% This is different from our analysis, which makes minimal assumptions in order 
% to extract cosmological information from 
% the BOSS galaxy clustering data.

Our analytic model for the galaxy power spectrum
is based on one-loop perturbation theory. 
It includes the non-linearities 
in the underlying dark matter field, bias expansion, and
redshift-space mapping. 
In addition, it properly takes into account the damping of the BAO produced by 
large-scale bulk flows, which is described by means of IR resummation. 
Finally, our model incorporates corrections due to backreaction of short-scale modes, 
which cannot be reliably modeled within perturbation theory itself. These effects are 
captured by a number of so-called ``counterterms,'' whose shape is fixed by
symmetries and whose amplitude is characterized by free coefficients, which are treated 
as nuisance parameters in this work. 
Another feature of this work is the use of 
the novel \texttt{FFTLog} algorithm \cite{Simonovic:2017mhp},
which made computations of the non-linear galaxy 
power spectra and related likelihoods highly efficient and robust. 
We implemented this algorithm in the Boltzmann code \texttt{CLASS} \cite{Blas:2011rf},
which enabled us to quickly produce theoretical templates for a given cosmology.
This code can be easily interfaced with common 
cosmological MCMC samplers like \texttt{Montepython} \cite{Audren:2012wb,Brinckmann:2018cvx} or 
\texttt{cobaya}\footnote{\href{https://github.com/CobayaSampler/cobaya}{
\textcolor{blue}{https://github.com/CobayaSampler/cobaya}}
}.
Thus, the present work is the first practical application of 
many recent efforts in large-scale structure theory.

The main outcome of our work is that the so-called shape priors 
are not actually necessary in the full-shape power spectrum analysis. 
The BOSS power spectrum data \textit{alone} can be used to constrain the 
late-time matter density and the Hubble parameter with precision 
similar to that of the Planck CMB measurements. 
Our study shows that the power spectrum shape contains 
a considerable amount of information in addition to the BAO wiggles
and the Alcock-Paczynski distortions, which were the main focus of
previous anisotropic galaxy clustering analyses.
We stress that even though our baseline analysis does not directly use the CMB data, 
it assumes informative priors on the power spectrum tilt,
the physical baryon density, and the total neutrino mass.
On the one hand, they can be seen as theoretical priors strongly motivated by the CMB measurements.
On the other hand, they can be viewed as a minimal input from the CMB,
which allows one to fix some degeneracies poorly constrained by the BOSS data itself.
In this regard, our baseline priors are similar, by spirit, to the FIRAS prior on the CMB 
monopole temperature, and the minimal neutrino mass allowed by the oscillation experiments,
which are the key external priors used to constrain the base Planck $\L$CDM model~\cite{Aghanim:2018eyx}.

The parameters of $\L$CDM measured in this work are consistent with 
the results of the Planck CMB observations \cite{Aghanim:2018eyx}
and the DES survey \cite{Abbott:2017wau}.
It would be interesting 
to see how much the cosmological parameter constraints can be improved
by combining the data from these 
experiments with our full-shape power spectrum likelihood.
Our method can also be easily applied to 
the eBOSS quasar clustering data \cite{Ata:2017dya,Gil-Marin:2018cgo}.

The main factor limiting the range of scales used in our analysis was the fingers-of-God effect. 
This effect forced us to increase the number of nuisance 
parameters and eventually prevented us from employing 
the modes with comoving wavenumbers bigger than $0.25\,h/\Mpc$. 
We believe that this problem can be alleviated by using the 
redshift-space wedges \cite{Grieb:2016uuo},
which can extend the regime of validity of our theoretical model
without having to compute higher-order corrections. 
Another aspect that requires improvement is the covariance matrix treatment.
Ultimately, it is desirable to use an analytic expression which can be easily recalculated 
for a new cosmology e.g. if the estimated cosmological parameters happen to be different 
from the ones used to generate the covariance matrix for the initial analysis. 
The validation of our results with different covariance matrices 
represents a necessary consistency check.
Finally, it would be interesting to see how the 
analysis can be improved by including the theoretical error, e.g. \cite{Baldauf:2016sjb,Chudaykin:2019ock}. These questions are left for future investigations.

\vspace{0.5cm} \textit{Note added.} 
When the \texttt{CLASS} module for fast perturbation theory calculations used in this paper 
was being developed, we became 
aware of the work \cite{Leonardo:2019me}, 
which was applying a similar theoretical model to analyze the BOSS data. 
This inspired us to use our code for the BOSS data analysis. 
We thank the authors of \cite{Leonardo:2019me}
for discussions and sharing with us their preliminary results. 
The methodology and theoretical model used in Ref.~\cite{Leonardo:2019me} are very similar to ours, 
but the numerical implementation is completely different. 
Note that compared to us, Ref.~\cite{Leonardo:2019me} 
uses slightly different data selections and prior choices. 
When overlap, our results agree.

\section*{Acknowledgments}

We are indebted to Florian Beutler for providing access to the BOSS DR12 measurements
and various related products,
and to Roman Scoccimarro for sharing with us the power spectra from the LasDamas N-body simulation.

We would like to thank Guido D'Amico, Jonathan Blazek, 
Anton Chudaykin, 
Elisabeth Krause, 
Julien Lesgourgues,
Antony Lewis,
Ariel Sanchez,
Alvise Raccanelli, 
Leonardo Senatore, 
Blake Sherwin, 
Zvonimir Vlah,
and Jay Wadekar for useful discussions. 
We are grateful to Colin Hill,  
Sergey Sibiryakov,
Uro$\check{\text{s}}$ Seljak, 
Masahiro Takada, 
and Benjamin Wallisch for comments on the preliminary version of this paper.
All numerical analyses of this work were performed on the 
Helios cluster at the Institute for Advanced Study. 
M.Z. is supported by NSF grants AST1409709, PHY-1521097 and PHY-1820775 the Canadian
Institute for Advanced Research (CIFAR) program on
Gravity and the Extreme Universe and the Simons Foundation Modern Inflationary Cosmology initiative.
M.I. thanks the CERN Theory Department and MIAPP Munich for hospitality during the completion of this work.
M.I. acknowledges
the support from the Swiss National Science Foundation 
and the RFBR grant 17-02-00651
at the initial stages of this work. 
M.I. is partially supported by the Simons Foundation's \textit{Origins of the Universe} program.

\appendix

\section{Theory Model}
\label{app:model}

Our model for galaxy power spectrum in redshift space is given by
\be
\begin{split}
P_{g}(k,\mu)= & Z^2_1(\k)
P_{\text{lin}}
(k)+ 2\int_{\q}Z^2_2(\q,\k-\q)
P_{\text{lin}}(|\k-\q|)
P_{\text{lin}}(q)\\
& + 6Z_1(\k)P_{\text{lin}}(k)\int_{\q}Z_3(\q,-\q,\k)P_{\text{lin}}(q)\\
& -2\tilde{c}_0 k^2 P_{\text{lin}}(k)
-2\tilde{c}_2 f \mu^2 k^2 P_{\text{lin}}(k)
-2\tilde{c}_4 f^2 \mu^4 k^2 P_{\text{lin}}(k)\,,\\
&-\tilde{c}f^4\mu^4 k^4 (b_1+f\m)^2 P_{\text{lin}}(k)  + P_{\text{shot}}\,,
\end{split}
\ee
where the redshift-space kernels are given by \cite{Bernardeau:2001qr},
\bseq 
\begin{align}
&Z_1(\k)  = b_1+f\mu^2\,,\\
&Z_2(\k_1,\k_2)  =\frac{b_2}{2}+b_{\mathcal{G}_2}\left(\frac{(\k_1\cdot \k_2)^2}{k_1^2k_2^2}-1\right)
+b_1 F_2(\k_1,\k_2)+f\mu^2 G_2(\k_1,\k_2)\notag\\
&\qquad\qquad\quad~~+\frac{f\mu k}{2}\left(\frac{\mu_1}{k_1}(b_1+f\mu_2^2)+
\frac{\mu_2}{k_2}(b_1+f\mu_1^2)
\right)
\,,
\end{align} 
\eseq
% \bseq 
% \begin{align}
\be
\begin{split}
&Z_3(\k_1,\k_2,\k_3)  =2b_{\G_3}\left[\frac{(\k_1\cdot
     (\k_2+\k_3))^2}{k_1^2(\k_2+\k_3)^2}-1\right]
\big[F_2(\k_2,\k_3)-G_2(\k_2,\k_3)\big]
% \notag
\\  
&\quad
+b_1 F_3(\k_1,\k_2,\k_3)+f\mu^2 G_3(\k_1,\k_2,\k_3)+\frac{(f\m k)^2}{2}(b_1+f \mu_1^2)\frac{\m_2}{k_2}\frac{\m_3}{k_3}
% \notag
\\
&\quad
+f\mu k\frac{\mu_3}{k_3}\left[b_1 F_2(\k_1,\k_2) + f \mu^2_{12} G_2(\k_1,\k_2)\right]
+f\m k (b_1+f \mu^2_1)\frac{\m_{23}}{k_{23}}G_2(\k_2,\k_3)
% \notag
\\
&\quad+b_2 F_2(\k_1,\k_2)+2b_{\mathcal{G}_2}\left[\frac{(\k_1\cdot (\k_2+\k_3))^2}{k_1^2(\k_2+\k_3)^2}-1\right]F_2(\k_2,\k_3)
+\frac{b_2f\mu k}{2}\frac{\mu_1}{k_1}
% \notag
\\
&\quad+b_{\mathcal{G}_2}f\mu k\frac{\m_1}{k_1}\left[\frac{(\k_2\cdot \k_3)^2}{k_2^2k_3^2}-1\right]
\,,
% \end{align} 
% \eseq
\end{split}
\ee
where $\k=\k_1+\k_2+\k_3$ and
the kernel $Z_3$ has to be symmetrized over its arguments.

Now let us discuss our implementation of IR resummation.
We follow the approach streamlined in Refs.~\cite{Blas:2016sfa,Ivanov:2018gjr}, which was developed 
in the context of time-sliced perturbation theory \cite{Blas:2015qsi}.
IR resummation  
splits the matter linear power spectrum
into the smooth and the wiggly parts,\footnote{In practice, we use the wiggly-smooth decomposition
technique introduced in Ref.~\cite{Hamann:2010pw}.}
\be
P_{\text{lin}}= P_{\text{nw}}(k)+P_\text{w}(k)\,,
\ee
where $P_{\text{nw}}$ is a broadband power-law function, 
and $P_\text{w}$ contains the BAO wiggles.
The IR resummed
anisotropic power sepectrum at leading order takes the following form,
\be
\label{Plo}
P_{\text{LO}}(k,\mu) \equiv  P_{\text{nw}}(k,\mu)+\e^{-k^2\Sigma^2_{tot}(\mu)}P_\text{w}(k,\mu)\,,
\ee
where we introduced the anisotropic damping factor,
\be
\label{eq:sigmatot}
\Sigma^2_\tot(\mu)=(1+f\mu^2(2+f))\Sigma^2+f^2\mu^2(\mu^2-1)\delta\Sigma^2\,,
\ee
that depends on the following contributions 
\be
\begin{split}
&\Sigma^2\equiv\frac{1}{6\pi^2}\int_0^{k_S}dqP_{\text{nw}}(q)\left[1-j_0\l\frac{q}{k_{osc}}\r+2j_2\l\frac{q}{k_{osc}}\r\right]\,,\\
&\delta\Sigma^2\equiv\frac{1}{2\pi^2}\int_0^{k_S}dqP_{\text{nw}}(q)j_2\l\frac{q}{k_{osc}}\r \,,
\end{split}
\ee
$k_{osc}$ is the BAO wavelenght $\sim 110\,h/$Mpc, $k_S$ is the separation
scale controlling the modes which are to be resummed,
and $j_n$ are the spherical Bessel function of order~$n$. 
In principle, $k_S$ is arbitrary, and any dependence on it should be
treated as a theoretical error. Following \cite{Blas:2016sfa} we define it 
to be $k_S=0.2\,h$/Mpc, which gives the same result as an alternative choice $k_S=k/2$,
adopted in \cite{Baldauf:2015xfa}.

In general, IR resummation in redshift space at next-to-leading (one-loop) order requires a computation of anisotropic loop integrals which cannot be reduced to one-dimensional ones. 
One can simplify these integrals by splitting the one-loop contribution itself into a smooth 
and wiggly part. More precisely, one first computes the one-loop integrals with a smooth 
part only. At a second step one evaluates these integrals with one insertion of the wiggly power spectrum and suppresses the output with a direction-dependent damping factor \eqref{eq:sigmatot} to get
\be
\begin{split}
	P_{g}(k,\mu) \to &  \quad P_{\text{nw},\,\text{lin}}(k,\mu) + P_{\text{nw},\,\text{1-loop}}(k,\mu) \\
	& \quad +
\e^{-k^2\Sigma^2_{\text{tot}}(\mu)}
	\left(P_{\text{w},\,\text{lin}}(k,\mu)(1+k^2\Sigma^2_{tot}(\mu))+ P_{\text{w},\,\text{1-loop}}(k,\mu)\right)\,,
\end{split}
\ee
where $P_{\text{1-loop}}[P_\lin]$ is treated as a functional of the input linear power spectrum, and
\be 
\begin{split}
& P_{\text{nw},\,\text{1-loop}}(k,\mu)\equiv P_{\text{1-loop}}[P_{\text{nw}}] \,,\\
& P_{\text{w},\,\text{1-loop}}(k,\mu)\equiv P_{\text{1-loop}}[P_{\text{nw}}+P_{\text{w}}]-P_{\text{1-loop}}[P_{\text{nw}}] \,.
\end{split}
\ee
The IR-resummed anisotropic power spectrum should then be used to compute the multipoles 
in Eq.~\eqref{eq:mult}.

To account for the AP effect one has to compute the observable galaxy power spectrum,
\be
P_\obs(k_\obs,\mu_\obs) = P_{g}(k_\true[k_\obs,\mu_\obs],\mu_\true[k_\obs,\mu_{\obs}]) \cdot \frac{D^2_{A,\fid}H_\true}{D^2_{A,\true}H_\fid},
\ee
where $k_{\text{true}}$ and $\mu_{\text{true}}$ are related to 
wavevectors and angles in the true cosmology, 
whereas $k_{\text{obs}}$ and $\mu_{\text{obs}}$ refer to quantities obtained 
for a given set of assumed cosmological parameters. 
The relation between the true and observed wavevectors is given by
\be
\begin{split}\label{AP_k_mu}
	k^2_\true&=k^2_\obs\left[\l\frac{H_\true}{H_\fid}\r^2\mu_\obs^2+\l\frac{D_{A,\fid}}{D_{A,\true}}\r^2(1-\mu_\obs^2)\right]\\
	\mu^2_\true&=\l\frac{H_\true}{H_\fid}\r^2\mu^2_\obs\left[\l\frac{H_\true}{H_\fid}\r^2\mu_\obs^2+\l\frac{D_{A,\fid}}{D_{A,\true}}\r^2(1-\mu_\obs^2)\right]^{-1}\,.
\end{split}
\ee
During MCMC analysis one tries to find $H_\true$ and $D_{A,\true}$ given
$H_\fid$ and $D_{A,\fid}$ that are fixed by the reference cosmological model used to 
create galaxy catalogs. 
The eventual galaxy multipoles with the AP effect are given by
\be 
P_{\ell,\text{AP}}(k) = \frac{2\ell+1}{2} \int_{-1}^1 d\mu_{\text{obs}}\, P_{\text{obs}}(k_{\text{obs}},\mu_{\text{obs}}) \cdot 
\mathcal{P}_\ell(\mu_{\text{obs}})\,.
\ee

\section{Tests on Mock Catalogs}
\label{app:mocks}

In this Appendix we show the tests of our pipeline on mock catalogs. 
First, we will apply our pipeline to the high-resolution mock catalogs based on the N-body simulation 
LasDamas, which are characterized by the gigantic volume of ($\sim 553$ (Gpc$/h$)$^3$). 
These are mocks of Luminous Red Galaxies that are desinged to match the sample observed by
SDSS
\cite{2009AAS...21342506M}.
% These are mocks of Luminous Red Galaxies, whose selection function is very close 
% to the one of the dataset actually observed by SDSS. The LasDamas mocks are 
% characterized by gigantic volume ($\sim 553$ (Gpc$/h$)$^3$). 
Second, we will test our pipeline on MultiDark \textsc{patchy} mock catalogs \cite{Kitaura:2015uqa}.
On the one hand, they are based on approximate gravity solvers and HOD models. 
On the other hand, they are designed to closely reproduce the 
data and have the same selection, window function, and fiber collision effects 
implemented.

\subsection{Tests on LasDamas N-body simulations}

We will fit the monopole and quadrupole of the galaxy power spectrum
of LasDamas Oriana simulations at redshift $z=0.34$. 
This redshift is lower than the ones 
used in our analysis and therefore it provides a more stringent test of our theoretical model
because the non-linear effects are stronger. 
The cosmological parameters used to generate mock catalogs are $h=0.7$, $\Omega_m=0.25$, 
$\Omega_b=0.04$, $\sigma_8=0.8~(A_s=2.22\cdot 10^{-9})$, $n_s = 1$, and $\sum m_\nu = 0$.
The details of LasDamas simulation can be found at.\footnote{\href{http://lss.phy.vanderbilt.edu/lasdamas/overview.html}{
\textcolor{blue}{http://lss.phy.vanderbilt.edu/lasdamas/overview.html}}\,
}

We fit the mean of power spectra extracted from 
40 independent simulation boxes, whose volume is $(2400~\text{Gpc}/h)^3$ each. 
This totals to $553$ (Gpc/$h$)$^3$ volume, which is almost 100 times bigger than the cumulative volume of BOSS.
However, the statistical error corresponding to this tremendous volume is so small that the 
two loop corrections supersede cosmic variance already on very large scales. 
The situation is different for low-volume surveys like BOSS, where the statistical error is expected 
to be bigger than the systematic one down to very high $k_{\rm max}$.
Hence, in order to be realistic,
we will assume a covariance that corresponds to the BOSS survey and not to the actual Las Damas volume.
Since 40 realizations are not enough to accurately estimate the covariance, 
we will use a theoretical prediction obtained in the Gaussian approximation (see, e.g.~\cite{Chudaykin:2019ock}),
\be
\begin{split}
& C^{(00)}_{ij}=\frac{2}{N_k}\left(P_0^2+\frac{1}{5}P_2^2\right)\delta_{ij}\,,\quad C^{(02)}_{ij}=C^{(20)}_{ij}=\frac{2}{N_k}\left(2P_0P_2+\frac{2}{7}P_2^2\right)\delta_{ij}\,,\\
& C^{(22)}_{ij}=\frac{2}{N_k}\left(
5 P_0^2+\frac{20 P_0 P_2}{7}+\frac{15 P_2^2}{7}\right)\delta_{ij}\,,
\end{split}
\ee
where we introduced the number of modes $N_k =4\pi k^2\Delta kV$, the binning step of Las Damas simulations 
$\Delta k = 0.0025$ $h$/Mpc and 
the survey volume $V$. 
Note that the monopole moment $P_0$ includes the shot-noise contribution, which is equal to $\bar{n}^{-1}=1.0\times 10^4$ [Mpc/h]$^3$ for the LasDamas mocks.
We will consider 
two particular choices,
\be
V_{\rm BOSS-like}=6~(\text{Gpc}/h)^3\quad \text{and}\quad V_{\rm 10\times BOSS-like}=60~(\text{Gpc}/h)^3\,.
\ee
$V_{\rm BOSS-like}$ is the total volume of the BOSS survey across all redshifts and sky parts, 
whereas 
$V_{\rm 10\times BOSS-like}$ is simply a 10 times bigger volume, 
which will be used to better pin down the theory systematic error.
Using an approximate Gaussian covariance also provides an additional 
challenge to our approach: neglecting the off-diagonal terms artificially reduces the error 
and makes one reject the true model more often than it should be. 

To make a closer contact to our analysis, we will keep $\omega_b, n_s$ and  $\sum m_\nu = 0$ fixed to the true values
and scan over $\omega_{cdm}$, $H_0$, $A^{1/2}=(A_s/A_{s,~{\rm fid}})^{1/2}$ in our analysis. 
We use the same nuisance parameters and assume the same priors for them as in our baseline analysis.

\begin{figure}[h!]
\centering 
\includegraphics[width=0.49\textwidth]{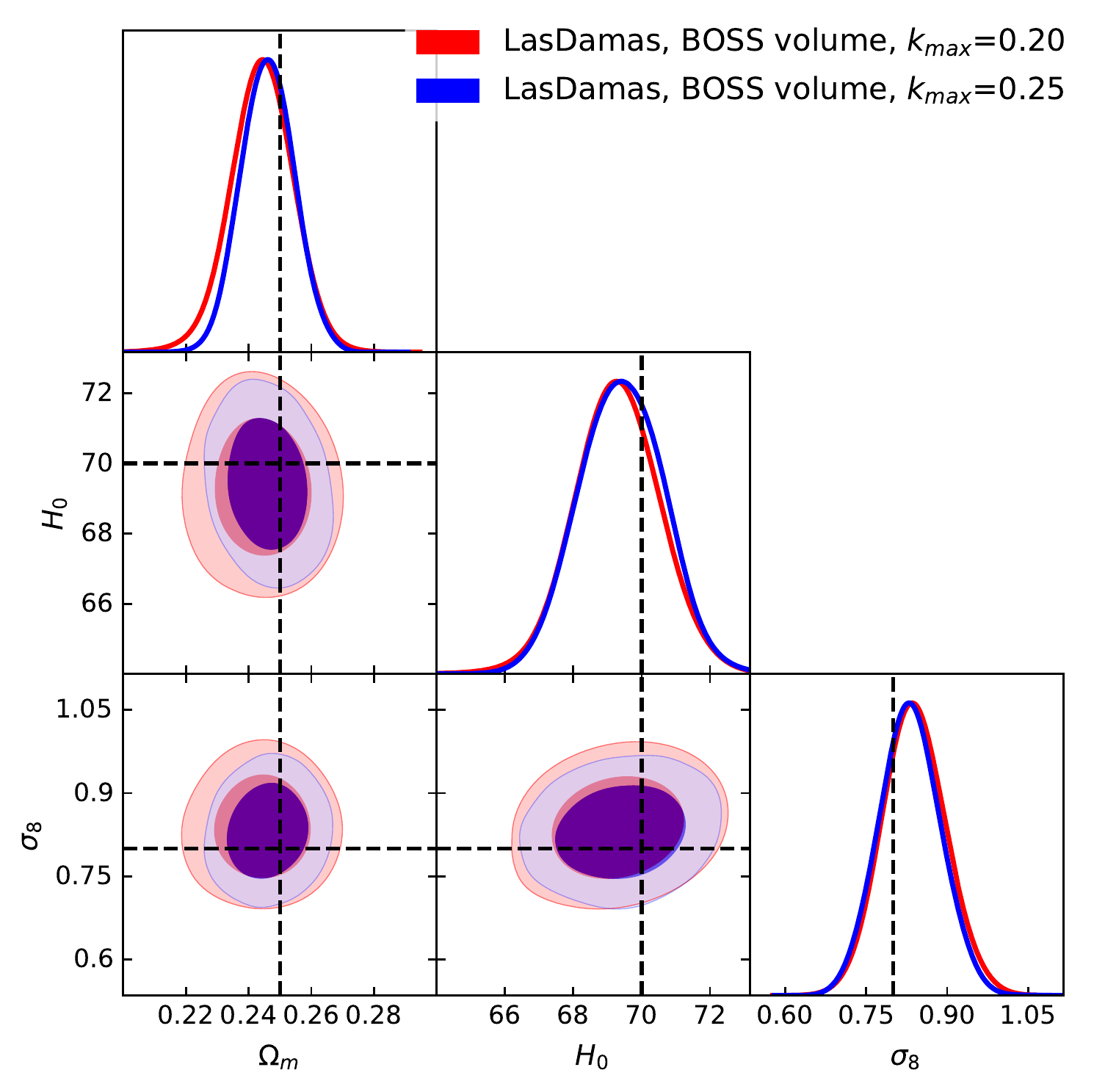}
\includegraphics[width=0.49\textwidth]{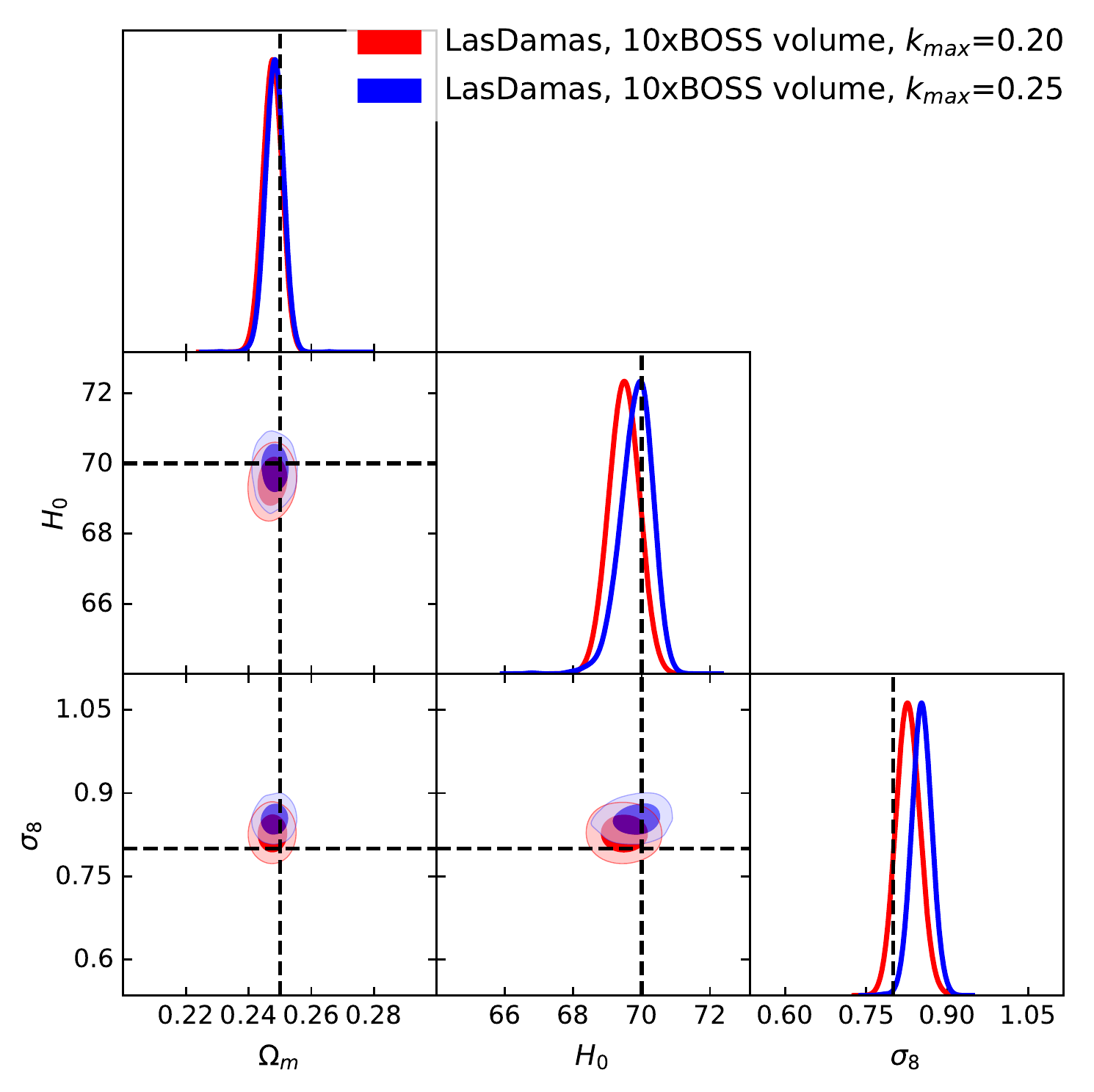}
\caption{Results of our analysis of the galaxy power spectrum of the LasDamas N-body simulations with the errorbars
scaled to the total BOSS volume (\textit{left panel}) and  10$\times$ the total BOSS volume (\textit{right panel}).
Dashed lines represent the true values used in simulations. 
The values of $k_{\text{max}}$ are quoted in units of $h/$Mpc, $H_0$ in units [km/s/Mpc].}  
\label{fig:LD}  
\end{figure}
\vspace{0.5cm}
\begin{table}[h!]
\begin{center}
 \begin{tabular}{|c|c|c|} 
 \hline
$V_{\rm BOSS}$ & best-fit & mean  $\pm 1\sigma$  \\ [0.5ex] 
 \hline\hline
 $A$ & $1.126$ & $1.15_{-0.17}^{+0.15}$  \\ \hline
 $H_0$  & $70.05$ & $69.4\pm 1.3$  \\ \hline
  $\omega_{cdm}$  & $0.1044$ & $0.09871_{-0.0056}^{+0.0053}$  \\
 \hline
 $b_1\times A^{1/2}$  & $2.161$ & $2.164_{-0.046}^{+0.047}$  \\  \hline
 $b_2\times A^{1/2}$ & $-1.418$& $-1.927_{-0.94}^{+0.69}$ \\  \hline
 $b_{\mathcal{G}_2}\times A^{1/2}$  & $ -0.1763$ & $-0.1598_{-0.21}^{+0.13}$  \\  \hline
 $c^2_0$ & $-0.0208$& $1.548_{-28}^{+34}$  \\  \hline
 $c^2_2$  &$38.2$ & $15.0_{-27}^{+39}$  \\  \hline
 $\tilde{c}$ & $681$ & $857_{-260}^{+210}$  \\ \hline
 $10^{-3}P_{\text{shot}}$ & $1.344$& $4.485_{-3.8}^{+2}$  \\ \hline\hline 
 $\sigma_8$ & $0.857$& $0.830\pm0.057$  \\ \hline
 $\Omega_m$ & $0.2527$& $0.2457_{-0.0087}^{+0.0085}$  \\ \hline
\end{tabular}
\begin{tabular}{|c|c|c|} 
 \hline
$10 \times V_{\rm BOSS}$ & best-fit & mean  $\pm 1\sigma$  \\ [0.5ex] 
 \hline\hline
 % $A^{1/2}$ & $1.026$ & $1.017\pm 0.102$  \\ \hline
 $A$ & $1.157$ & $1.160\pm 0.059$  \\ \hline
 $H_0$  & $69.95$ & $69.86_{-0.4}^{+0.46}$  \\
 \hline
  $\omega_{cdm}$  & $0.1024$ & $0.1017_{-0.0018}^{+0.0025}$  \\\hline
 $b_1\times A^{1/2}$  & $2.173$ & $2.165_{-0.015}^{+0.017}$  \\  \hline
 $b_2\times A^{1/2}$ & $-1.462$& $-1.664_{-0.3}^{+0.38}$  \\  \hline
 $b_{\mathcal{G}_2}\times A^{1/2}$  & $-0.2252$ & $-0.194_{-0.05}^{+0.035}$  \\  \hline
 $c^2_0$ & $-2.0$& $-3.125_{-8.8}^{+9.7}$  \\  \hline
 $c^2_2$  &$41.6$ & $32^{+15}_{-10}$  \\  \hline
 $\tilde{c}$ & $355$& $ 541_{-288}^{+200}$  \\ \hline
 $10^{-3}P_{\text{shot}}$ & $0.9192$& $2.094_{-1.9}^{+0.7}$  \\ \hline \hline
 $\sigma_8$ & $0.861$& $0.853\pm 0.020$  \\ \hline
 $\Omega_m$ & $0.2494$& $0.2486_{-0.0028}^{+0.0033}$  \\ \hline
\end{tabular}
\caption{
The results of our MCMC analysis for
the LasDamas mock data 
with $k_{\text{max}}=0.25\,h^{-1}$Mpc. 
$H_0$ is quoted in units [km/s/Mpc]. The parameters $c_0^2$
and $c_2^2$ are quoted in units $[\Mpc/h]^2$, 
$\tilde{c}$ in units $[\Mpc/h]^4$, $P_{\text{shot}}$ in units $[\Mpc/h]^3$.
The fiducial values for cosmological parameters used in the simulations are $H_0=70,\,
\Omega_{m}=0.25\,(\omega_{cdm} = 0.1029),\,\sigma_8=0.8~(A=1)$.
}
 \label{tab:LD}
\end{center}
\end{table}

Our results are presented in Fig.~\ref{fig:LD} and in Table~\ref{tab:LD}. 
In Fig.~\ref{fig:LD} we show the contours obtained for two choices 
of $k_{\rm max}=0.2~h$/Mpc and $k_{\rm max}=0.25~h$/Mpc. 
Table~\ref{tab:LD} displays the marginalized one-dimensional limits
for $k_{\rm max}=0.25~h$/Mpc, which will be eventually selected as a baseline data cut.

Let us focus on the case corresponding to the total BOSS volume $V_{\rm BOSS-like}$ (left panel of Fig.~\ref{tab:LD}).
One can see that our pipeline correctly extracts the cosmological parameters within $1\sigma$ for both choices of 
$k_{\rm max}$. 
Remarkably, the errobars are very similar to the ones obtained in the analysis of the real data. 
The difference between the two choices of $k_{\rm max}$
is marginal, which merely reflects the fact that the errorbars 
cannot be further improved due to the shot noise. 
Given that the results for $k_{\rm max}=0.25$ $h$/Mpc are somewhat better, 
we prefer to adopt it as our standard cut. 
We have checked that going to $k_{\rm max}=0.30$ $h$/Mpc gives 
very minor improvement on the errorbars and produces posteriors that are more shifted w.r.t. the true values. Given this reason, we prefer to stick to 
$k_{\rm max}=0.25$ $h$/Mpc in order to be more conservative.

% Even if we naively take the difference between the extracted best-fit values and the truth as an estimate of the 
% theoretical systematic bias at $k_{\rm max}=0.25$ $h$/Mpc, we find the following shifts,
% in units of the 1$\sigma$ expected dispersion for each parameter 
% $\{\Delta \omega_{cdm},\Delta H_0,\Delta A~|~\Delta \Omega_m, \Delta \sigma_8\}$, $\{-0.5,\}$. 
% Alternatively, one can compute the shifts between the means and the true values: $\{-0.5,\}$.
% The difference between the two is introduced by parameter volume (marginalization) effects.
% % These can be contrasted with the shifts for $k_{\rm max}=0.2$ $h$/Mpc: $\{-0.5,\}$.
% In either case the shifts are tolerable by the standards adopted in other cosmological analyses, i.e. in Planck \cite{Aghanim:2018eyx}.

Finally, to better understand the validity of our model we 
have repeated our analysis with a covariance reduced by a factor of 10.
The results are shown in the right panel of Fig.~\ref{fig:LD}
and Table~\ref{tab:LD}. One can see that even in this case our model correctly
reproduces the input parameters of the simulations. 
At $k_{\rm max}=0.2~h$/Mpc all the parameters are recovered within 1$\sigma$
of the reduced errors, whereas for $k_{\rm max}=0.25~h$/Mpc 
we observe a 2$\sigma$ shift in $\sigma_8$, while the $H_0$ 
and $\Omega_m$ are accurately recovered. 
However, one may notice that at $k_{\rm max}=0.2~h$/Mpc the  
means of the posteriors for $H_0$ and $\Omega_m$ are more shifted with respect to
the true values as compared to the $k_{\rm max}=0.25~h$/Mpc case. 
At the same time, the best-fit parameters are very close to the true ones.
This implies that the observed shifts of the posterior means are caused by parameter marginalization
(parameter volume) effects.
Comparing this with the results for the actual BOSS volume we see that 
the means of the distributions are even further shifted w.r.t. the true values.
This shows that for the BOSS errorbars the marginalization 
effects are more significant than the theory-systematic error.

\subsection{Tests on Patchy Mocks}

\begin{figure}[h!]
\centering 
\includegraphics[width=0.49\textwidth]{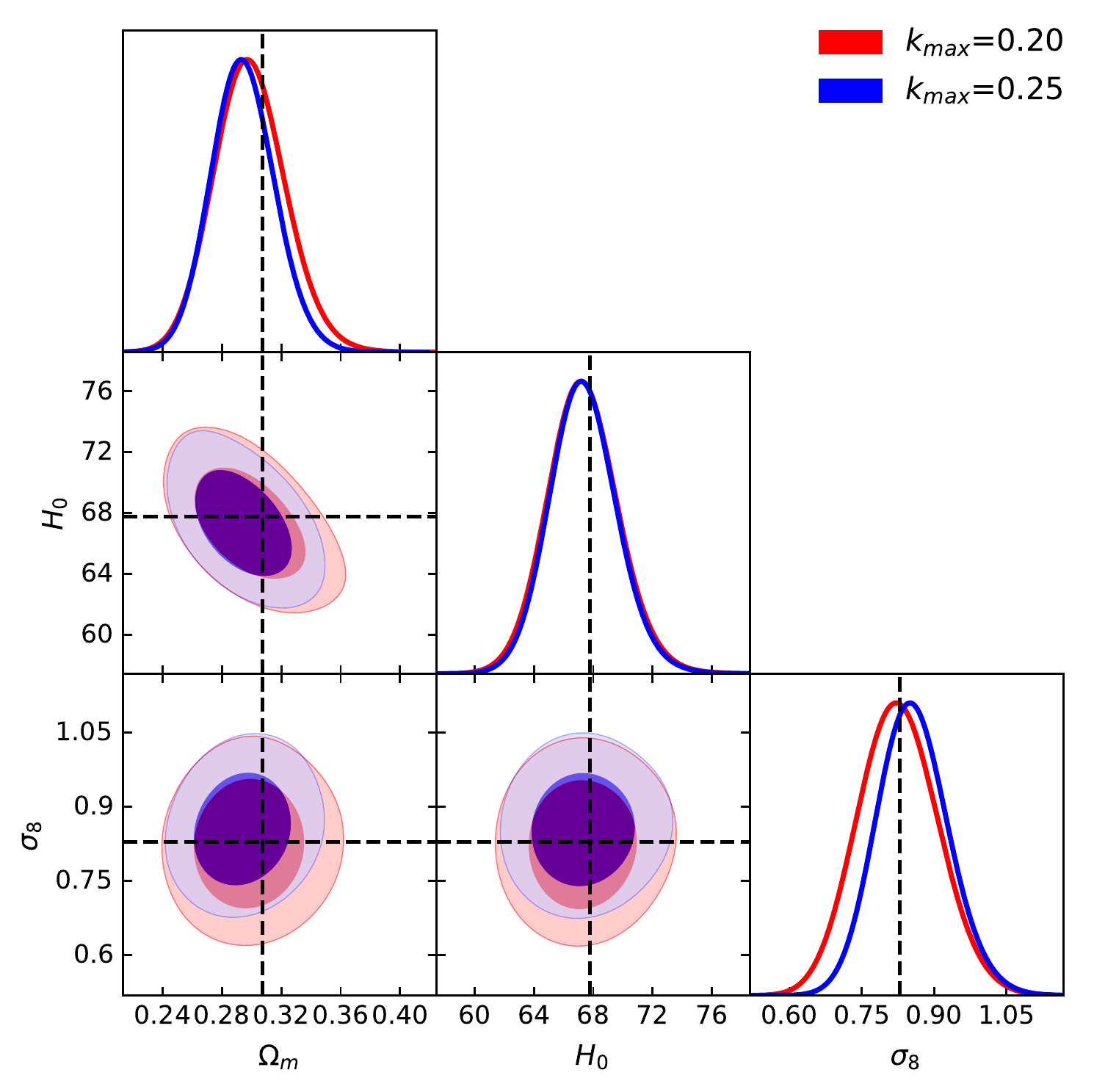}
\includegraphics[width=0.49\textwidth]{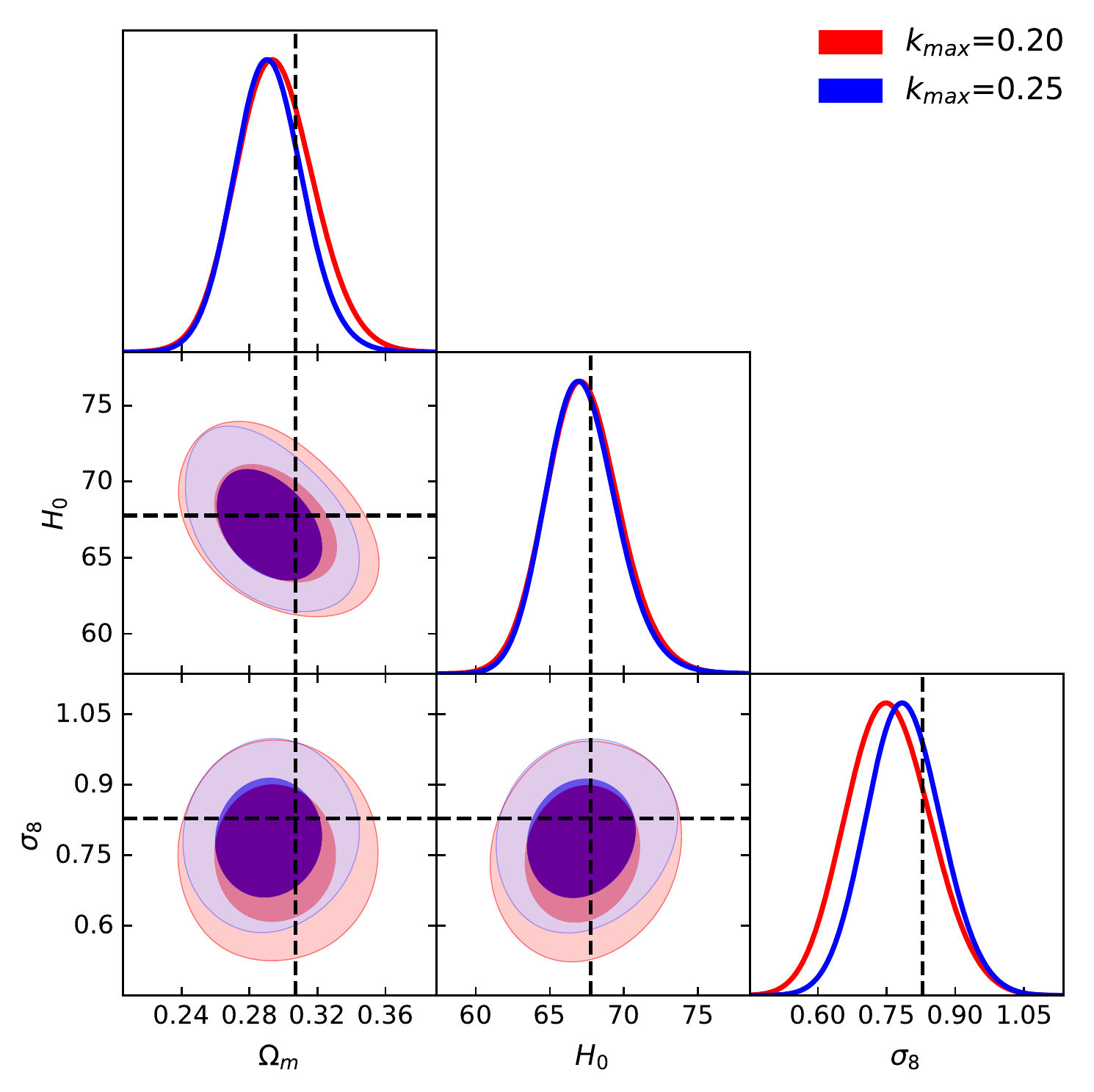}
\caption{Mocks for high-z (left panel) and low-z (right panel) NGC samples: 2d posterior contours and 1d marginalized distribution for cosmological parameters. Dashed lines represent the true values used in simulations. 
The values of $k_{\text{max}}$ are quoted in units of $h/$Mpc, $H_0$ in units [km/s/Mpc].}  
\label{mocks_plots}  
\end{figure}
\vspace{0.5cm}
\begin{table}[h!]
\begin{center}
 \begin{tabular}{|c|c|c|} 
 \hline
high-z NGC & best-fit & mean  $\pm 1\sigma$  \\ [0.5ex] 
 \hline\hline
 % $A^{1/2}$ & $1.1098$ & $1.084\pm 0.102$  \\ \hline
 $A$ & $1.300$ & $1.186_{-0.249}^{+0.192}$  \\ \hline
 $H_0$  & $67.0$ & $67.3 \pm 2.3$  \\
 \hline
 $10^2\omega_{b}$  & $2.213$ & $2.214 \pm 0.015$  \\
 \hline
  $\omega_{cdm}$  & $0.1104$ & $0.1111 \pm 0.0095$  \\
 \hline
 $b_1\times A^{1/2}$  & $1.9412$ & $1.951\pm 0.065$  \\  \hline
 $b_2\times A^{1/2}$ & $-1.99$& $-1.81_{-1.77}^{+0.81}$ \\  \hline
 $b_{\mathcal{G}_2}\times A^{1/2}$  & $ -0.13$ & $-0.0014_{-0.43}^{+0.25}$  \\  \hline
 $c^2_0$ & $17.2$& $17.4_{-30.2}^{+37.8}$  \\  \hline
 $c^2_2$  &$28.6$ & $21.3_{-29.9}^{+52.7}$  \\  \hline
 $\tilde{c}$ & $230$ & $ 286_{-164}^{+124}$  \\ \hline
 $10^{-3}P_{\text{shot}}$ & $4.39$& $4.54_{-3.00}^{+2.44}$  \\ \hline\hline 
 $\sigma_8$ & $0.900$& $0.855\pm0.074$  \\ \hline
 $\Omega_m$ & $0.295$& $0.294\pm0.022$  \\ \hline
\end{tabular}
\begin{tabular}{|c|c|c|} 
 \hline
low-z NGC & best-fit & mean  $\pm 1\sigma$  \\ [0.5ex] 
 \hline\hline
 % $A^{1/2}$ & $1.026$ & $1.017\pm 0.102$  \\ \hline
 $A$ & $1.110$ & $1.031_{-0.250}^{+0.200}$  \\ \hline
 $H_0$  & $67.7$ & $67.2 \pm 2.4$  \\
 \hline
 $10^2\omega_{b}$  & $2.211$ & $2.214 \pm 0.015$  \\\hline
  $\omega_{cdm}$  & $0.1132$ & $0.1093 \pm 0.0100$  \\\hline
 $b_1\times A^{1/2}$  & $1.775$ & $1.791\pm 0.068$  \\  \hline
 $b_2\times A^{1/2}$ & $-1.197$& $-1.467^{+0.93}_{-1.62}$  \\  \hline
 $b_{\mathcal{G}_2}\times A^{1/2}$  & $-0.129$ & $-0.016_{-0.375}^{+0.203}$  \\  \hline
 $c^2_0$ & $22.4$& $22.3^{+34.5}_{-30.3}$  \\  \hline
 $c^2_2$  &$39.2$ & $16.9^{+55.0}_{-26.9}$  \\  \hline
 $\tilde{c}$ & $355$& $ 541_{-288}^{+200}$  \\ \hline
 $10^{-3}P_{\text{shot}}$ & $3.28$& $4.92_{-2.96}^{+2.74}$  \\ \hline \hline
 $\sigma_8$ & $0.847$& $0.788\pm 0.081$  \\ \hline
 $\Omega_m$ & $0.296$& $0.291\pm 0.020$  \\ \hline
\end{tabular}
\caption{
The results of our MCMC analysis for
the high-z (left table) and low-z (right table)
NGC \textsc{patchy} mocks data samples 
with $k_{\text{max}}=0.25\,h^{-1}$Mpc. 
$H_0$ is quoted in units [km/s/Mpc]. The parameters $c_0^2$
and $c_2^2$ are quoted in units $[\Mpc/h]^2$, 
$\tilde{c}$ in units $[\Mpc/h]^4$, $P_{\text{shot}}$ in units $[\Mpc/h]^3$.
The fiducial values for cosmological parameters used in the simulations are $H_0=67.77,\,
\Omega_{m}=0.307115\,(\omega_{cdm} = 0.118911),10^2\omega_{b}=2.214,\,\sigma_8=0.8288~(A=1)$.
}
 \label{tab:mocks_post}
\end{center}
\end{table}

Now let us focus on Patchy mocks and consider the NGC mock datasets, which have bigger volumes. 
We fit the mean of 2048 mock power spectra with the covariance matrix of a single simulation box. This allows us to significantly reduce the statistical scatter among different realizations. 
For the analysis we assumed the same base $\Lambda$CDM priors as the ones discussed above 
(see Tab.~\ref{tab:priors}), along with the Gaussian prior
$\omega_b = 0.02214 \pm 0.00015$.
Note that we excluded the neutrino masses from the fit as the simulations were run for massless neutrinos. 
The multipoles of the mock catalogs were produced assuming a fiducial cosmology with $\Omega_m=0.31$, which is different from the true value used in the simulations. 
This is designed to introduce an additional anisotropy to be constrained through the AP effect. 

We focused on four different choices of $k_{\text{max}}=0.15,0.20,0.25,0.30$ $h/\Mpc$. 
Similarly to the case of LasDamas, 
our analysis suggests that at $k_{\text{max}}=0.3$ $h/\Mpc$ the systematic error 
becomes comparable to the statistical one, whereas at $k_{\text{max}}=0.15$ $h/\Mpc$ our model 
has too much freedom, and thus requires more narrow priors on the nuisance parameters
in order to reduce the eventual errorbars. Given these reasons, we focus on 
$k_{\text{max}}= 0.20,0.25$ $h/\Mpc$ in what follows.
The posterior 
distribution obtained with our MCMC analysis is displayed in Fig.~\ref{mocks_plots}.
The marginalized limits for the
cosmological and bias parameters obtained in our mock catalog
analysis for $k_{\text{max}}=0.25$ $h/\Mpc$ (which is used in our baseline analysis) 
are displayed in Table.~\ref{tab:mocks_post}.

One observes that for $k_{\text{max}}=0.25$ $h/\Mpc$
the best-fit and mean values of
the inferred cosmological parameters are within $1\sigma$ from the true values,
but some $\sim 0.5\,\sigma$ shifts w.r.t the true value are clearly visible.
There are two sources of these shifts. First, there is 
a parameter projection effect, 
which can drive the mean values away from the best-fit 
along degeneracy directions. 
Put simply, these effects reflect that fact that the statistical error of the data is not good enough
to break certain degeneracies among model parameters. We stress that this effect is somewhat different from 
the so-called prior volume effect. This effect takes place if the constraints on some parameters 
are prior-dominated, so that the mean values shift in certain directions allowed by the priors.

To study the projection effect we have run the same analysis with 
the survey volume of the mock covariances increased by a factor of 9.
Just like in the LasDamas case, we have found the inferred means of $H_0$ and $\Omega_m$ to be 
much closer to the true values, 
but still offset at the level $\sim 1\sigma$
of the new variance, which is reduced by a factor of $3$ compared to the actual 
BOSS volume. 
Another test was described in Section~\ref{sec:info}, where we analyzed mock datavectors
generated with our theoretical model. 
Although the best-fit parameters obtained with our MCMC scans coincide with the input values,
the means of some parameters (e.g. $\Omega_m$ and $\sigma_8$) were noticeably shifted.
This suggests that the parameter projection effects are inevitable 
for the BOSS covariance, but can be reduced in future surveys with bigger volumes.

The second effect responsible for the shifts is a real systematic error 
related to higher-order corrections omitted in our theoretical model.
We have found that our theoretical model can correctly recover the true cosmology
of the mock data at $k_{\text{max}}=0.20$ $h/\Mpc$ even for survey volumes $\sim 10$ times
bigger than the actual BOSS survey. 
However, it gives a biased estimate of $\sigma_8$ if we go to higher $k_{\text{max}}$'s. 
This shift reaches $\sim 5\%$ at $k_{\text{max}}=0.25$ $h/\Mpc$, which is still 
marginally smaller than our final statistical error on this parameter obtained by combining all the BOSS data samples.
The systematic shifts observed in the estimated $\Omega_m$ and $H_0$ are negligible 
(see the discussion above). 
Given these reasons, 
we decided to stick to $k_{\text{max}}=0.25$ $h/\Mpc$ because in this case the 
total marginalized error (statistical + systematic, added in quadratures)
on the cosmological parameters is smaller 
than the similar error at $k_{\text{max}}=0.20$ $h/\Mpc$, which is dominated 
by the statistical component. 

Note that once we inflate the error to match the actual BOSS volume, the 
systematic error couples with volume effects, 
which shift the inferred value of $\Omega_m$
instead of $\sigma_8$ along the degeneracy direction between them.
This correlation
explains why the shifts of $\sigma_8$ are negligible in Fig.~\ref{mocks_plots}, but
become a leading systematic effect once we increase the survey volume in the covariance for the mock catalogs.
The observed picture is, essentially, the same at $k_{\text{max}}=0.20\,h/\Mpc$
for both redshift bins of the BOSS data. 

All in all, we believe that the choice of $k_{\text{max}}=0.25,h/\Mpc$ represents a good balance between systematic and statistical errors. 
We emphasize that the 1d marginalized limits presented in this paper
should not be over-interpreted beyond the level of $\sim 1\sigma$ uncertainty 
related to the inaccuracies of the theoretical modeling and parameter projection effects. 
Our tests on LasDamas mocks with higher volumes suggest that the shifts in the full parameter space 
(before marginalization)
are actually much smaller than $1\sigma$ for the BOSS covariance. 

\begin{figure}[h!]
\centering
\includegraphics[width=1.0\textwidth]{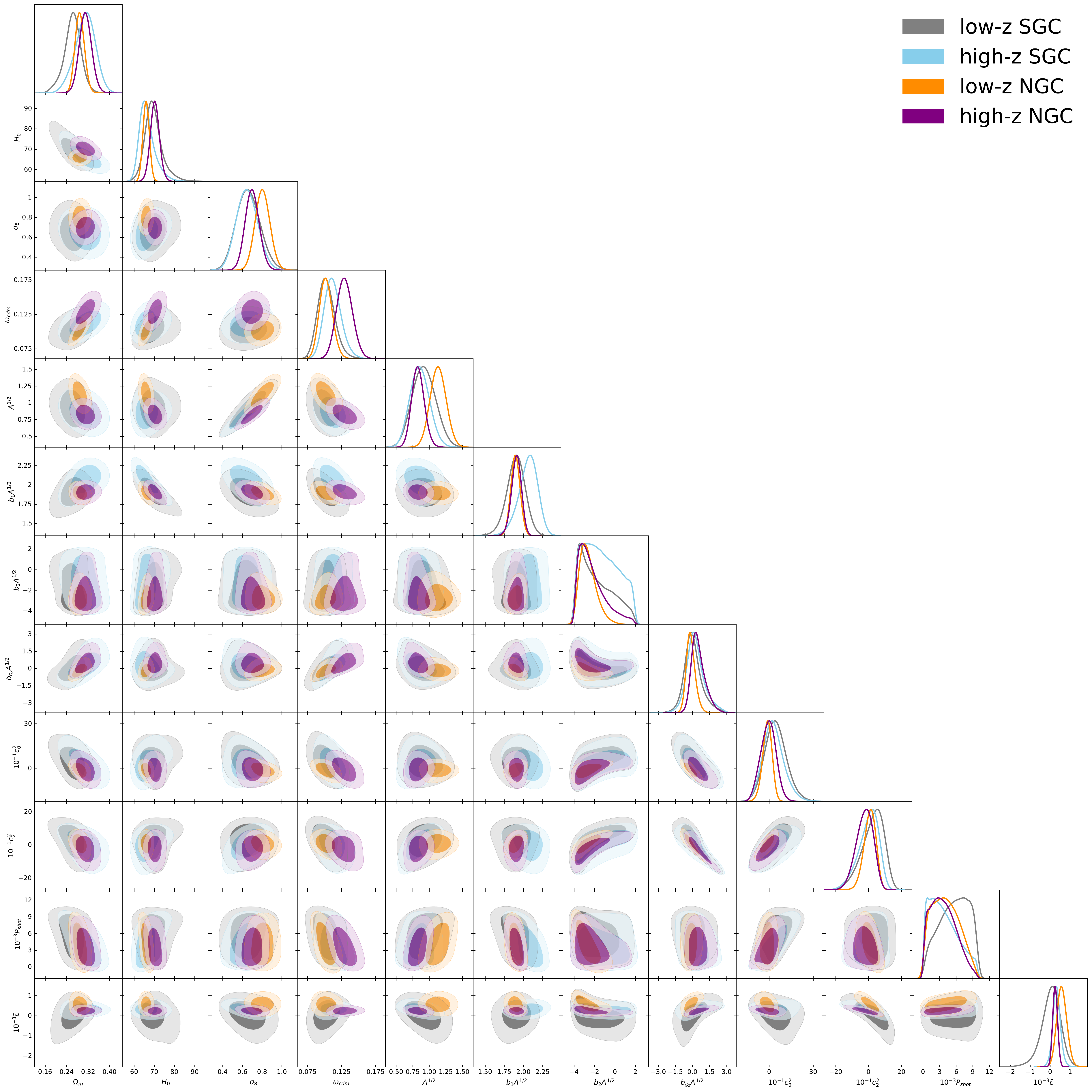}
\caption{
The triangle plot for cosmological and nuisance parameters of four independent 
BOSS datasets. 
\label{fig:monster}
}    
\end{figure}
\begin{figure}[h!]
\includegraphics[width=0.49\textwidth]{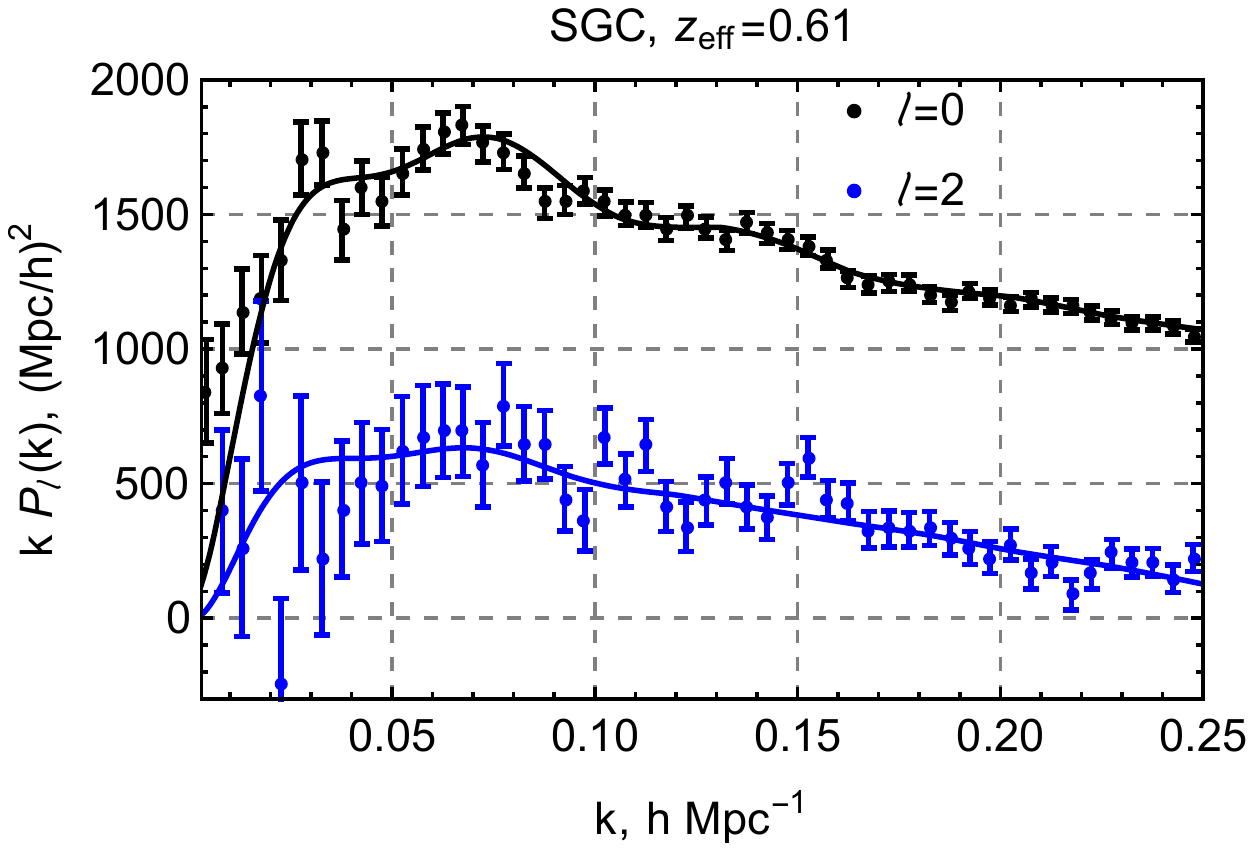}
\includegraphics[width=0.49\textwidth]{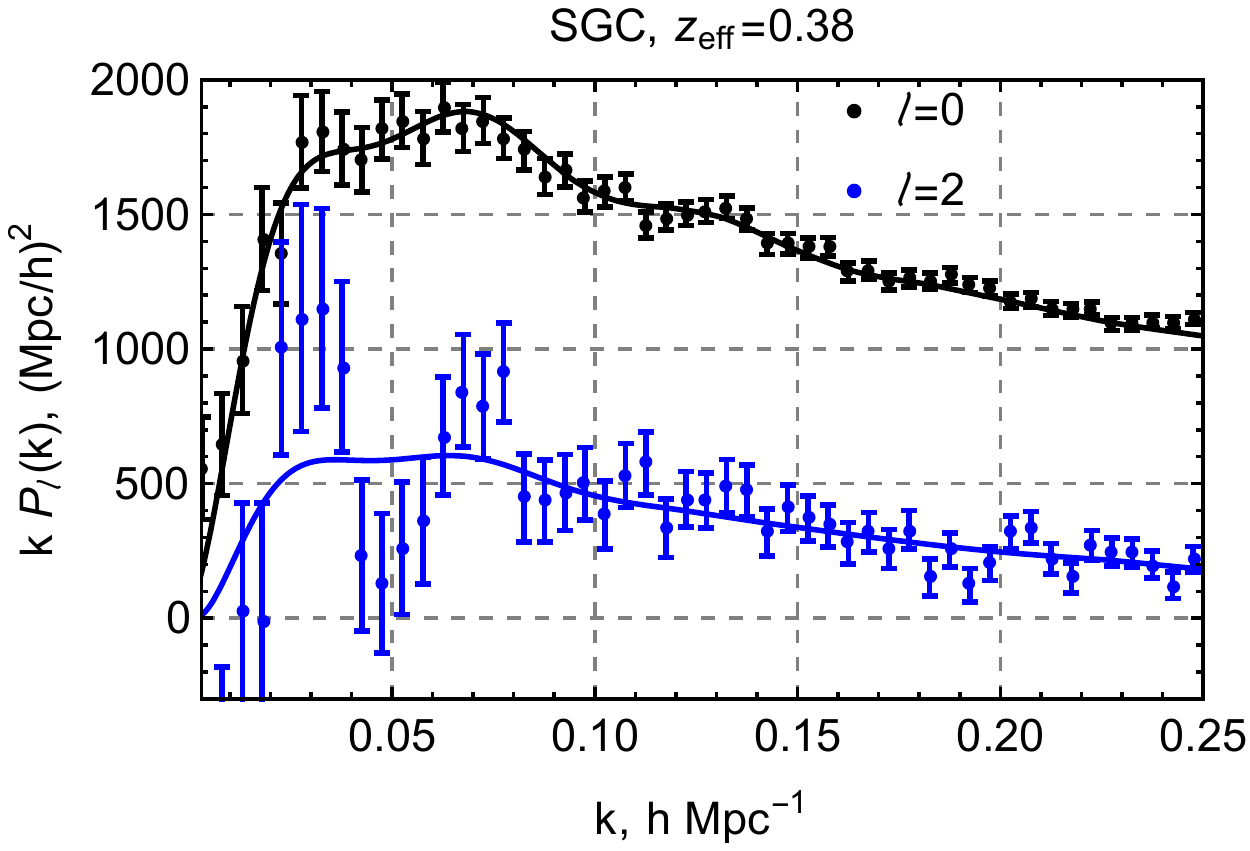}
\caption{Comparison of the data for the monopole and the quadrupole with the best-fit models, whose parameters are listed in Table~\ref{tab:datangc}. The goodness of fit can be assessed by 
the reduced $\chi^2$ given in Table~\ref{tab:datangc}.
\label{fig:bf}
}    
\end{figure}
\begin{table}[ht!]
% \vspace{0.5cm}
\begin{center}
 \begin{tabular}{|c|c|c|} 
 \hline
high-z NGC & best-fit & mean  $\pm 1\sigma$  \\ [0.5ex] 
 \hline\hline
 % $norm$ & $0.948$ & $0.880 \pm 0.088$  \\  \hline
 $A$ & $0.744$ & $0.697^{+0.123}_{-0.178}$  \\ \hline
 $h$  & $0.704$ & $0.703 \pm 0.023$  \\\hline
 $10^2\omega_{b}$  & $2.242$ & $2.237 \pm 0.015$  \\\hline
 $\omega_{cdm}$  & $0.1334$ & $0.1294 \pm 0.012$  \\\hline
 $m_\nu$  & $0.122$ & $0.119^{+0.033}_{-0.057}$  \\\hline
 $b_1\times A^{1/2}$  & $1.926$ & $1.907^{+0.068}_{-0.058}$  \\  \hline
 $b_2\times  A^{1/2}$ & $-2.77$& $-2.27^{+0.40}_{-1.70}$  \\  \hline
 $b_{\mathcal{G}_2}\times  A^{1/2}$  & $ 0.47$ & $0.49^{+0.42}_{-0.71}$  \\  \hline
 $c^2_0$ & $-53.44$& $20.5^{+55.0}_{-49.3}$  \\  \hline
 $c^2_2$  &$-21.0$ & $-22.5^{+59.3}_{-43.0}$  \\  \hline
 $\tilde{c}$ & $187$ & $ 243\pm 123$  \\ \hline
 $10^{-3}P_{\text{shot}}$ & $1.32$& $3.78^{+2.38}_{-3.06}$  \\ \hline \hline 
 $\sigma_8$ & $ 0.744$ & $0.699\pm 0.070$  \\ \hline
 $\Omega_m$ & $ 0.320$ & $0.310\pm 0.023$  \\ \hline
  \multicolumn{3}{|c|}{ $\chi^2_{\text{best-fit}}/N_{\text{dof}}=106.9/(100-12)=1.21$ }\\  \hline
\end{tabular}
%\vspace{0.1cm}
\begin{tabular}{|c|c|c|} 
 \hline
low-z NGC & best-fit & mean  $\pm 1\sigma$  \\ [0.5ex] 
 \hline\hline
 $A$ & $1.442$ & $1.289_{-0.302}^{+0.231}$  \\ \hline
 $h$  & $0.662$ & $0.661 \pm 0.016$  \\ \hline
 $10^2\omega_{b}$  & $2.240$ & $2.237 \pm 0.015$  \\\hline
 $\omega_{cdm}$  & $0.1054$ & $0.1033 \pm 0.0097$  \\ \hline
 $m_\nu$  & $0.154$ & $0.120 \pm 0.040$  \\\hline
 $b_1\times  A^{1/2}$  & $1.895$ & $1.891 \pm 0.060$  \\  \hline
 $b_2\times  A^{1/2}$ & $-2.57$& $-2.64^{+0.54}_{-1.0}$  \\  \hline
 $b_{\mathcal{G}_2}\times  A^{1/2}$  & $-0.15$ & $-0.12^{+0.32}_{-0.44}$  \\  \hline
 $c^2_0$ & $-22.9$& $-14.7^{+34.2}_{-29.2}$  \\  \hline
 $c^2_2$  &$15.8$ & $7.06^{+40.8}_{-31.1}$  \\  \hline
 $\tilde{c}$ & $479$& $ 579^{+224}_{-263}$  \\ \hline
 $10^{-3}P_{\text{shot}}$ & $2.68$& $4.15^{+1.79}_{-3.44}$  \\ \hline \hline 
 $\sigma_8$ & $ 0.866$ & $0.808  \pm  0.073$  \\ \hline
  $\Omega_m$ & $ 0.296$ & $ 0.290 \pm 0.017 $  \\ \hline
  \multicolumn{3}{|c|}{$\chi^2_{\text{best-fit}}/N_{\text{dof}}=126.7/(100-12)=1.44$}\\  \hline
\end{tabular}
 \begin{tabular}{|c|c|c|} 
 \hline
high-z SGC & best-fit & mean  $\pm 1\sigma$  \\ [0.5ex] 
 \hline\hline
 $A$ & $0.934$ & $0.753^{+0.181}_{-0.302}$  \\ \hline
 $h$  & $0.639$ & $0.665_{-0.047}^{+0.022}$  \\ \hline
 $10^2\omega_{b}$  & $2.234$ & $2.237 \pm 0.015$  \\\hline
 $\omega_{cdm}$  & $0.1135$ & $0.1120_{-0.0163}^{+0.0119}$  \\\hline
 $m_\nu$  & $0.077$ & $0.120 \pm 0.041$  \\\hline
 $b_1\times  A^{1/2}$  & $2.109$ & $2.059_{-0.099}^{+0.140}$  \\  \hline
 $b_2\times  A^{1/2}$ & $-1.61$& $-1.32^{+0.93}_{-2.63}$  \\  \hline
 $b_{\mathcal{G}_2}\times  A^{1/2}$  & $ 0.13$ & $0.26^{+0.58}_{-0.86}$  \\  \hline
 $c^2_0$ & $-14.1$& $29.2^{+61.7}_{-77.0}$  \\  \hline
 $c^2_2$  &$23.0$ & $-0.17^{+76.8}_{-43.0}$  \\  \hline
 $\tilde{c}$ & $ 203$ & $ 319\pm 195$  \\ \hline
 $10^{-3}P_{\text{shot}}$ & $0.97$& $3.80^{+1.10}_{-3.80}$  \\ \hline \hline 
  $\sigma_8$ & $ 0.744$ & $ 0.646 \pm 0.107 $  \\ \hline
  $\Omega_m$ & $ 0.334$ & $ 0.309^{+0.041}_{-0.032} $  \\ \hline
   \multicolumn{3}{|c|}{$\chi^2_{\text{best-fit}}/N_{\text{dof}}=130.2/(100-12)=1.48$}\\  \hline
\end{tabular}
\begin{tabular}{|c|c|c|} 
 \hline
low-z SGC & best-fit & mean  $\pm 1\sigma$  \\ [0.5ex] 
 \hline\hline
 $A$ & $0.996$ & $0.875^{+0.229}_{-0.385}$  \\ \hline
 $h$  & $0.683$ & $0.697^{+0.029}_{-0.048}$  \\ \hline
 $10^2\omega_{b}$  & $2.236$ & $2.237 \pm 0.015$  \\ \hline
 $\omega_{cdm}$  & $0.1082$ & $0.1026^{+0.0100}_{-0.0136}$  \\\hline
 $m_\nu$  & $0.170$ & $0.122_{-0.027}^{+0.055}$  \\ \hline
 $b_1\times  A^{1/2}$  & $1.885$ & $1.904_{-0.108}^{+0.120}$  \\  \hline
 $b_2\times  A^{1/2}$ & $-3.00$& $-1.90^{+0.65}_{-2.10}$  \\  \hline
 $b_{\mathcal{G}_2}\times  A^{1/2}$  & $0.43$ & $0.61^{+0.56}_{-0.78}$  \\  \hline
 $c^2_0$ & $-18.1$& $39.0^{+62.3}_{-74.3}$  \\  \hline
 $c^2_2$  &$-12.2$ & $25.0^{+80.3}_{-46.9}$  \\  \hline
 $\tilde{c}$ & $209$& $ 414^{+496}_{-388}$  \\ \hline
 $10^{-3}P_{\text{shot}}$ & $5.56$& $5.56_{-1.99}^{+3.72}$  \\ \hline \hline 
 $\sigma_8$ & $ 0.734$ & $ 0.658^{+0.106}_{-0.126} $  \\ \hline
 $\Omega_m$ & $ 0.284$ & $ 0.262^{+0.031}_{-0.026}$  \\ \hline
 \multicolumn{3}{|c|}{$\chi^2_{\text{best-fit}}/N_{\text{dof}}=95.1/(100-12)=1.08$}\\  \hline
\end{tabular}
\caption{
The results of our MCMC analysis for
different data samples. 
The neutrino mass is quoted in units of [eV],
$H_0$ in [km/s/Mpc],
parameters $c_0^2$
and $c_2^2$ are quoted in units $[\Mpc/h]^2$, 
$\tilde{c}$ in units $[\Mpc/h]^4$, $P_{\text{shot}}$ in units $[\Mpc/h]^3$.
Note that the limits on $\omega_b$, $m_\nu$, $b_2 A^{1/2}$ 
and $P_{\text{shot}}$ are prior-dominated.
% We show asymmetric limits if they are different by more than $10\%$.
}
 \label{tab:datangc}
\end{center}
\end{table}

\section{Supplementary Material}
\label{app:sup}

In this Appendix we present some additional material. It includes full parameter constrain tables 
and corner plots for the baseline analysis, along with the results of the extended analyses 
that waived priors on the primordial power spectrum tilt $n_s$
and the neutrino mass. Finally, we show that are constraints are not sensitive to the data on very 
large scales, which are susceptible to systematics. 

\subsection{Full Triangle Plot and Constraint Tables}
\label{app:full}

Let us present some additional material related to the 
baseline analysis with the Planck prior on $\omega_b$. The results for the BBN
priors are the same.
The full triangle plot for four 
non-overlapping BOSS data chunks can be found in Fig.~\ref{fig:monster}. 
We do not show the contours for $m_\nu$ and $\omega_b$ 
as they are prior-dominated. The corresponding 1d marginalized
limits can be found in Table~\ref{tab:datangc}.
For completeness, we also show the spectra for the SGC datasets along with the 
best-fit theoretical curves in Fig.~\ref{fig:bf}. Similar plots for the 
NGC data were shown in Fig.~\ref{fig:final}.

Note that the reduced $\chi^2$ is a very inaccurate metric for the goodness of fit.
First, it does not include the covariance between different $k$-bins.
Second, the naive reduced $\chi^2$ does not take into account that the  cosmological constraints are always driven
by the biggest wavenumbers used in the analysis.
This is important to keep in mind when interpreting our results. 
Indeed, some of the values quoted in Table~\ref{tab:datangc} (e.g. for the high-z SGC sample) 
are noticeably bigger than unity,
which naively implies a bad fit. 
However, if we compute the reduced $\chi^2$ for the same parameters but
using, e.g.~$k_{\text{min}}=0.05\,h$/Mpc instead of $0.0025\,h$/Mpc employed in our analysis, 
we find different numbers:
$61.7/(80-12)=0.91$, 
$97.4/(80-12)=1.43$, 
$75.5/(80-12)=1.11$, 
$69.6/(80-12)=1.02$
for the high-z NGC, low-z NGC, high-z SGC, and low-z SGC samples, respectively.
Note a significant improvement for the high-z datasamples.

Note that the choice of $k_{\text{min}}$ within some reasonable range
has a very mild effect on the parameter estimates (less than 1$\sigma$).
This illustrates that the values 
quoted in Table~\ref{tab:datangc} only give a very rough idea on
the quality of the fit and hence
should be taken with a grain of salt.

\begin{table}[ht!]
\begin{center}
 \begin{tabular}{|c|c|c|} 
 \hline
$z_{\text{eff}}=0.61$ & best-fit & mean  $\pm 1\sigma$  \\ [0.5ex] \hline\hline
 $f\sigma_8(z_{\text{eff}})$ & $ 0.47135$ & $ 0.4689^{+0.0070}_{-0.0045}$  \\ \hline
 $H(z_{\text{eff}})$ & $ 95.58$ & $ 95.16_{-0.29}^{+0.55}$  \\ \hline
 $D_A(z_{\text{eff}})$ & $ 1425.4$ & $ 1438.9^{+8.9}_{-17.1} $  \\ \hline
 $F_{\text{AP}}(z_{\text{eff}})$ & $ 0.7317$ & $ 0.7353_{-0.0045}^{+0.0025} $  \\ \hline
 $D_V(z_{\text{eff}})$ & $ 2160.0$ & $ 2176.7^{+11.1}_{-21.4} $  \\ \hline
\end{tabular}
\vspace{0.2cm}
\begin{tabular}{|c|c|c|} 
 \hline
$z_{\text{eff}}=0.38$ & best-fit & mean  $\pm 1\sigma$  \\ [0.5ex] 
 \hline\hline
 $f\sigma_8(z_{\text{eff}})$ & $ 0.4769$ & $ 0.4766^{+0.0062}_{-0.0053}$  \\ \hline
 $H(z_{\text{eff}})$ & $ 83.319$ & $ 82.69^{+0.80}_{-0.43}$  \\ \hline
 $D_A(z_{\text{eff}})$ & $ 1102.49$ & $ 1114.81^{+8.17}_{-15.60} $  \\ \hline
 $F_{\text{AP}}(z_{\text{eff}})$ & $ 0.42284$ & $ 0.42429_{-0.0018}^{+0.0010} $  \\ \hline
 $D_V(z_{\text{eff}})$ & $ 1468.2$ & $ 1478.6_{-18.6}^{+9.77} $  \\ \hline
\end{tabular}
\caption{
Planck results for distances to the BOSS galaxy samples in
the base $\Lambda$CDM with massive neutrinos. The values of $H$ are quoted in units of [km/s/Mpc], $D_A$ and $D_V$
in [Mpc].
}
\label{tab:distPlanck}
\end{center}
\end{table}

\begin{figure}[h!]
\includegraphics[width=0.49\textwidth]{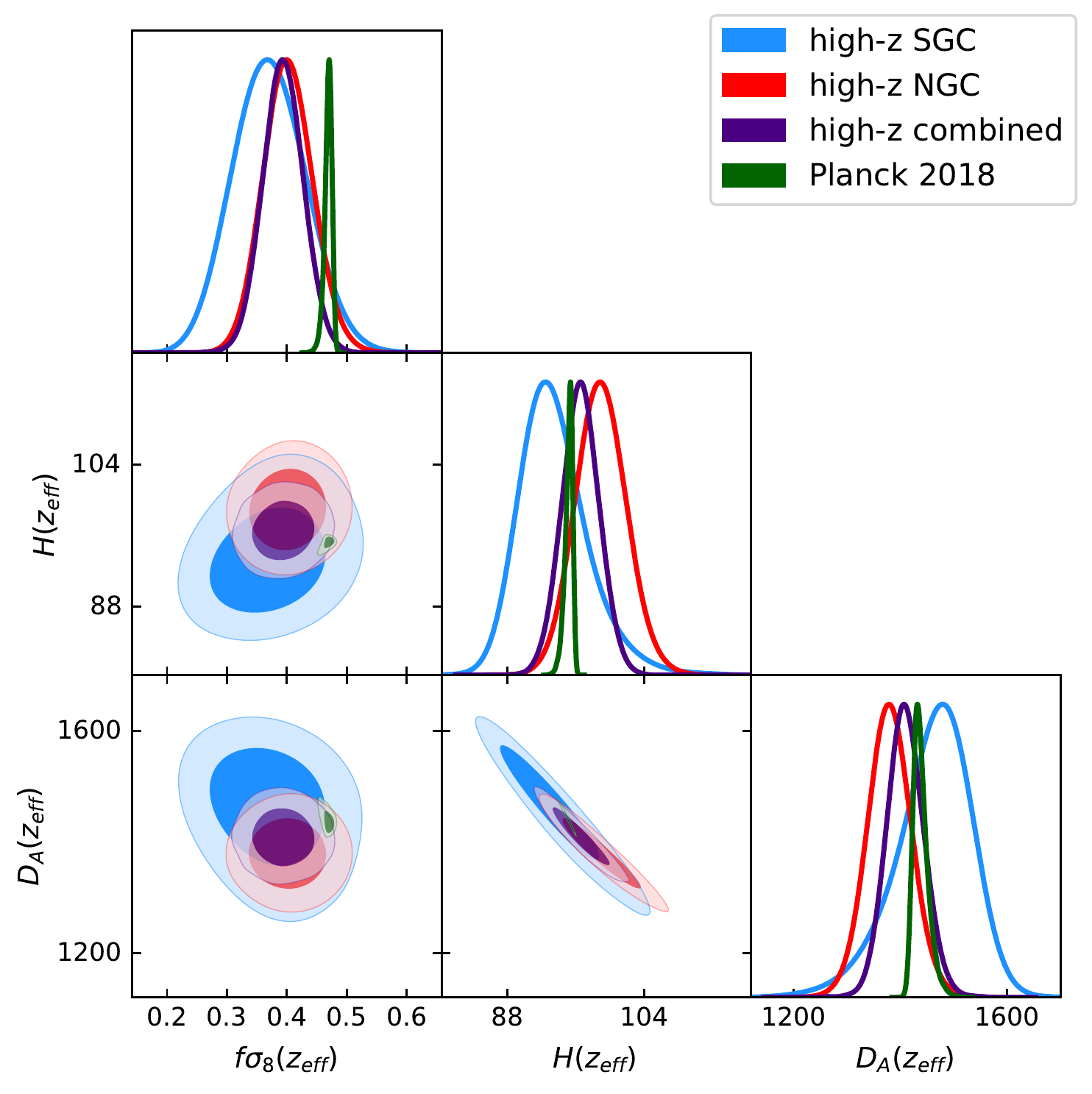}
\includegraphics[width=0.49\textwidth]{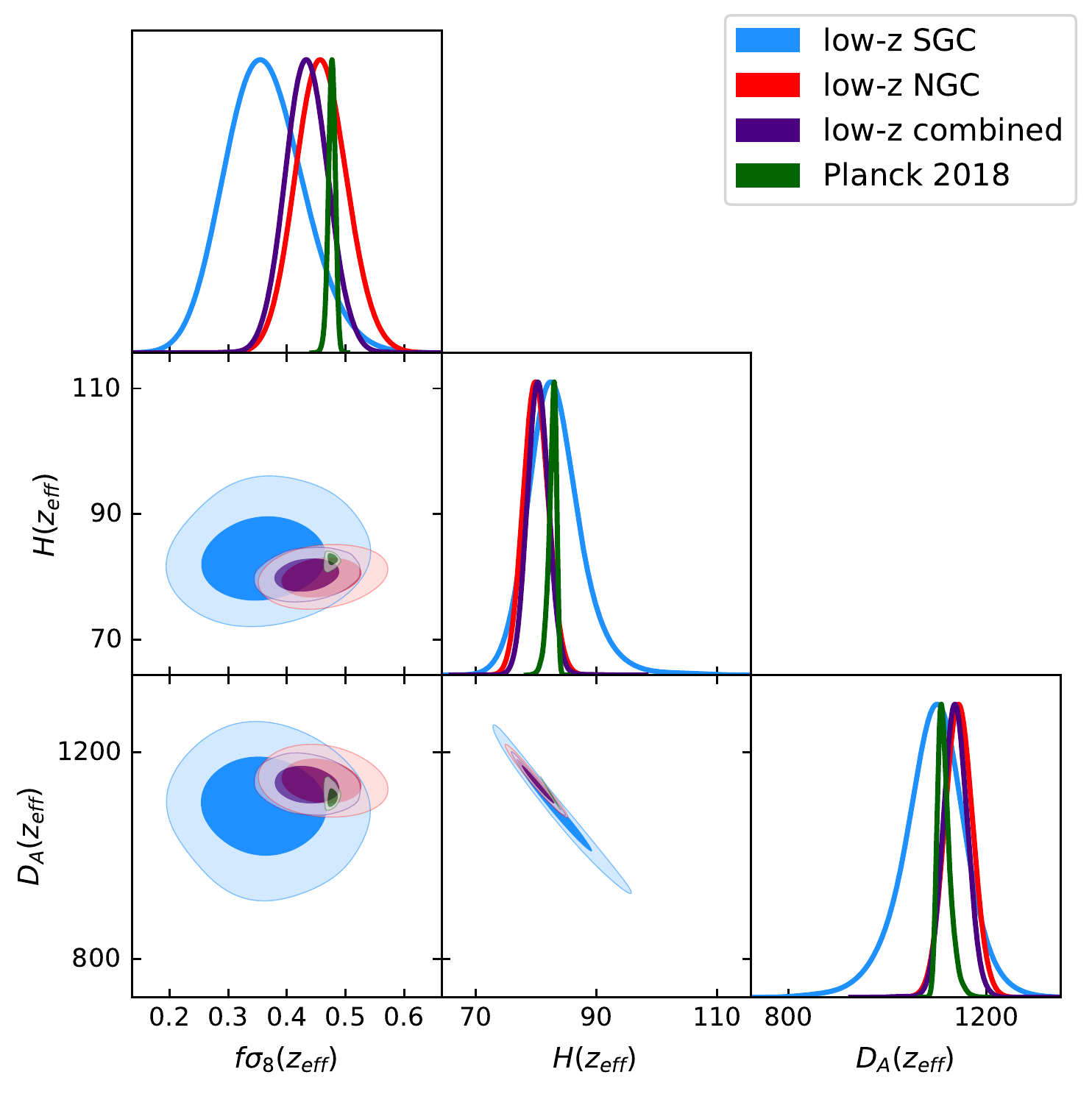}
% \begin{figure}[h!]
\includegraphics[width=0.49\textwidth]{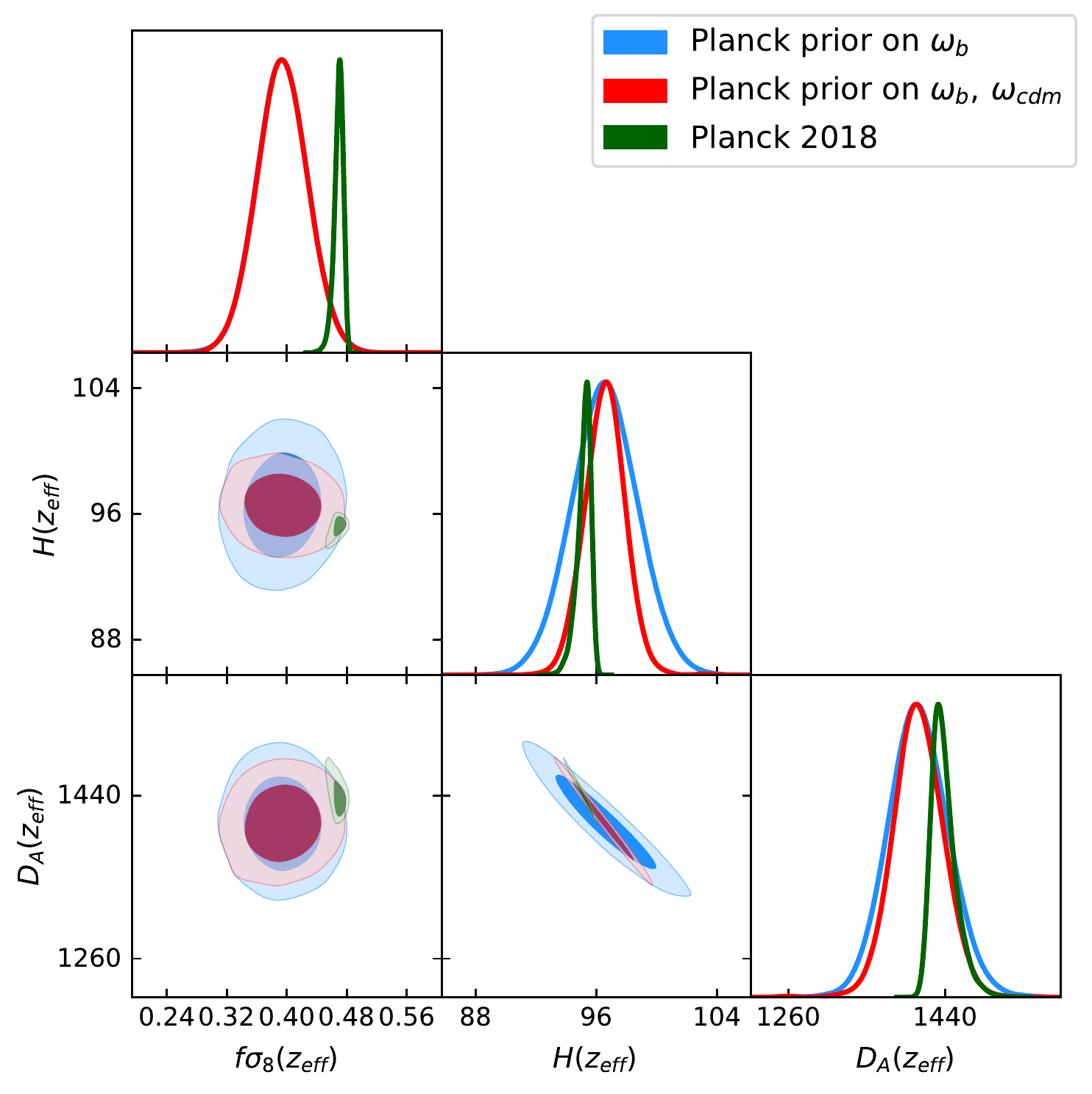}
\includegraphics[width=0.49\textwidth]{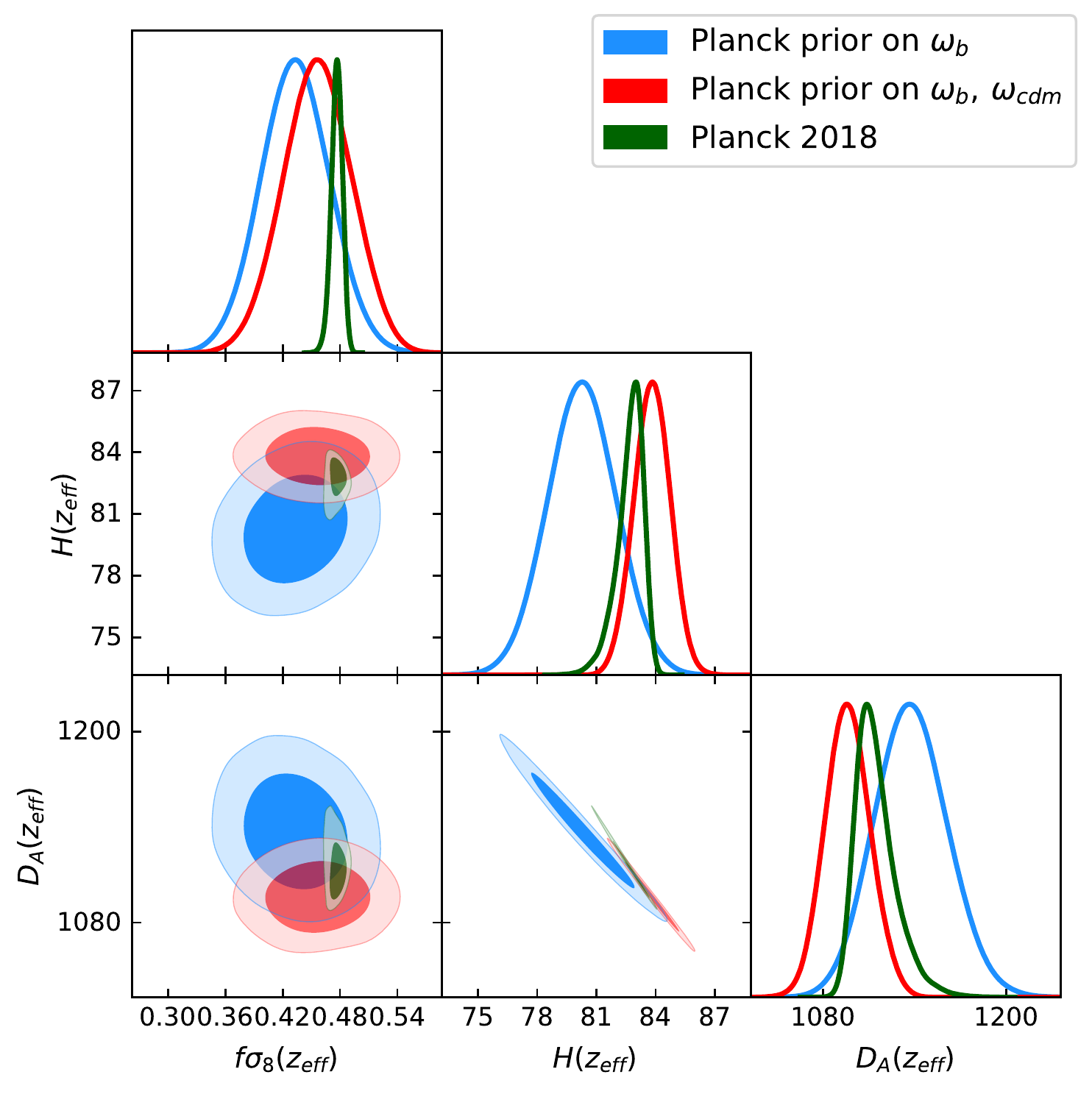}
% \caption{The posteriors for $f\sigma_8-D_A-H$
% at $z_{\text{eff}}=0.61$ (left panel) and 
% $z_{\text{eff}}=0.38$ (right panel).
% \label{fig:fs8-hz-da-rd}
% }    
% \end{figure}
\caption{The posteriors for $f\sigma_8-D_A-H$
as extracted from our baseline MCMC chains with Planck priors on $\omega_b$ (upper panels)
and Planck priors on both $\omega_b$ and $\omega_{cdm}$ (lower panels)
at $z_{\text{eff}}=0.61$ (left panels) and 
$z_{\text{eff}}=0.38$ (right panels). 
The values of $H$ are quoted in units of [km/s/Mpc], $D_A$
in [Mpc].
\label{fig:fs8-hz-da-comb}
}    
\end{figure}

\begin{table}[ht!]
\begin{center}
 \begin{tabular}{|c|c|c|} 
 \hline
high-z & best-fit & mean  $\pm 1\sigma$  \\ [0.5ex] \hline\hline
 $f\sigma_8(z_{\text{eff}})$ & $ 0.393$ & $ 0.394 \pm  0.034$  \\ \hline
 $H(z_{\text{eff}})$ & $ 96.53$ & $ 96.84 \pm  2.33$  \\ \hline
 $D_A(z_{\text{eff}})$ & $ 1409$ & $ 1405 \pm 36 $  \\ \hline
 $F_{\text{AP}}(z_{\text{eff}})$ & $ 0.7307$ & $ 0.7303 \pm 0.0057 $  \\ \hline
 $D_V(z_{\text{eff}})$ & $ 2137$ & $ 2130 \pm 53 $  \\ \hline
\end{tabular}
\vspace{0.2cm}
\begin{tabular}{|c|c|c|} 
 \hline
low-z & best-fit & mean  $\pm 1\sigma$  \\ [0.5ex] 
 \hline\hline
 $f\sigma_8(z_{\text{eff}})$ & $ 0.4308$ & $ 0.434 \pm  0.038$  \\ \hline
 $H(z_{\text{eff}})$ & $ 80.23$ & $ 80.35 \pm  1.8$  \\ \hline
 $D_A(z_{\text{eff}})$ & $ 1138$ & $ 1137 \pm 25 $  \\ \hline
 $F_{\text{AP}}(z_{\text{eff}})$ & $ 0.4202$ & $ 0.4203 \pm 0.0018 $  \\ \hline
 $D_V(z_{\text{eff}})$ & $ 1487$ & $ 1486 \pm 33 $  \\ \hline
\end{tabular}
\caption{
The distances and the fluctuation growth parameter for
the high-z (left table), low-z (right table) 
data samples for the base $\Lambda$CDM with the Planck prior on $\omega_b$. 
The values of $H$ are quoted in units of [km/s/Mpc], $D_A$ and $D_V$
in [Mpc].
}
\label{tab:daHfs8}
\end{center}
\end{table}

\begin{table}[ht!]
\begin{center}
 \begin{tabular}{|c|c|c|} 
 \hline
high-z & best-fit & mean  $\pm 1\sigma$  \\ [0.5ex] \hline\hline
 $\Omega_m$ & $ 0.3007$ & $ 0.3026 \pm  0.0172$  \\ \hline
 $H_0$ & $ 69.13$ & $ 69.00 \pm  1.93$  \\ \hline
 $\sigma_8$ & $ 0.702$ & $ 0.686 \pm  0.060$  \\ \hline \hline 
 $f\sigma_8(z_{\text{eff}})$ & $ 0.403$ & $ 0.394 \pm  0.035$  \\ \hline
 $H(z_{\text{eff}})$ & $ 96.64$ & $ 96.57 \pm  1.38$  \\ \hline
 $D_A(z_{\text{eff}})$ & $ 1406$ & $ 1409 \pm 30 $  \\ \hline
 $F_{\text{AP}}(z_{\text{eff}})$ & $ 0.7300$ & $ 0.7305 \pm 0.0052 $  \\ \hline
 $D_V(z_{\text{eff}})$ & $ 2133$ & $ 2136 \pm 40 $  \\ \hline
\end{tabular}
\vspace{0.2cm}
\begin{tabular}{|c|c|c|} 
 \hline
low-z & best-fit & mean  $\pm 1\sigma$  \\ [0.5ex] \hline\hline
 $\Omega_m$ & $ 0.3032$ & $ 0.3057 \pm  0.0082$  \\ \hline
 $H_0$ & $ 68.99$ & $ 68.46 \pm  1.07$  \\ \hline
 $\sigma_8$ & $ 0.913$ & $ 0.783 \pm  0.061$  \\ \hline \hline 
 $f\sigma_8(z_{\text{eff}})$ & $ 0.530$ & $ 0.456 \pm  0.035$  \\ \hline
 $H(z_{\text{eff}})$ & $ 84.32$ & $ 83.82 \pm  0.91$  \\ \hline
 $D_A(z_{\text{eff}})$ & $ 1089$ & $ 1096 \pm 15 $  \\ \hline
 $F_{\text{AP}}(z_{\text{eff}})$ & $ 0.4225$ & $ 0.4229 \pm 0.0011 $  \\ \hline
 $D_V(z_{\text{eff}})$ & $ 1419$ & $ 1428 \pm 17 $  \\ \hline
\end{tabular}
\end{center}
\caption{ 
The distances and the fluctuation growth parameter for
the high-z (left table), low-z (right table) 
data samples for the base $\Lambda$CDM with the Planck priors on $\omega_b$ and $\omega_{cdm}$. 
The values of $H$ are quoted in units of [km/s/Mpc], $D_A$ and $D_V$
in [Mpc].
}
\label{dist_rd}
\end{table}

It is instructive to convert our results into the triplet
$f\sigma_8-D_A-H$ at $z_{\text{eff}}$ commonly used
in the large-scale structure literature. 
We focus on the results obtained with the 
Planck prior on $\omega_b$.
The corresponding 2d posterior distribution
projections are displayed in Fig.~\ref{fig:fs8-hz-da-comb} (upper panels), 1d marginalized limits are given in Table.~\ref{tab:daHfs8}. For comparison, in Table.~\ref{tab:distPlanck}
we quote the limits extracted from the Planck MCMC chains run 
for $\L$CDM with a free neutrino mass. 
Overall, we see that the BOSS distance information is superseded by Planck, which 
gives much better constraints on $H(z_{\text{eff}})$ and $D_A(z_{\text{eff}})$. One can notice a significant correlation between $D_A$ and $H$
in the corresponding panels. This degeneracy direction simply reflects 
the fact that in $\Lambda$CDM these two quantities are related by 
definition, see Eq.~\eqref{eq:Dadef},
such that the product $D_A H$ is nearly constant. 
We stress that the constraints on the distance parameters shown in Table.~\ref{tab:daHfs8}
do not use the Planck prior on $r_d$.

Finally, we show distance measurements in the case of the 
joint Planck prior on $\omega_b$ and $\omega_{cdm}$, which are 
presented in Table.~\ref{dist_rd} and displayed in Fig.~\ref{fig:fs8-hz-da-comb} (lower panels).

\subsection{Full Likelihood including the Power Spectrum Tilt}
\label{app:tilt}

\begin{table}[ht!]
\begin{center}
 \begin{tabular}{|c|c|c|} 
 \hline
BBN $\omega_b$& best-fit & mean  $\pm 1\sigma$  \\ [0.5ex] \hline\hline
 $\omega_{cdm}$ & $ 0.1267$ & $ 0.1268 \pm  0.0099$  \\ \hline
  $H_0$ & $ 68.61$ & $ 68.55 \pm  1.47$  \\ \hline
  $n_s$ & $ 0.874$ & $ 0.876 \pm  0.076$  \\ \hline
 \hline 
 $\sigma_8$ & $ 0.724$ & $ 0.728 \pm  0.052$  \\ \hline
  $\Omega_m$ & $ 0.320$ & $ 0.321 \pm  0.018$  \\ \hline
\end{tabular}
\caption{
The results for cosmological parameters from the full BOSS likelihoods
with \textit{all} relevant cosmological parameters varied.
$H_0$ is quoted in units [km/s/Mpc]. We do not show the limits on $\omega_b$
and $\sum m_\nu$ as they are prior-dominated.
}
 \label{tab:ns}
\end{center}
\end{table}

\begin{figure}[ht!]
\centering \includegraphics[width=0.6\textwidth]{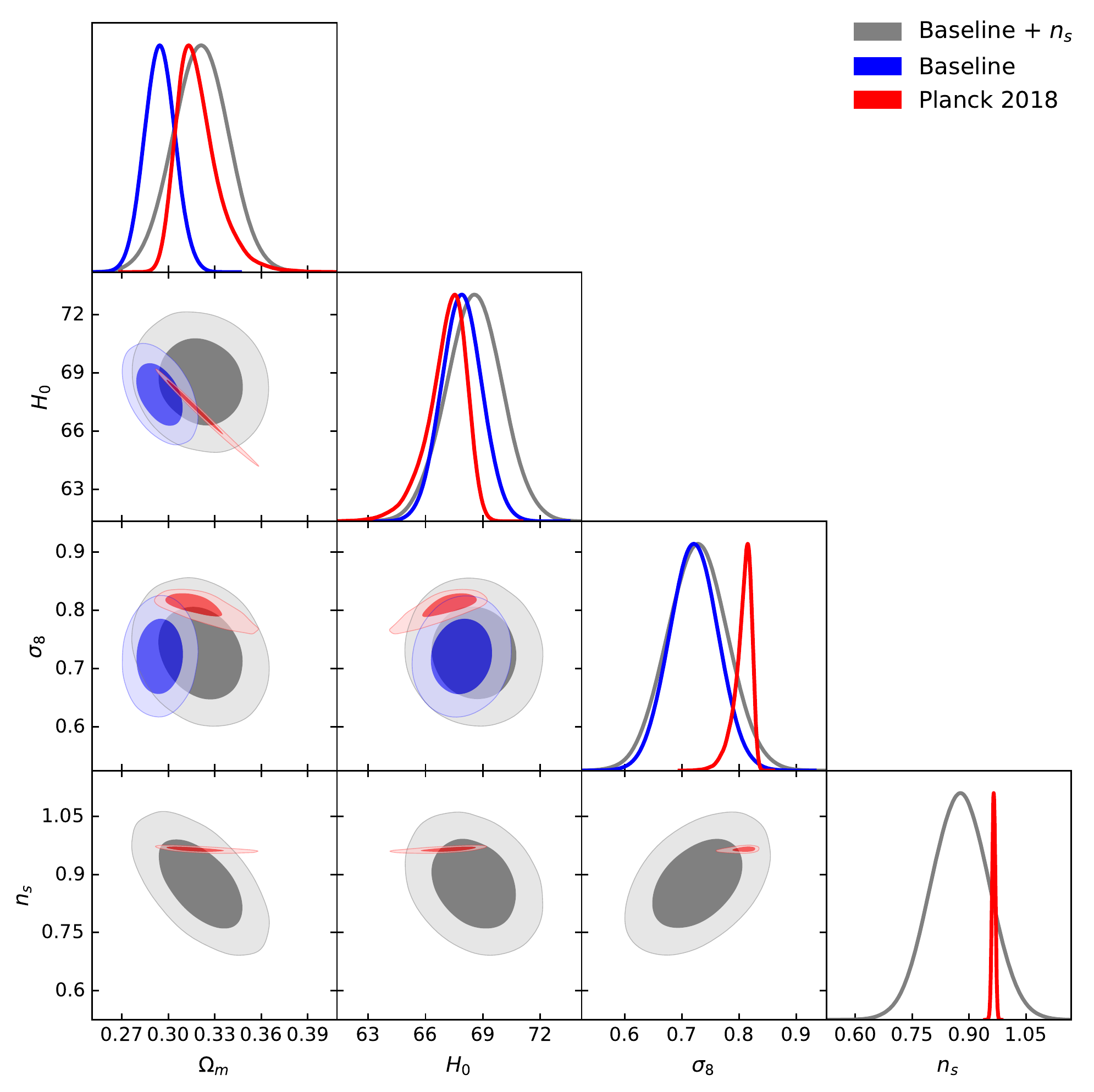}
\caption{2d posterior distribution and 1d marginalized curves for $\Omega_m, H_0,\sigma_8$
and $n_s$ (gray contours) obtained with the BBN prior on $\omega_b$ and the tight prior
on $\sum m_\nu$. Analogous contours obtained for a fixed $n_s=0.9649$ are shown in blue. They correspond to our baseline analysis. For comparison, we also display the Planck 2018 CMB results (in red) for the same cosmological model ($\L$CDM with varied neutrino masses).}    
\label{fig:ns}
\end{figure}

Our baseline analysis was performed for the fixed power spectrum tilt $n_s$. In this Appendix 
we present the results of varying the full power spectrum likelihood. As in the baseline analysis, we keep the BBN prior on $\omega_b$ \eqref{eq:omega_b_BBN} and the prior on the neutrino masses \eqref{eq:mnuprior}, but do not assume any prior on $n_s$ whatsoever (i.e. we use a plat prior in the range ($-\infty,\infty$)). The results are displayed in Table~\ref{tab:ns}
and in Fig.~\ref{fig:ns}. 

One observes that including $n_s$ to the fit notably worsens the 
precision of the $\omega_{cdm}$ measurement. This is to be expected as both these parameters
are extracted from the power spectrum slope. The correlation between $\omega_{cdm}$ and $n_s$ backfires on the posterior distribution for 
$\Omega_m$, which shifts to a higher value almost by 1$\sigma$ w.r.t.~our baseline analysis (with fixed $n_s$). 
In turn, $\Omega_m$ pulls $H_0$ up and somewhat widens its marginalized posterior.
Overall, the shift in $H_0$ is not very significant ($\lesssim 0.5\sigma$).
Note that the independent measurement of $n_s$ is consistent within 1$\sigma$
with the Planck CMB result. 

\subsection{Effect of Neutrino Masses}
\label{app:mnu}

\begin{table}[ht!]
\begin{center}
 \begin{tabular}{|c|c|c|} 
 \hline
low-z NGC, Baseline + $M_{\rm tot}$& best-fit & mean  $\pm 1\sigma$  \\ [0.5ex] \hline\hline
$A$ & $ 1.68$ & $ 1.38^{+0.23}_{-0.34}$  \\ \hline
 $\omega_{cdm}$ & $ 0.0986$ & $ 0.1061^{+0.09}_{-0.011}$  \\ \hline
  $H_0$ & $ 65.26$ & $ 66.6 \pm  1.7$  \\ \hline
  $M_{\rm tot}$ & $ 0.262$ & $ <1.17$~($95\%$ CL)  \\ \hline
 \hline 
 $\sigma_8$ & $ 0.862$ & $ 0.786 \pm  0.077$  \\ \hline
  $\Omega_m$ & $ 0.291$ & $ 0.3013^{+0.019}_{-0.023}$  \\ \hline
  $\Omega_{cb}$ & $ 0.285$ & $ 0.290 \pm  0.018$  \\ \hline
\end{tabular}
\caption{
The results for cosmological parameters 
obtained in our baseline analysis without imposing a prior on the total neutrino mass $M_{\rm tot}$ (in eV).
$H_0$ is quoted in units [km/s/Mpc]. $\Omega_{cb}$ is the current fractional density of the cold dark matter 
and baryons. For comparison, 
its value inferred in our baseline analysis with the tight prior on $M_{\rm tot}$
is $\Omega_{cb}=0.288\pm 0.017$ (best-fit $\Omega_{cb}=0.292$). 
}
 \label{tab:mnu}
\end{center}
\end{table}

\begin{figure}[ht!]
\centering \includegraphics[width=0.99\textwidth]{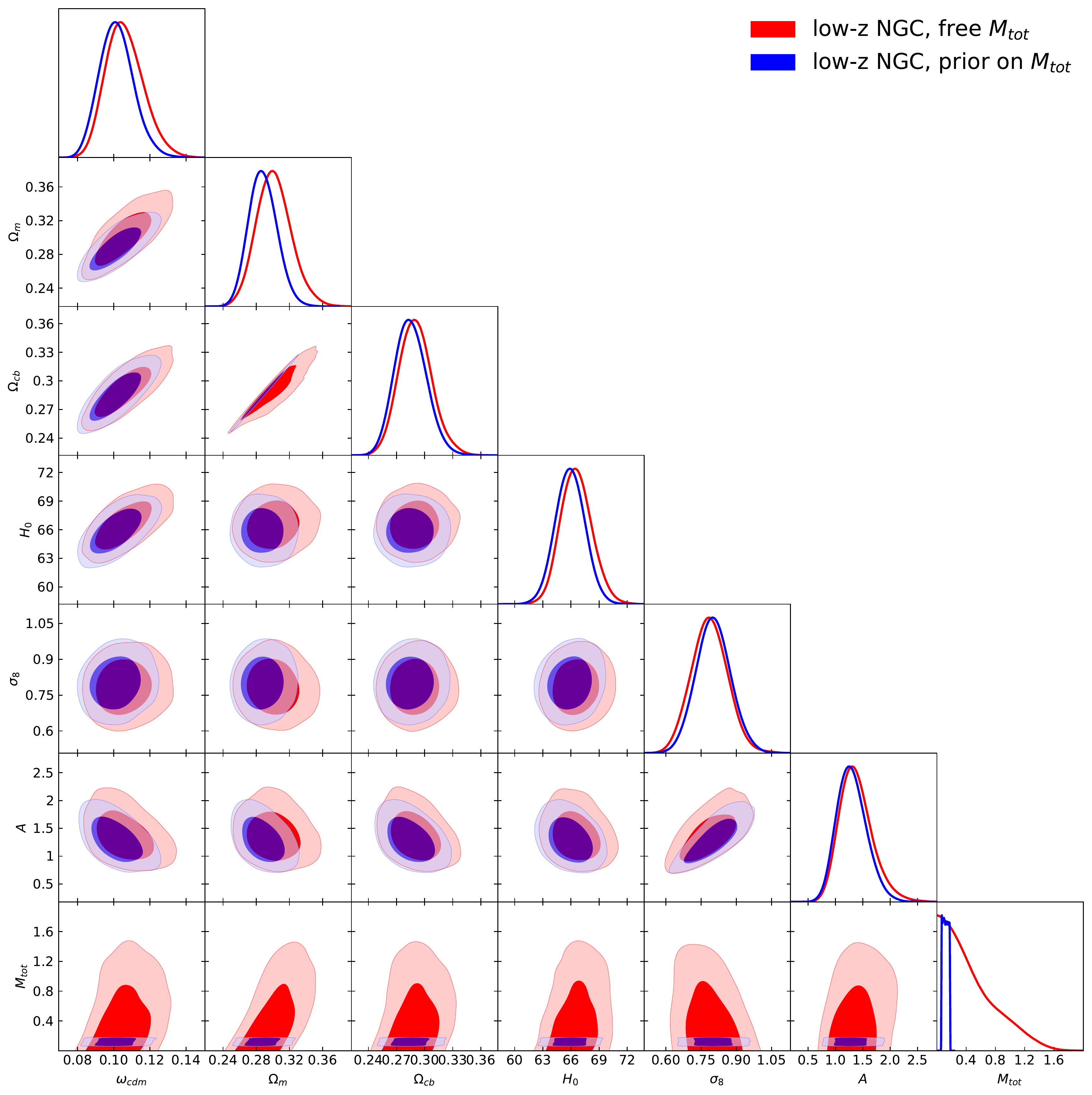}
\caption{2d posterior distribution and 1d marginalized curves for the cosmological parameters of $\L$CDM 
obtained from our baseline analysis without imposing a prior on the neutrino mass.
}    
\label{fig:mnu}
\end{figure}

In this Appendix we present results of our analysis of the low-z NGC datasample without informative 
priors on the neutrino mass $\sum m_\nu \equiv M_{\rm tot}$. Other than that, our methodology and priors are the same as in the baseline analysis. In particular, we assume the BBN prior on $\omega_b$
and fix $n_s$ to the Planck best-fit value.
The results are presented in Table~\ref{tab:mnu} and in Fig.~\ref{fig:mnu}.

The first relevant observation is that our constraints on $H_0$ and $\sigma_8$ are almost insensitive 
to the neutrino mass. This must be contrasted with the Planck constraints on these parameters \cite{Aghanim:2018eyx}, 
which depend strongly on the neutrino masses. 
The second observation is that $\Omega_m$ is, obviously, correlated with the neutrino mass and in this sense cannot be treated 
as an independent parameter. 
However, our limit on the late-time cold dark matter and baryon 
density fraction $\Omega_{cb}$ is almost the same in our baseline analysis and in the 
analysis with totally free $M_{\rm tot}$.
Thus, the measurements of ${\Omega_{cb},H_0,\sigma_8}$ 
from the BOSS data are quite robust w.r.t. the priors on the neutrino 
mass. 

Overall, we conclude that BOSS alone are not very sensitive to the neutrino mass,
and even very large $M_{\rm tot}$ around 1 eV is allowed.
These values are already excluded by other cosmological probes, 
e.g. by the Ly$\alpha$-forest data alone (which sets a limit $M_{\rm tot}<0.71$ eV \cite{Palanque-Delabrouille:2019iyz}),
as well as by the particle physics laboratory experiments like KATRIN \cite{Aker:2019uuj}.

\subsection{Effect of Large Scales }

\begin{figure}[ht!]
\centering \includegraphics[width=0.49\textwidth]{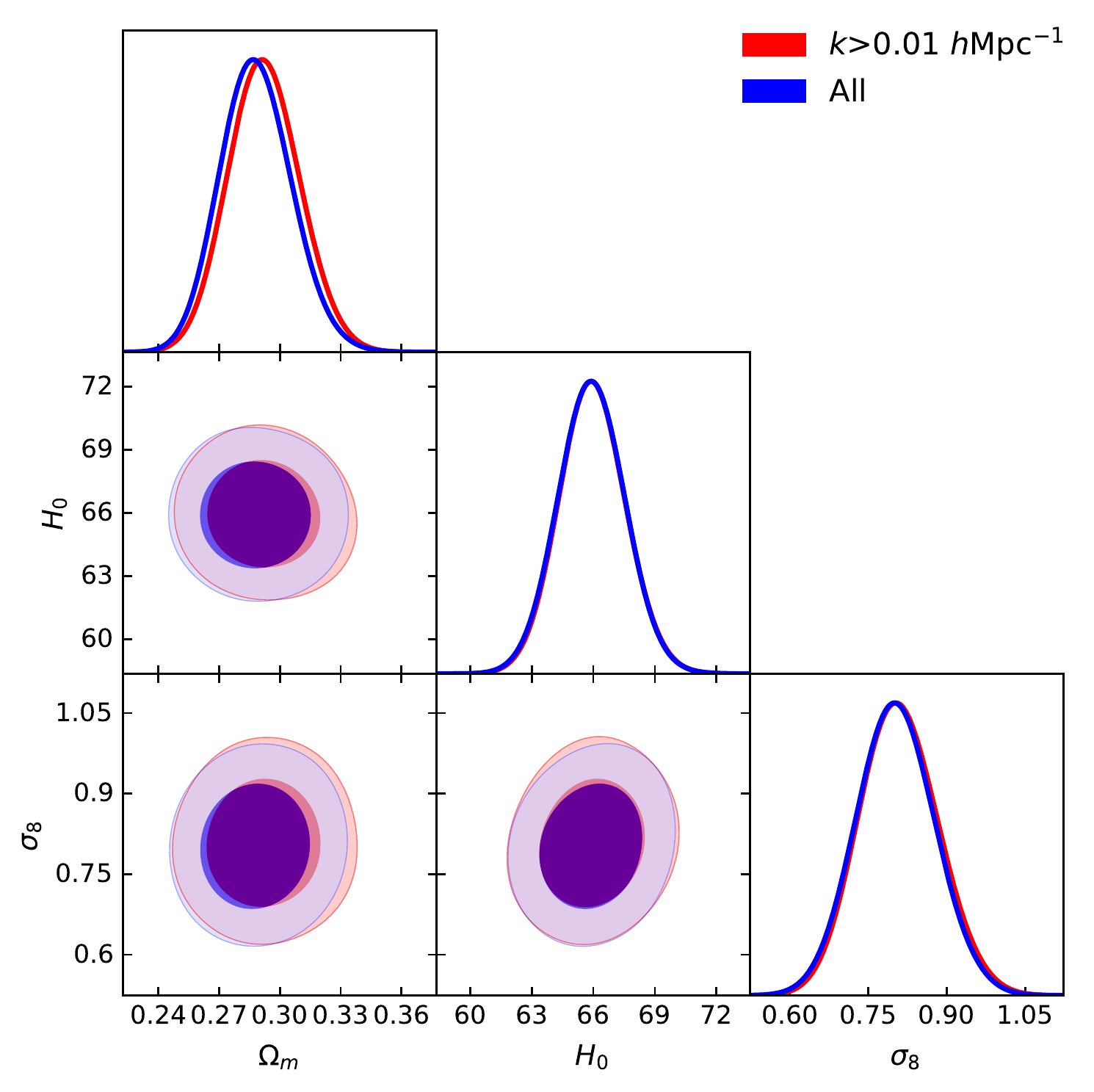}
\caption{Corner plot for the cosmological parameters of $\L$CDM 
obtained in our baseline analysis. 
The shown are results from the data with all $k$-bins (in blue) and with the 
momentum cut $k>0.01~h$Mpc$^{-1}$ (in red). 
}    
\label{fig:kmin}
\end{figure}

The BOSS spectra feature some spurious excess of power on scales \mbox{$k<0.01~h$Mpc$^{-1}$}. 
In Fig.~\ref{fig:kmin} we show the results obtained from two analyses of the low-z NGC data:
using all the $k$-bins and having imposed the cut $k>0.01~h$Mpc$^{-1}$. 
The marginalized posteriors for $H_0$ and $\sigma_8$ are identical to the ones obtained in our 
baseline analysis, whereas the mean value of $\Omega_m$ is shifted upwards by $0.1\sigma$ 
when imposing the cut. 
Clearly, the signal is dominated by wavenumbers $k>0.01~h$Mpc$^{-1}$.

\section{Scaling Parameter Analysis}
\label{app:alphas}

In this Section we give some details on our analysis in which 
we followed the standard parameterization and parameter estimation
routine adopted in the previous BOSS FS analyses.
We try to reproduce
the analyses performed in Refs.~\cite{Beutler:2016arn,Gil-Marin:2015sqa}
as close as we could without a drastic modification of our pipeline.
Our $\alpha$-analysis performed in this paper is only meant to capture some main
qualitative features of the standard pipeline. It is not aimed at accurately reproducing 
the previous results.

\begin{figure}[ht!]
\centering \includegraphics[width=0.49\textwidth]{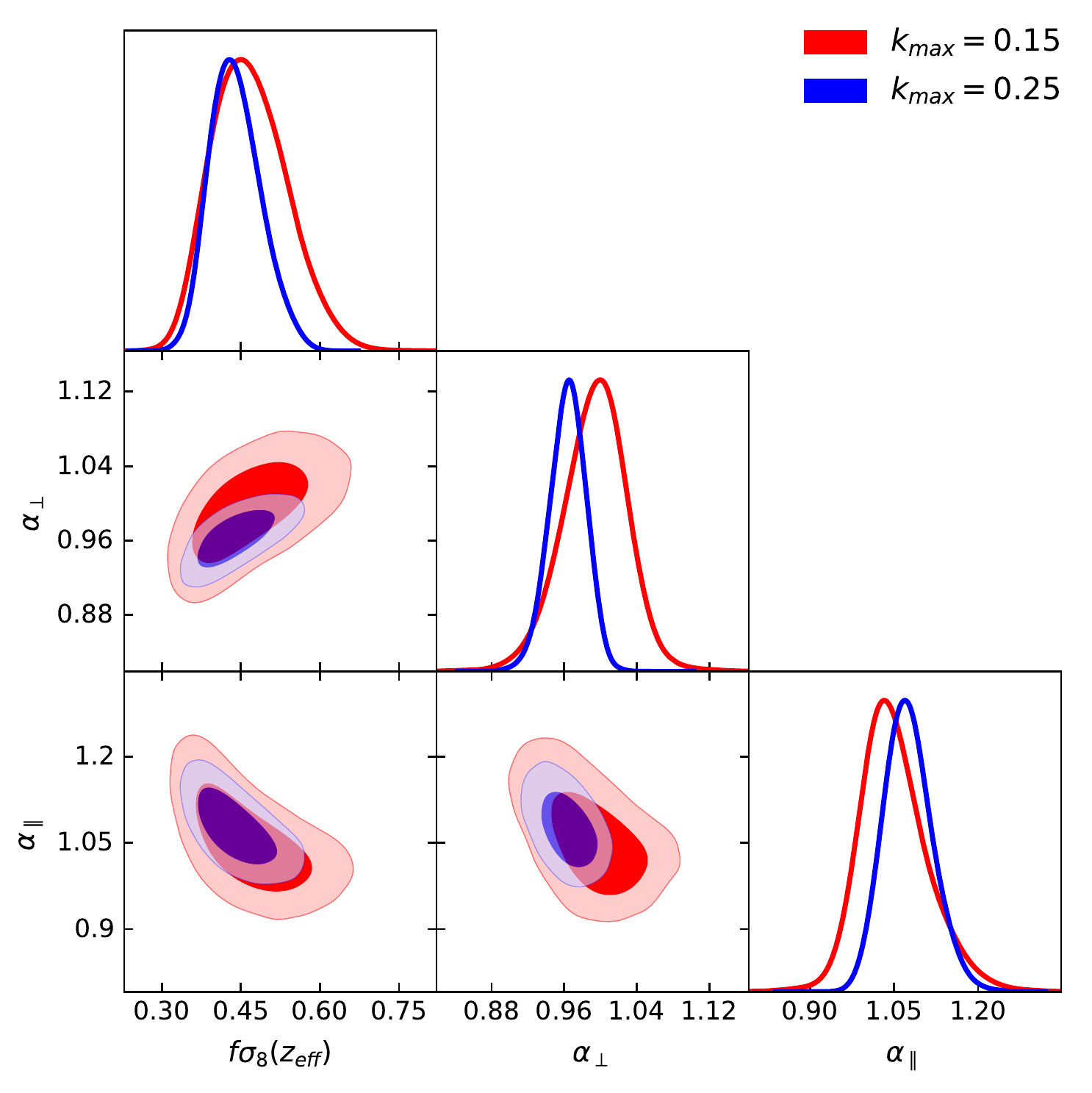}
\centering \includegraphics[width=0.49\textwidth]{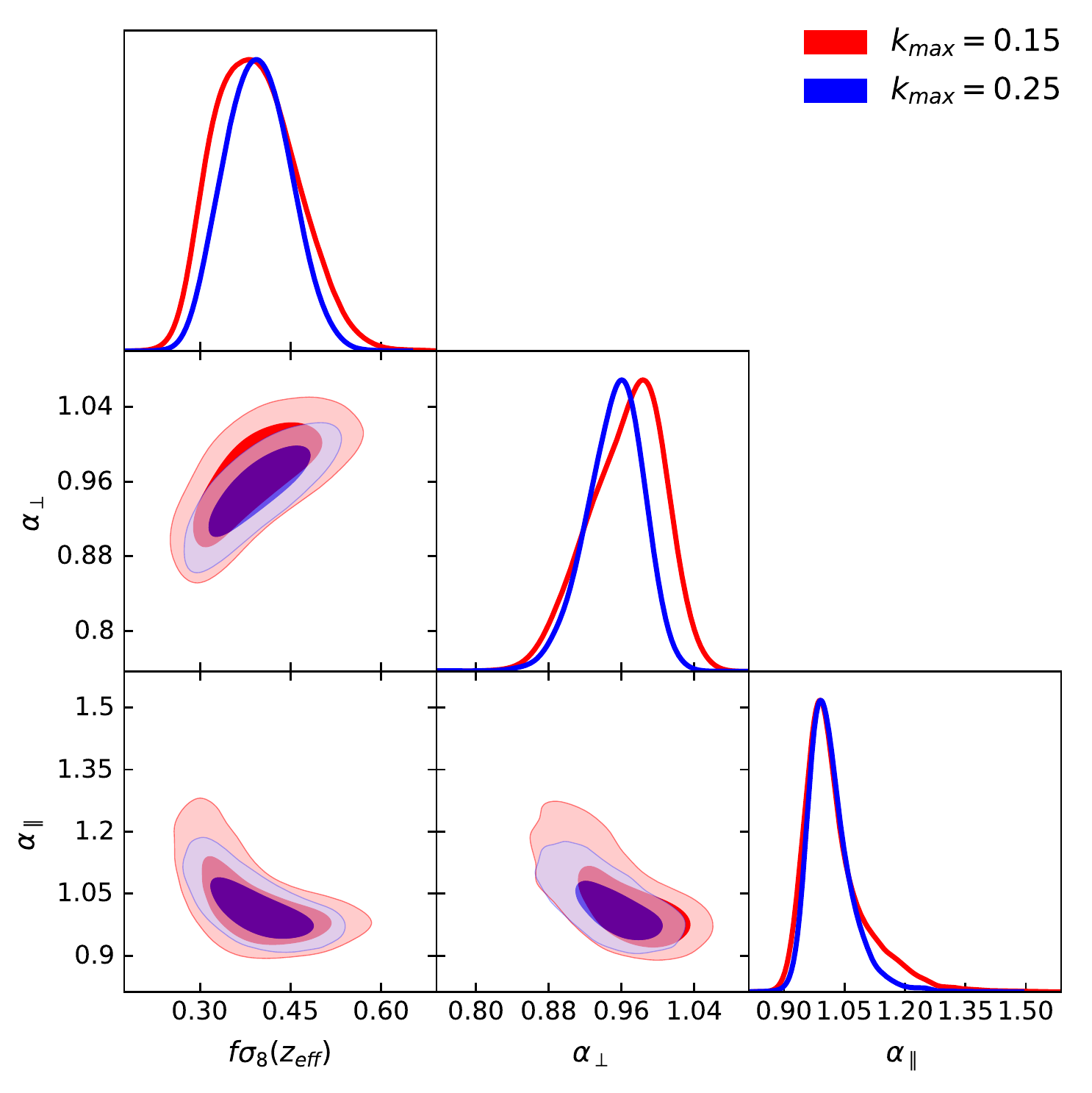}
\caption{ The results of our $\alpha$-analysis for the low-z (left) and high-z (right)
NGC BOSS samples. The values of $k_{\text{max}}$ are quoted in units [$h$/Mpc].}    
\label{fig:alphas}
\end{figure}

\begin{table}[ht!]
\begin{center}
 \begin{tabular}{|c|c|c|} 
 \hline
high-z & best-fit & mean  $\pm 1\sigma$  \\ [0.5ex] \hline\hline
 \multicolumn{3}{|c|}{$k_{\text{max}}=0.15$}\\  \hline
 $f\sigma_8(z_{\text{eff}})$ & $ 0.391$ & $ 0.392 \pm  0.068$  \\ \hline
 $\a_\parallel$ & $ 1.019$ & $ 1.027\pm  0.082$  \\ \hline
 $\a_\perp$ & $ 0.964$ & $ 0.964 \pm  0.042$  \\ \hline \hline
 $F_{\text{AP}}(z_{\text{eff}})$ & $ 0.688$ & $ 0.693 \pm 0.072 $  \\ \hline
 $H(z_{\text{eff}})$ & $ 93.42$ & $ 93.22 \pm  6.76$  \\ \hline
 $D_A(z_{\text{eff}})$ & $ 1370.89$ & $ 1382 \pm 60 $  \\ \hline
 $D_V(z_{\text{eff}})$ & $ 2120.62$ & $ 2135 \pm 48 $  \\ \hline
 \multicolumn{3}{|c|}{$k_{\text{max}}=0.25$}\\  \hline
 $f\sigma_8(z_{\text{eff}})$ & $ 0.430$ & $ 0.395 \pm  0.054$  \\ \hline
 $\a_\parallel$ & $ 0.985$ & $ 1.015\pm  0.057$  \\ \hline
 $\a_\perp$ & $ 0.972$ & $ 0.952 \pm  0.033$  \\ \hline \hline
 $F_{\text{AP}}(z_{\text{eff}})$ & $ 0.723$ & $  0.690 \pm 0.054 $  \\ \hline
 $H(z_{\text{eff}})$ & $ 96.69$ & $ 94.11 \pm  4.98$  \\ \hline
 $D_A(z_{\text{eff}})$ & $ 1393$ & $ 1364 \pm 47 $  \\ \hline
 $D_V(z_{\text{eff}})$ & $ 2118$ & $ 2109 \pm 40 $  \\ \hline
\end{tabular}
\vspace{0.2cm}
\begin{tabular}{|c|c|c|} 
 \hline
low-z & best-fit & mean  $\pm 1\sigma$  \\ [0.5ex] 
 \hline\hline
 \multicolumn{3}{|c|}{$k_{\text{max}}=0.15$}\\  \hline
 $f\sigma_8(z_{\text{eff}})$ & $ 0.478$ & $ 0.467 \pm  0.072$  \\ \hline
 $\a_\parallel$ & $ 1.026$ & $ 1.051 \pm  0.063$  \\ \hline
 $\a_\perp$ & $ 1.004$ & $ 0.994 \pm  0.037$  \\ \hline \hline
 $F_{\text{AP}}(z_{\text{eff}})$ & $ 0.414$ & $ 0.402 \pm 0.033 $  \\ \hline
 $H(z_{\text{eff}})$ & $ 80.86$ & $ 79.19 \pm  4.66$  \\ \hline
 $D_A(z_{\text{eff}})$ & $ 1112.71$ & $ 1101.77 \pm 40.56 $  \\ \hline
 $D_V(z_{\text{eff}})$ & $ 1492.12$ & $ 1493.22\pm 34.88 $  \\ \hline
 \multicolumn{3}{|c|}{$k_{\text{max}}=0.25$}\\  \hline
 $f\sigma_8(z_{\text{eff}})$ & $ 0.452$ & $ 0.440 \pm  0.048$  \\ \hline
 $\a_\parallel$ & $ 1.063$ & $ 1.075 \pm  0.044$  \\ \hline
 $\a_\perp$ & $ 0.967$ & $ 0.964 \pm  0.020$  \\ \hline \hline
 $F_{\text{AP}}(z_{\text{eff}})$ & $ 0.385$ & $ 0.380 \pm 0.020 $  \\ \hline
 $H(z_{\text{eff}})$ & $ 78.04$ & $ 77.24 \pm  3.12$  \\ \hline
 $D_A(z_{\text{eff}})$ & $ 1072$ & $ 1069 \pm 23 $  \\ \hline
 $D_V(z_{\text{eff}})$ & $ 1473$ & $ 1475\pm 22 $  \\ \hline
\end{tabular}
\caption{
The results of our $\alpha$-analysis for
the high-z NGC (left panel, $z_{\text{eff}}=0.61$) and low-z NGC (right panel, $z_{\text{eff}}=0.38$)
data samples. The values of $H$ are quoted in units of [km/s/Mpc], $D_A$ and $D_V$
in [Mpc].
}
\label{tab:alphas}
\end{center}
\end{table}

To match the standard methodology 
we modified our theoretical model 
to agree with the one used in the previous analyses. 
Specifically, we use
\be
\label{eq:tns}
P_g(k,\mu) = e^{-(f \mu \sigma_v k)^2 } P^{\text{1-loop, SPT, IR resummed}}_{g}(k,\mu)\,,
\ee
and do not introduce any RSD counterterms. The fingers-of-God effect 
is then described by only one parameter - the velocity dispersion
$\sigma_v$.
We have computed the theoretical power spectra 
for a fiducial cosmology with 
\be
\begin{split}
& n_s=0.96\quad \sigma_8 = 0.8\,, \quad h=0.676\,,\\
& \Omega_b h^2=0.022\,,\quad \Omega_m = 0.31\,, \quad \sum m_\nu = 0.06\,\eV\,,
\end{split}
\ee
which matches the parameters used in the most recent BOSS FS
analysis \cite{Beutler:2016arn}.
We account for the AP effect by means of the scaling parameters $\alpha_\parallel,\alpha_\perp$, 
defined as 
\be
\alpha_\parallel = \frac{H_\fid}{H}\Bigg|_{z_{\text{eff}}}\,,\qquad \a_\perp = \frac{D_A}{D_{A,\,\fid}}\Bigg|_{z_{\text{eff}}}\,.
\ee
Overall, the cosmology-dependence of this model is parameterized by 
$\alpha_\parallel,\alpha_\perp$ and $f\sigma_8$. 
The non-linear bias and redshift-space distortion effects are 
parameterized by coefficients $b_1\sigma_8$, $b_2\sigma_8$,
$P_{\text{shot}}$ and the velocity dispersion $\sigma_v$. 
We fix the tidal bias to the value suggested by the coevolution model \cite{Desjacques:2016bnm},
\be
b_{\mathcal{G}_2}=-\frac{4}{7}(b_1-1)\,. 
\ee
We use the same priors on the bias parameters as in our main analysis, except for $P_{\text{shot}}$, which is allowed to vary in the range $[-10^4,10^4]$ [Mpc/$h$]$^3$. This is done in order to agree with the analysis of Ref.~\cite{Beutler:2016arn},
which finds preferred values of the shot noise to be negative. 

We studied two different choices of $k_{\text{max}}$: $0.15$ and 
$0.25$ $h/\Mpc$. The results of these analyses are shown in Fig.~\ref{fig:alphas}, where we display the marginalized posterior 
contours for low-z (left panel) and high-z (right panel) NGC samples.
The 1d marginalized limits are presented in Table~\ref{tab:alphas}.
One clearly sees that the inferred distance parameters 
become shifted w.r.t. the fiducial values at $k_{\text{max}}=0.25\,h$/Mpc.
Moreover, the inferred values of the $H(z_{\text{eff}}),D_A(z_{\text{eff}})$ are more than 1$\sigma$-away 
from the Planck mean values. 
The distance measurements obtained with our $\alpha$-parametrization 
should be compared with the analysis of the DDE model,
which found $H$, $D_A$ to be very close to the Planck values.
We believe that this difference is mainly produced 
by our choice of priors and the use of a different theoretical model.

\bibliographystyle{JHEP}
\bibliography{short}

\end{document}